\documentclass[draft,tightenlines,nofootinbib,preprint,aps,eqsecnum,amsmath,amssymb]{revtex4}

\newcommand{\beq}{\begin{equation}}
\newcommand{\eeq}{\end{equation}}
\newcommand{\bea}{\begin{eqnarray}}
\newcommand{\eea}{\end{eqnarray}}
\newcommand{\cir}{{\buildrel \circ \over =}}
\newcommand{\sgn}{\epsilon}
\newcommand{\eo}{{}^4{\buildrel \circ \over E}}

\begin{document}

\title{Dust  in the York Canonical Basis of ADM Tetrad Gravity: the Problem of Vorticity.}

\medskip

\author{David Alba}

\affiliation{Sezione INFN di Firenze\\Polo Scientifico, via Sansone 1\\
 50019 Sesto Fiorentino, Italy\\
 E-mail alba@fi.infn.it}

\author{Luca Lusanna}

\affiliation{ Sezione INFN di Firenze\\ Polo Scientifico, Via Sansone 1\\
50019 Sesto Fiorentino (FI), Italy\\ E-mail: lusanna@fi.infn.it}

\today

\begin{abstract}

Brown's formulation of dynamical perfect fluids in Minkowski
space-time is extended to ADM tetrad gravity in globally hyperbolic,
asymptotically Minkowskian space-times. For the dust we get the
Hamiltonian description in closed form in the York canonical basis,
where we can separate the inertial gauge variables of the
gravitational field in the non-Euclidean 3-spaces of global
non-inertial frames from the physical tidal ones. After writing the
Hamilton equations of the dust, we identify the sector of
irrotational motions and the gauge fixings forcing the dust 3-spaces
to coincide with the 3-spaces of the non-inertial frame. The role of
the inertial gauge variable York time (the remnant of the clock
synchronization gauge freedom) is emphasized. Finally the
Hamiltonian Post-Minkowskian linearization is studied. The future
application of this formalism will be the study of cosmological
back-reaction (as an alternative to dark energy) in the York
canonical basis.

\end{abstract}

\maketitle

\vfill\eject

\section{Introduction}

By using Dirac's theory of constraints in a series of papers
\cite{1,2,3,4} we developed the Hamiltonian formulation of ADM
tetrad gravity in suitable globally hyperbolic, asymptotically
Minkowskian space-times admitting global 3+1 splittings (i.e. global
non-inertial frames) with dynamical matter represented by
positive-energy charged scalar point particles and by the
electro-magnetic field in the radiation gauge. By means of a
Shanmugadhasan canonical transformation, implementing the York map
\cite{5}, we found a York canonical basis adapted to ten of the
fourteen first class constraints of tetrad gravity and we could
identify the gauge variables (the {\it inertial effects}) and the
physical degrees of freedom (the {\it tidal effects}) of the
gravitational field in  global non-inertial frames (the only ones
allowed by the equivalence principle).
\medskip

In Refs.\cite{3,4} we also developed the Hamiltonian
Post-Minkowskian (HPM) linearization  in a family of (non-harmonic)
3-orthogonal Schwinger time gauges. In particular it turns out
\cite{4} that at least part of the astrophysical and cosmological
{\it dark matter} could arise as a {\it relativistic inertial
effect} determined by the trace of the extrinsic curvature (York
time) of the instantaneous 3-spaces as 3-manifolds embedded into
space-time. This inertial gauge variable is the general relativistic
remnant of the special relativistic gauge freedom in clock
synchronization as a definition of instantaneous 3-space. In
canonical general relativity the 3-spaces are dynamically determined
modulo this gauge freedom \cite{6}.
\bigskip

With the exception of the three gauge variables describing the
freedom in the choice of the 3-coordinates inside the instantaneous
3-spaces of the non-inertial frame all the other inertial and tidal
effects are 3-scalar fields on the 3-spaces in this Hamiltonian
formulation. The  3-scalar nature of the tidal variables and of many
of the inertial ones opens the possibility to study the averaging
over 3-volumes of most of the Hamilton equations (and not of only a
small subset as in the standard approaches), extending the results
of Buchert approach \cite{7} to cosmology in which the {\it
back-reaction} arising from the volume-average of the non-linear
Einstein equations is proposed as an alternative to {\it dark
energy}. It will be interesting to see which is the role of the York
time in a framework using neither 3-spaces comoving with an
irrotational fluid nor comoving coordinates (see Ref.\cite{8} for
the reformulation of the approach of Ref.\cite{7} in arbitrary
coordinate systems).\medskip

But before doing this attempt we need to include dynamical perfect
fluids in our framework, because this is the standard matter (very
often only test matter) in cosmology. Moreover we want to avoid the
use of irrotational test fluids in cosmological comoving
coordinates, because this simplification does not allow to make
explicit the gauge freedom in the York time.\medskip

In this paper we define the general relativistic extension of
Brown's approach \cite{9} to dynamical perfect fluids in Minkowski
space-time. Our starting point will be Refs.\cite{10,11}, where we
reformulated this approach in the framework of the inertial rest
frames, a subcase of the general treatment of special relativistic
non-inertial frames developed in Ref.\cite{12}.\medskip

We consider in detail dust, because, as shown in Refs.\cite{10},
only in this case (and in few others including the photon gas) we
can get a closed form for the Hamiltonian mass density of the fluid
\footnote{In general the transition from velocities to momenta in
this approach leads to a trascendental equation, whose explicit
solution is known in few cases.}.

\bigskip

In Section II we give a review of the 3+1 point of view which allows
to define global non-inertial frames with well defined instantaneous
3-spaces necessary for the formulation of the Cauchy problem of
field equations. After the special relativistic formulation based on
Ref.\cite{12}, we give its extension to the class of space-times
used in ADM tetrad gravity \cite{1,2} together with the definitions
of the tetrad and cotetrad fields. In both cases there are two
congruences of time-like observers associated with each 3+1
splitting of space-time: i) the congruence of the Eulerian
observers, whose 4-velocity is the unit normal to the 3-spaces; ii)
the skew, in general non surface-forming, congruence, whose
4-velocity is the evolution vector field associated with the 3+1
splitting of space-time.\medskip

After having clarified this kinematical scenario we review in
Section III Brown's approach \cite{9} to perfect fluids in Minkowski
space-time in the formulation of Ref.\cite{10} employing the
rest-frame instant form of the dynamics of isolated systems
\cite{12}. We show that we can get the explicit Hamiltonian
expression of the energy-momentum tensor and of the unit 4-velocity
of the fluid only for special equations of state, in particular for
dust. In the rest of the paper only the dust will be considered. The
dust is not irrotational so that its 4-velocity is not
surface-forming.\medskip

Then in Section IV we study the action of ADM tetrad gravity coupled
to dynamical matter used in Refs.\cite{1,2} with the matter action
of a dust. After having obtained the Hamiltonian formulation we make
the canonical transformation to the York canonical basis with the
resulting expressions for the 4-velocity, the energy-momentum tensor
and the dust Hamilton equations. Also the expressions in the family
of (non-harmonic) 3-orthogonal Schwinger time gauges, where the
spatial metric is diagonal, is given. Finally we present the
Hamiltonian Post-Minkowskian linearization of Refs.\cite{3,4} in the
case of dust.\medskip

\medskip

Section V contains a study of the skew congruence of time-like
observers defined by the unit 4-velocity of the dust (their
world-lines are the flux lines). It is shown that a gauge fixing of
the inertial shift functions of the gravitational field (it is a
restriction on the 3-coordinates on the 3-spaces) is needed to make
it to coincide with the skew congruence of the 3+1 splitting of
space-time. After analyzing the properties of the dust flux lines,
we introduce the Eulerian point of view for the description of the
dust following Ref.\cite{11}. Then we introduce the acceleration,
the expansion, the shear and the vorticity of the dust. We find
which are the two first class Hamiltonian constraints to be added by
hand to restrict the dust motions to the irrotational ones. These
motions are described only by one pair of canonical variables and
their 4-velocity is surface-forming: it can be made to coincide with
the unit normal to the 3-spaces of the 3+1 splitting by adding a
gauge fixing on the inertial lapse function (it is a restriction on
the York time, i.e. on the clock synchronization freedom). In this
gauge the irrotational dust is comoving with the 3-space and the
flux lines are the world-lines of the Eulerian observers. We studied
all these properties in detail, because usually these topics are not
discussed in the literature.
\medskip

In the Conclusions we make some comments on the future use of these
result in cosmology.
\medskip

In Appendix A there is a review of the York canonical basis of ADM
tetrad gravity \cite{1,2}, followed by the definition of the
acceleration, expansion and shear of the irrotational Eulerian
observers whose 4-velocity is the unit normal to the instantaneous
3-spaces.\medskip

In Appendix B there the discussion on the relation between
Lagrangian velocity and Hamiltonian momenta as a function of the
equation of state of the perfect fluid.

 \vfill\eject

\section{The 3+1 Point of View and Global Non-Inertial Frames
in Special and General Relativity}

In this Section we review some aspects of the theory of global
non-inertial frames, the only ones admitted in general relativity
due to the equivalence principle. Their formulation in Minkowski
space-time was given in Ref.\cite{12}. Then it was extended to
globally hyperbolic, asymptotically Minkowskian Einstein space-times
without super-translations in Refs.\cite{1,2}.

\subsection{Minkowski Space-Time}

As shown in Ref.\cite{12} we now have a metrology-oriented
description of non-inertial frames in special relativity. This can
be done with the {\it 3+1 point of view} and the use of
observer-dependent Lorentz-scalar radar 4-coordinates. Let us give
the world-line $x^{\mu}(\tau)$ of an arbitrary time-like observer
carrying a standard atomic clock: $\tau$ is an arbitrary
monotonically increasing function of the proper time of this clock.
Then we give an admissible 3+1 splitting of Minkowski space-time,
namely a nice foliation with space-like instantaneous 3-spaces
$\Sigma_{\tau}$: it is the mathematical idealization of a protocol
for clock synchronization (all the clocks in the points of
$\Sigma_{\tau}$ sign the same time of the atomic clock of the
observer). On each 3-space $\Sigma_{\tau}$ we choose curvilinear
3-coordinates $\sigma^r$ having the observer as origin. These are
the {\it radar 4-coordinates} $\sigma^A = (\tau; \sigma^r)$. If
$x^{\mu} \mapsto \sigma^A(x)$ is the coordinate transformation from
the Cartesian 4-coordinates $x^{\mu}$ of a reference inertial
observer to radar coordinates, its inverse $\sigma^A \mapsto x^{\mu}
= z^{\mu}(\tau ,\sigma^r)$ defines the {\it embedding} functions
$z^{\mu}(\tau ,\sigma^r)$ describing the 3-spaces $\Sigma_{\tau}$ as
embedded 3-manifold into Minkowski space-time. From now on we shall
denote the curvilinear 3-coordinates $\sigma^r$ with the notation
$\vec \sigma$ for the sake of simplicity.\medskip

The induced 4-metric on $\Sigma_{\tau}$ is the following functional
of the embedding: ${}^4g_{AB}(\tau , \vec \sigma) = [z^{\mu}_A\,
\eta_{\mu\nu}\, z^{\nu}_B](\tau , \vec \sigma)$, where $z^{\mu}_A =
\partial\, z^{\mu}/\partial\, \sigma^A$ and ${}^4\eta_{\mu\nu}$ is
the flat metric. The 4-metric ${}^4g_{AB}$ has signature $\sgn\,
(+---)$ with $\sgn = \pm$ (the particle physics, $\sgn = +$, and
general relativity, $\sgn = -$, conventions); the flat Minkowski
metric is $\eta_{\mu\nu} = \sgn\, (+---)$.\medskip

While the 4-vectors $z^{\mu}_r(\tau , \vec \sigma)$ are tangent to
$\Sigma_{\tau}$, so that the unit normal $l^{\mu}(\tau , \vec
\sigma)$ is proportional to $\epsilon^{\mu}{}_{\alpha \beta\gamma}\,
[z^{\alpha}_1\, z^{\beta}_2\, z^{\gamma}_3](\tau , \vec \sigma)$, we
have $z^{\mu}_{\tau}(\tau , \vec \sigma) = [N\, l^{\mu} + n^r\,
z^{\mu}_r](\tau , \vec \sigma)$ for the so-called evolution
4-vector, where $N(\tau , \vec \sigma) = 1 + n(\tau, \vec \sigma) =
\sgn\, [z^{\mu}_{\tau}\, l_{\mu}](\tau, \vec \sigma)$ and $n_r(\tau,
\vec \sigma) = - \sgn\, g_{\tau r}(\tau, \vec \sigma) =
[{}^3g_{rs}\, n^s](\tau, \vec \sigma)$ are the lapse and shift
functions. We also have $|det\, {}^4g| = (1 + n)\, \sqrt{\gamma}$;
$\sqrt{\gamma} = \sqrt{det\, {}^3g}$ with ${}^3g_{rs} = - \sgn\,
{}^4g_{rs}$ of positive signature.\medskip

The conditions for having an admissible 3+1 splitting of space-time
are:\hfill\break a) $1 + n(\tau ,\vec \sigma) > 0$ everywhere (the
instantaneous 3-spaces never intersect each other, so that there are
no coordinate singularities like it happen with Fermi
coordinates);\hfill\break b) the M$\o$ller conditions \cite{12},
which imply\hfill\break i) $\sgn\, {}^4g_{\tau\tau} > 0$, i.e. $(1 +
n)^2 > \sum_r\, n_r\, n^r$ (the rotational velocity never exceeds
the velocity of light $c$, so that there no coordinate singularities
like it happens with the rotating disk); \hfill\break ii) $\sgn\,
{}^4g_{rr} = - {}^3g_{rr} < 0$ (satisfied by the signature of
${}^3g_{rs}$), ${}^4g_{rr}\, {}^4g_{ss} - ({}^4g_{rs})^2
> 0$ and $ det\, \sgn\, {}^4g_{rs} = - det\, {}^3g_{rs} < 0$
(satisfied by the signature of ${}^3g_{rs}$) so that $det\,
{}^4g_{AB} < 0$ (these conditions imply that ${}^3g_{rs}$ has three
definite positive eigenvalues $\lambda_r = \Lambda_r^2$ in the
non-degenerate case without Killing symmetries, the only one we
consider).

\medskip

In this 3+1 point of view the embedding functions $z^{\mu}(\tau,
\vec \sigma)$ describe the inertial effects present in the given
non-inertial frame. The 4-metric ${}^4g_{AB}(\tau, \vec \sigma)$ is
the potential for the induced inertial effects. For instance the
extrinsic curvature ${}^3K_{rs}(\tau, \vec \sigma) = \Big({1\over
{2\, (1 + n)}}\, (n_{r|s} + n_{s|r} - \partial_{\tau}\,
{}^3g_{rs})\Big)(\tau, \vec \sigma)$ of the non-Euclidean 3-spaces
$\Sigma_{\tau}$ ($|r$ denotes the covariant derivative in it) is one
of these induced inertial effects. All these inertial effects derive
from the gauge freedoms of clock synchronization and choice of the
3-coordinates.

\medskip

In Ref.\cite{12} there is a complete description of isolated systems
in non-inertial frames by means of {\it parametrized Minkowski
theories} where there is a well defined action principle containing
the embeddings $z^{\mu}(\tau, \vec \sigma)$ as Lagrangian variables
and implying their gauge nature.

\medskip

In special relativity we can restrict ourselves to inertial frames
and define the {\it inertial rest-frame instant form of dynamics for
isolated systems} by choosing the 3+1 splitting corresponding to the
intrinsic inertial rest frame of the isolated system centered on an
inertial observer: the instantaneous 3-spaces, named Wigner 3-space
due to the fact that the 3-vectors inside them are Wigner spin-1
3-vectors \cite{12}, are orthogonal to the conserved 4-momentum
$P^{\mu}$ of the configuration. In Ref.\cite{12} there is the
extension to admissible {\it non-inertial rest frames}, where
$P^{\mu}$ is orthogonal to the asymptotic space-like hyper-planes to
which the instantaneous 3-spaces tend at spatial infinity. This
non-inertial family of 3+1 splittings is the only one admitted by
the asymptotically Minkowskian space-times covered by the canonical
gravity formulation of Refs.\cite{1,2,3,4}.

\subsection{Globally Hyperbolic, Asymptotically Minkowskian Einstein
Space-times}

In the globally hyperbolic, asymptotically Minkowskian space-times
without super-translations of Ref.\cite{1} we can use radar
4-coordinates $\sigma^A = (\sigma^{\tau} = \tau ; \sigma^r)$, $A =
\tau ,r$, adapted to an admissible 3+1 splitting of the space-time
and centered on an arbitrary time-like observer like in special
relativity: they define a non-inertial frame centered on the
observer. The absence of super-translations implies \cite{13} that
the instantaneous 3-spaces  $\Sigma_{\tau}$ are orthogonal to the
conserved ADM 4-momentum at spatial infinity, i.e. they are
non-inertial rest frames of the 3-universe. Therefore at spatial
infinity there is an {\it asymptotic Minkowski background 4-metric}
and there are asymptotic inertial observers to be identified with
the fixed stars of the star catalogues.

\medskip

The 3-spaces $\Sigma_{\tau}$ are identified by the embeddings
$x^{\mu} = z^{\mu}(\tau, \vec \sigma)$, but now the quantities
$z^{\mu}_A(\tau, \vec \sigma) = {{\partial\, z^{\mu}(\tau, \vec
\sigma)}\over {\partial\, \sigma^A}}$ are the transition
coefficients from the adapted radar 4-coordinates $\sigma^A = (\tau,
\sigma^r)$ to  world 4-coordinates $x^{\mu}$ (${}^4g_{AB} =
z^{\mu}_A\, {}^4g_{\mu\nu}\, z^{\nu}_B$). As in special relativity
the space-like 4-vectors $z^{\mu}_r(\tau, \vec \sigma)$ are tangent
to the 3-spaces, the unit normal to them is $l^{\mu}(\tau, \vec
\sigma) = \Big(z^{\mu}_A\, l^A\Big)(\tau, \vec \sigma) =
\Big({{{}^4g^{\mu\nu}\, \sqrt{|det\, {}^4g_{\rho\sigma}|}}\over
{\tilde \phi}}\, \epsilon_{\mu\alpha\beta\gamma}\, z^{\alpha}_1\,
z^{\beta}_2\, z^{\gamma}_3\Big)(\tau, \vec \sigma)$ and the
time-like evolution 4-vector $z^{\mu}_{\tau}(\tau, \vec \sigma)$ has
the decomposition $z^{\mu}_{\tau}(\tau, \vec \sigma) = \Big((1 +
n)\, l^{\mu} + n^r\, z^{\mu}_r\Big)(\tau, \vec \sigma)$, with
$v^{\mu}(\tau, \vec \sigma) = \Big(z^{\mu}_A\, v^A\Big)(\tau, \vec
\sigma) = \Big(z^{\mu}_{\tau}/\sqrt{\sgn\, {}^4g_{\mu\nu}\,
z^{\mu}_{\tau}\, z^{\nu}_{\tau}}\Big)(\tau, \vec \sigma)$  the
associated unit time-like 4-vector (skew with respect to the
foliation with 3-spaces).

\bigskip

In general relativity, where the 3-spaces are dynamically determined
\cite{6} except for the York time ${}^3K$ (the general relativistic
remnant of the special relativistic gauge freedom in clock
synchronization; differently from special relativity ${}^3K(\tau,
\vec \sigma)$ is an independent inertial gauge variable and not an
induced inertial effect), it is convenient to use the embedding
$z^{\mu}(\tau, \vec \sigma) = x^{\mu}_o + \epsilon^{\mu}_{\tau}\,
\tau + \epsilon^{\mu}_r\, \sigma^r$, where $\epsilon^{\mu}_A$ are
asymptotic flat tetrads ($\epsilon^{\mu}_{\tau}$ is orthogonal to
the asymptotic flat Euclidean 3-space, being proportional to the
conserved ADM 4-momentum).\bigskip

In Refs. \cite{1,2,3,4}(see also Appendix A) we introduced ADM
tetrad gravity by considering the ADM action as a functional of
cotetrads ${}^4E^{(\alpha)}_A(\tau, \vec \sigma) = z^{\mu}_A(\tau,
\vec \sigma)\, {}^4E^{(\alpha)}_{\mu}(\tau, \vec \sigma)$
\footnote{Flat indices $(\alpha )$, $\alpha = o, a$, are raised and
lowered by the flat Minkowski metric ${}^4\eta_{(\alpha )(\beta )} =
\sgn\, (+---)$. We define ${}^4\eta_{(a)(b)} = - \sgn\,
\delta_{(a)(b)}$ with a positive-definite Euclidean 3-metric.} by
using the following decomposition of the 4-metric in radar
4-coordinates

\bea
  {}^4g_{AB} &=& {}^4E^{(\alpha )}_A\, {}^4\eta_{(\alpha
 )(\beta )}\, {}^4E^{(\beta )}_B = \eo^{(\alpha )}_A\, {}^4\eta_{(\alpha
 )(\beta )}\, \eo^{(\beta )}_B,\nonumber \\
 &&{}\nonumber \\
 {}^4E^{(\alpha)}_A &=& L^{(\alpha)}{}_{(\beta)}(\varphi_{(c)})\,
 \eo^{(\beta)}_A = L^{(\alpha)}{}_{(o)}(\varphi_{(c)})\,
 {}^4{\buildrel \circ \over {\bar E}}_A^{(o)} +
 L^{(\alpha)}_{(a)}(\varphi_{(c)})\, R^T_{(a)(b)}(\alpha_{(d)})\,
 {}^4{\buildrel \circ \over {\bar E}}_A^{(b)}.\nonumber \\
 &&{}
 \label{2.1}
 \eea

\bigskip

The cotetrads $\eo^{(\alpha)}_A$ are adapted to the 3+1 splitting
(the time-like adapted tetrad is the unit normal $l^A$ to
$\Sigma_{\tau}$) by a point-dependent standard Lorentz boost for
time-like orbits acting on the flat indices parametrized by
$\varphi_{(a)}(\tau, \vec \sigma)$ \footnote{As shown in
Ref.\cite{1,13}, the flat indices $(a)$ of the adapted tetrads and
cotetrads and of the triads and cotriads on $\Sigma_{\tau}$
transform as Wigner spin-1 indices under the point-dependent SO(3)
Wigner rotations  associated with Lorentz transformations
$\Lambda^{(\alpha )}{}_{(\beta )}(z)$ in the tangent plane to the
space-time in the given point of $\Sigma_{\tau}$. Instead the index
$(o)$ of the adapted tetrads and cotetrads is a local Lorentz scalar
index.}. The adapted tetrads and cotetrads  (corresponding to the so
called {\it Schwinger time gauges}) have the expression

\bea
 \eo^A_{(o)} &=& {1\over {1 + n}}\, (1; - n_{(a)}\,
 {}^3e^r_{(a)}) = l^A,\qquad \eo^A_{(a)} = (0; {}^3e^r_{(a)}), \nonumber \\
 &&{}\nonumber  \\
 \eo^{(o)}_A &=& (1 + n)\, (1; \vec 0) = \sgn\, l_A,\qquad \eo^{(a)}_A
= (n_{(a)}; {}^3e_{(a)r}),
 \label{2.2}
 \eea

\noindent where ${}^3e^r_{(a)}(\tau, \vec \sigma)$ and
${}^3e_{(a)r}(\tau, \vec \sigma)$ are triads and cotriads on
$\Sigma_{\tau}$ and $n_{(a)}(\tau, \vec \sigma) = \Big(n_r\,
{}^3e^r_{(a)}\Big)(\tau, \vec \sigma) = \Big(n^r\,
{}^3e_{(a)r}\Big)(\tau, \vec \sigma)$ \footnote{Since we use the
positive-definite 3-metric $\delta_{(a)(b)} $, we shall use only
lower flat spatial indices. Therefore for the cotriads we use the
notation ${}^3e^{(a)}_r\,\, {\buildrel {def}\over =}\, {}^3e_{(a)r}$
with $\delta_{(a)(b)} = {}^3e^r_{(a)}\, {}^3e_{(b)r}$.} are adapted
shift functions vanishing at spatial infinity.

\bigskip

The adapted tetrads $\eo^A_{(a)}$ and triads ${}^3e^r_{(a)}$ are
defined modulo SO(3) rotations $\eo^A_{(a)} =
R_{(a)(b)}(\alpha_{(e)})\, {}^4{\buildrel \circ \over {\bar
E}}^A_{(b)}$, ${}^3e^r_{(a)} = R_{(a)(b)}(\alpha_{(e)})\, {}^3{\bar
e}^r_{(b)}$, where $\alpha_{(a)}(\tau ,\vec \sigma )$ are three
point-dependent Euler angles. After having chosen an arbitrary
point-dependent origin $\alpha_{(a)}(\tau ,\vec \sigma ) = 0$, we
arrive at the following adapted tetrads and cotetrads [${\bar
n}_{(a)} = \sum_b\, n_{(b)}\, R_{(b)(a)}(\alpha_{(e)})\,$]

\bea
 {}^4{\buildrel \circ \over {\bar E}}^A_{(o)}
 &=& \eo^A_{(o)} = {1\over {1 + n}}\, (1; - {\bar n}_{(a)}\,
 {}^3{\bar e}^r_{(a)}) = l^A,\qquad {}^4{\buildrel \circ \over
 {\bar E}}^A_{(a)} = (0; {}^3{\bar e}^r_{(a)}), \nonumber \\
 &&{}\nonumber  \\
 {}^4{\buildrel \circ \over {\bar E}}^{(o)}_A
 &=& \eo^{(o)}_A = (1 + n)\, (1; \vec 0) = \sgn\, l_A,\qquad
 {}^4{\buildrel \circ \over {\bar E}}^{(a)}_A
 = {}^4{\buildrel \circ \over {\bar E}}_{(a)A}
 = ({\bar n}_{(a)}; {}^3{\bar e}_{(a)r}).
 \label{2.3}
 \eea

\bigskip

The future-oriented unit normal to $\Sigma_{\tau}$ and the projector
on $\Sigma_{\tau}$ are\medskip

\bea
 l_A &=& \sgn\, (1 + n)\, \Big(1;\, 0\Big),\qquad {}^4g^{AB}\, l_A\, l_B =
\sgn ,\nonumber \\
 &&{}\nonumber \\
 l^A &=& \sgn\, (1 + n)\, {}^4g^{A\tau} = {1\over {1 + n}}\, \Big(1;\, - n^r\Big) =
{1\over {1 + n}}\, \Big(1;\, - {\bar n}_{(a)}\, {}^3{\bar e}_{(a)}^r\Big),\nonumber \\
 &&{}\nonumber \\
 {}^3h^B_A &=& \delta^B_A - \sgn\, l_A\, l^B,\qquad
 {}^3h^{\tau}_{\tau} = {}^3h^{\tau}_r = 0,\quad {}^3h^r_{\tau} =
 {\bar n}_{(a)}\, {}^3{\bar e}^r_{(a)},\quad {}^3h^r_s = \delta^r_s,\nonumber \\
 &&{}\nonumber \\
 &&{}^3h_{\tau\tau} = - \sgn\, \sum_a\, {\bar n}^2_{(a)},\quad {}^3h_{\tau
 r} = - \sgn\, {\bar n}_{(a)}\, {}^3{\bar e}_{(a)r},\quad {}^3h_{rs} = - \sgn\,
 {}^3{\bar e}_{(a)r}\, {}^3{\bar e}_{(a)s},\nonumber \\
 &&{}^3h^{\tau\tau} = {}^3h^{\tau r} = 0,\qquad {}^3h^{rs} = -
 \sgn\, {}^3{\bar e}^r_{(a)}\, {}^3{\bar e}^s_{(a)}.\nonumber \\
 &&{}
 \label{2.4}
 \eea

\bigskip

The 4-metric has the following expression (the lapse and shift
function are independent inertial gauge variables)

 \bea
 {}^4g_{\tau\tau} &=& \sgn\, [(1 + n)^2 - {}^3g^{rs}\, n_r\,
 n_s] = \sgn\, [(1 + n)^2 - \sum_a\, n^2_{(a)}] =
 \sgn\, [(1 + n)^2 - \sum_a\, {\bar n}^2_{(a)}],\nonumber \\
 {}^4g_{\tau r} &=& - \sgn\, n_r = -\sgn\, n_{(a)}\,
 {}^3e_{(a)r} = - \sgn\, {\bar n}_{(a)}\, {}^3{\bar e}_{(a)r},\nonumber \\
  {}^4g_{rs} &=& -\sgn\, {}^3g_{rs},\nonumber \\
 &&{}\nonumber \\
 &&{}^3g_{rs} = {}^3e_{(a)r}\, {}^3e_{(a)s}
 = {}^3{\bar e}_{(a)r}\, {}^3{\bar e}_{(a)s},\qquad {}^3g^{rs} =
 {}^3h^{rs} = {}^3e^r_{(a)}\, {}^3e^s_{(a)} = {}^3{\bar e}^r_{(a)}\,
 {}^3{\bar e}^s_{(a)},\nonumber \\
 &&{}\nonumber \\
 {}^4g^{\tau\tau} &=& {{\sgn}\over {(1 + n)^2}},\qquad
  {}^4g^{\tau r} = -\sgn\, {{n^r}\over {(1 + n)^2}} = -\sgn\, {{{}^3e^r_{(a)}\,
 n_{(a)}}\over {(1 + n)^2}} = -\sgn\, {{{}^3{\bar e}^r_{(a)}\,
 {\bar n}_{(a)}}\over {(1 + n)^2}},\nonumber \\
 {}^4g^{rs} &=& -\sgn\, ({}^3g^{rs} - {{n^r\, n^s}\over
 {(1 + n)^2}}) = -\sgn\, {}^3e^r_{(a)}\, {}^3e^s_{(b)}\, (\delta_{(a)(b)} -
 {{n_{(a)}\, n_{(b)}}\over {(1 + n)^2}}) =\nonumber \\
 &=& -\sgn\, {}^3{\bar e}^r_{(a)}\, {}^3{\bar e}^s_{(b)}\, (\delta_{(a)(b)} -
 {{{\bar n}_{(a)}\, {\bar n}_{(b)}}\over {(1 + n)^2}}),\qquad {}^3g^{rs} = h^{rs},\nonumber \\
 &&{}\nonumber \\
 &&{}^4g^{\tau\tau}\, {}^4g^{rs} - {}^4g^{\tau r}\, {}^4g^{\tau s} = -
{{{}^3g^{rs}}\over {(1 + n)^2}},\qquad {}^3g = \gamma =
({}^3e)^2,\quad {}^3e = det\, {}^3e_{(a)r},\nonumber \\
 &&\sqrt{- g } = \sqrt{|{}^4g|} = {{\sqrt{{}^3g}}\over {\sqrt{\sgn\,
{}^4g^{\tau\tau}}}} = \sqrt{\gamma}\, (1 + n) = {}^3e\, (1 + n).
 \label{2.5}
 \eea

\bigskip

The 3-metric ${}^3g_{rs}$ has signature $(+++)$, so that we may put
all the flat 3-indices {\it down}. We have ${}^3g^{ru}\, {}^3g_{us}
= \delta^r_s$, $\partial_A\, {}^3g^{rs} = - {}^3g^{ru}\,
{}^3g^{sv}\, \partial_A\, {}^3g_{uv}$.

\subsection{The Two Congruences of Time-like Observers associated
with Non-Inertial Frames}

Each 3+1 splitting of space-time, i.e. each global non-inertial
frame, has two associated congruences of time-like
observers:\hfill\break

i) The congruence of the Eulerian observers with the unit normal
$l^{\mu}(\tau, \vec \sigma) = \Big(z^{\mu}_A\, l^A\Big)(\tau, \vec
\sigma)$ to the 3-spaces as unit 4-velocity. The world-lines of
these observers are the integral curves of the unit normal and in
general are not geodesics. In adapted radar 4-coordinates the
contro-variant ($l^A(\tau, \vec \sigma)$, ${}^4{\buildrel \circ
\over {\bar E}}^A_{(a)}(\tau, \vec \sigma)$) and covariant
($l_A(\tau, \vec \sigma)$, ${}^4{\buildrel \circ \over {\bar
E}}_{(a)A}(\tau, \vec \sigma)$) orthonormal tetrads carried by the
Eulerian observers are given in Eqs.(\ref{2.3}).

\bigskip

ii) The skew congruence with unit 4-velocity $v^{\mu}(\tau, \vec
\sigma) = \Big(z^{\mu}_A\, v^A\Big)(\tau, \vec \sigma)$ (in general
it is not surface-forming, i.e. it has a non-vanishing vorticity).
The observers of the skew congruence have the world-lines (integral
curves of the 4-velocity) defined by $\sigma^r = const.$ for every
$\tau$, because the unit 4-velocity tangent to the flux lines
$x^{\mu}_{{\vec \sigma}_o}(\tau) = z^{\mu}(\tau, {\vec \sigma}_o)$
is $v^{\mu}_{{\vec \sigma}_o}(\tau) = z^{\mu}_{\tau}(\tau, {\vec
\sigma}_o)/\sqrt{\sgn\, {}^4g_{\tau\tau}(\tau, {\vec \sigma}_o)}$.
They carry the adapted contro-variant orthonormal tetrads (${\cal
V}^A_{(a)}(\tau, \vec \sigma)$ are not tangent to the 3-spaces
$\Sigma_{\tau}$ like ${}^4{\buildrel \circ \over {\bar
E}}^A_{(a)}(\tau, \vec \sigma)$ of Eqs.(\ref{2.3}))

\bea
 v^A(\tau, \vec \sigma) &=& {{(1; 0)}\over {\sqrt{(1 + n)^2 - \sum_a\,
 {\bar n}^2_{(a)}}}}(\tau, \vec \sigma),\nonumber \\
 {\cal V}^A_{(a)}(\tau, \vec \sigma) &=&  \Big({{{\bar
 n}_{(a)}}\over {(1 + n)^2}}; (\delta_{(a)(b)} - {{{\bar n}_{(a)}\,
 {\bar n}_{(b)}}\over {(1 + n)^2}})\, {}^3{\bar e}^r_{(b)}\Big)(\tau,
 \vec \sigma).
 \label{2.6}
 \eea

\noindent The covariant version of these tetrads is

\bea
 \sgn\, v_A(\tau, \vec \sigma) &=&  \Big(\sqrt{(1 + n)^2 - \sum_c\,
 {\bar n}^2_{(c)}}; {{- {\bar n}_{(a)}\, {}^3{\bar e}_{(a)r}}\over
 {\sqrt{(1 + n)^2 - \sum_c\, {\bar n}^2_{(c)}}}}\Big)(\tau, \vec
 \sigma),\nonumber \\
  {\cal V}_{(a)A}(\tau, \vec \sigma) &=& \Big(0; {}^3{\bar
 e}_{(a)r}\Big)(\tau, \vec \sigma).
 \label{2.7}
 \eea

In each point there is a Lorentz transformation connecting these
tetrads to the ones of the Eulerian observer present in this point.
\medskip

When there is a perfect fluid with unit time-like 4-velocity
$U^A(\tau, \vec \sigma)$, there is also the congruence of the
time-like flux curves: in general it is not surface-forming and it
is independent from the previous two congruences. If $\Big(U^A(\tau,
\vec \sigma); {\cal U}^A_{(a)}(\tau, \vec \sigma)\Big)$ is an
orthonormal tetrad carried by a flux line, the connection of these
4-vectors to the orthonormal tetrad of the Eulerian observers
\footnote{Analogous expressions hold in terms of the tetrads
(\ref{2.6}) of the observers of the skew congruence. See
Eqs.(\ref{5.3}) for the gauge fixings implying $U^A(\tau, \vec
\sigma) \approx v^A(\tau, \vec \sigma)$.} is

\bea
 U^A(\tau, \vec \sigma) &=& \Gamma\, \Big( l^A + \sum_a\, \beta_{(a)}\,
 {}^4{\buildrel \circ \over {\bar E}}^A_{(a)}\Big)(\tau, \vec
 \sigma),\nonumber \\
 {\cal U}^A_{(a)}(\tau, \vec \sigma) &=& \Big(t_{(a)}\, l^A + \sum_b\,
 \gamma_{(a)(b)}\, {}^4{\buildrel \circ \over
 {\bar E}}^A_{(b)}\Big)(\tau, \vec \sigma),
 \label{2.8}
 \eea

\noindent with $t_{(a)}(\tau, \vec \sigma) = \Big( \sum_b\,
\gamma_{(a)(b)}\, \beta_{(b)}\Big)(\tau, \vec \sigma)$ and
$\Big[\sum_{cd}\, \Big(\delta_{(c)(d)} - \beta_{(c)}\,
\beta_{(d)}\Big)\, \gamma_{(a)(c)}\, \gamma_{(b)(d)}\Big](\tau, \vec
\sigma) = \delta_{(a)(b)}$. See Section V for the gauge fixings
implying $U^A(\tau, \vec \sigma) \approx l^A(\tau, \vec \sigma)$.

\medskip

In Ref.\cite{8} (see also Refs. \cite{14}) the results of
Ref.\cite{7} are reformulated in arbitrary coordinates with the
following change of notation

\bea
 U^A &=& \Gamma(w)\, \Big(l^A + m^A\Big) = \Gamma(w)\, \Big(1;
 - [\sum_a\, {\bar n}_{(a)}\, {}^3{\bar e}^r_{(a)} + (1 + n)\,
 w^r]\Big),\nonumber \\
 &&m^A = \sum_a\, \beta_{(a)}\,
 {}^4{\buildrel \circ \over {\bar E}}^A_{(a)} = \Big(0; - w^r =
 \sum_a\, \beta_{(a)}\, {}^3{\bar e}^r_{(a)}\Big),\nonumber \\
 &&{}\nonumber \\
 U_A &=& \Gamma(w)\, \Big(l_A + m_A\Big) = \sgn\, \Gamma(w)\,
 \Big(1 + n + \sum_{as}\, {\bar n}_{(a)}\, {}^3{\bar e}^s_{(a)}\,
  w_s; w_r\Big),\nonumber \\
  &&m_A = \sgn\, \Big(\sum_{as}\, {\bar n}_{(a)}\, {}^3{\bar e}^s_{(a)}\,
  w_s; w_r\Big), \qquad m_A\, l^A = 0,\nonumber \\
  &&\Gamma(w)=\frac{1}{\sqrt{1 - {}^3g^{rs}w_r\,w_s}} = \frac{1}{\sqrt{1
  - {}^3g_{rs}w^r\,w^s}} = \sqrt{1 + {}^3g^{rs}\, U_r\, U_s},\nonumber \\
  &&w_r = \sgn\, {{U_r}\over {\sqrt{1 + {}^3g^{rs}\, U_r\, U_s}}},
  \qquad w^r = {}^3g^{rs}\, w_s.\nonumber \\
  &&{}
 \label{2.9}
 \eea

\bigskip

When the vorticity of the fluid vanishes, so that its 4-velocity is
surface forming, there is a 3+1 splitting of space-time determined
by the irrotational fluid. While in special relativity we can always
choose a global non-inertial frame coinciding with these 3+1
splitting, in general relativity we have to show that there is a
gauge fixing on the inertial gauge variable ${}^3K(\tau, \vec
\sigma)$ (the York time) allowing this identification (see Section
V).

\medskip

Let us remark that Eqs.(\ref{2.8}) establish the bridge between our
3+1 point of view and the 1+3 point of view of Refs.\cite{15}, where
one describes both the gravitational field and the matter as seen by
a generic family of observers with 4-velocity $U^A(\tau, \vec
\sigma)$. Most of the results in cosmology (see for instance
Refs.\cite{16}) are presented in the 1+3 framework. However, in the
1+3 point of view vorticity is an obstruction to formulate the
Cauchy problem (3-spaces are not existing; each observer uses as
rest frame the tangent 3-space orthogonal to the 4-velocity) and
there is no natural way to identify the inertial gravitational gauge
variables of the Hamiltonian formalism based on Dirac's constraint
theory (see also Appendix A of paper I for the induced treatment of
the non-Hamiltonian ADM equations).

\vfill\eject

\section{Perfect Fluids in Minkowski Space-Time}

In this Section we will review Brown approach \cite{9} to perfect
fluids in Minkowski space-time. In Ref.\cite{9} a set of space-time
scalar fields ${\tilde \alpha}^i(x)$, $i=1,2,3,$ was interpreted as
{\it Lagrangian (or comoving) coordinates for the fluid}. They label
the fluid flow lines (physically determined by the average particle
motions) passing through the points inside the boundary. In an
allowed 3+1 splitting of Minkowski space-time, we use the three
scalar fields $\alpha^i(\tau ,\vec \sigma ) = {\tilde \alpha}^i(z
(\tau ,\vec \sigma )) $ for the Lagrangian coordinates of the fluid
in the instantaneous 3-space $\Sigma_{\tau}$. Now either the
$\alpha^i(\tau ,\vec \sigma )$'s have a compact boundary
$V_{\alpha}(\tau ) \subset \Sigma_{\tau}$ or have boundary
conditions at spatial infinity. For each value of $\tau$, one could
invert $\alpha^i=\alpha^i(\tau ,\vec \sigma )$ to $\sigma^r =
\sigma^r(\tau ,\alpha^i)$ and use the $\alpha^i$'s as a special
coordinate system on $\Sigma_{\tau}$ inside the support
$V_{\alpha}(\tau )\subset \Sigma_{\tau}$: $z^{\mu}(\tau ,\vec \sigma
(\tau ,\alpha^i))={\check z}^{\mu}(\tau ,\alpha^i)$.

\medskip

The {\it unit time-like 4-velocity field} $U^A(\tau, \vec \sigma)$
of the fluid is a derived function of the Lagrangian coordinates of
the fluid. In this way, as shown in Ref.\cite{9}, one has not to add
new quantities (like the Clebsch potentials) with Lagrange
multipliers to the action to implement the particle number
conservation and the absence of entropy exchange between neighboring
flow lines: they are automatically satisfied as a consequence of the
comoving nature of these Lagrangian coordinates, which implies

\beq
 \Big(U^A\, \partial_A\, \alpha^i\Big)(\tau, \vec \sigma) = 0.
 \label{3.1}
 \eeq
\medskip

Instead of the 4-velocity it is convenient to introduce a {\it
material current} \cite{9,10} $J^A(\alpha^i(\tau, \vec \sigma)) =
\sqrt{|{}^4g(\tau, \vec \sigma)|}\, \tilde n(\tau, \vec \sigma)\,
U^A(\tau, \vec \sigma)$, $\Big(J^A\, \partial_A\,
\alpha^i\Big)(\tau, \vec \sigma) = 0$, where $\tilde n(\tau, \vec
\sigma)$ is the particle number density. This material current has
the following form \footnote{Here $J^{\tau}$ has the opposite sign
with respect to Ref.\cite{10}. We have $\epsilon^{ruv}\,
\epsilon_{ijk}\, \partial_m\, \alpha^i\, \partial_u\, \alpha^j\,
\partial_v\, \alpha^k = 2\, \delta^r_m\, det\, (\partial_u\,
\alpha^k) = 2\, J^{\tau}\, \delta^r_m$ and $\epsilon^{ruv}\,
\epsilon_{ijk}\, \partial_r\, \alpha^h\, \partial_u\, \alpha^j\,
\partial_v\, \alpha^k = 2\, J^{\tau}\, \delta^h_i$.}

\bea
 J^{A}(\alpha^i(\tau ,\vec \sigma ))&=&[ (1 + n)\, \sqrt{\gamma}\,
 \tilde n\, U^{ A}](\tau ,\vec \sigma ),\qquad [J^{\tau}\, \partial_{\tau}\, \alpha^i
 + J^r\, \partial_r\, \alpha^i](\tau, \vec \sigma) = 0,
 \nonumber \\
  &&{}\nonumber \\
 J^{\tau}(\alpha^i(\tau ,\vec \sigma ))&=&[\epsilon^{ruv}\,
 \partial _{r}\, \alpha^1\, \partial_{u}\, \alpha^2\, \partial_{v}\,
 \alpha^3](\tau ,\vec \sigma ) =\nonumber \\
 &=& {1\over 6}\,\epsilon^{ruv}\, \epsilon_{ijk}\, [\partial_r\, \alpha^i\,
 \partial_u\, \alpha^j\, \partial_v\, \alpha^k](\tau, \vec \sigma) =
 det\, ( \partial_{r}\,\alpha^i )(\tau, \vec \sigma ),\nonumber \\
 J^{ r}(\alpha^i(\tau ,\vec \sigma ))&=& -
 {1\over 2}\, \epsilon^{ruv}\, \epsilon_{ijk}\, [\partial_{\tau}\,
  \alpha^i\, \partial_{u}\, \alpha^j\, \partial_{v}\, \alpha^k](\tau
  ,\vec \sigma ),\nonumber \\
  &&{}\nonumber \\
  {\buildrel {(\ref{3.1})}\over \Rightarrow}
  && \partial_{\tau}\, \alpha^i =  - {{J^r}\over
  {J^{\tau}}}\, \partial_r\, \alpha^i = - {{U^r}\over {U^{\tau}}}\,
  \partial_r\, \alpha^i,
 \label{3.2}
  \eea

\noindent with ${\cal N} = \int_{V_{\alpha}(\tau )} d^3\sigma\,
J^{\tau}(\alpha^i(\tau ,\vec \sigma ))$ giving the conserved
particle number and with $\int_{V _{\alpha}(\tau )} d^3\sigma (s\,
J^{\tau})(\tau ,\vec \sigma )$ giving the conserved entropy per
particle.\medskip

This shows that the fluid flow lines, whose tangent vector field is
the fluid 4-velocity time-like vector field $U^A$, are identified by
$ \alpha^i = const.$ and that the particle number conservation is
automatic. Moreover, if the entropy for particle is a function only
of the fluid Lagrangian coordinates, $s = s( \alpha^i)$, the assumed
form of $J^A$ also implies automatically the absence of entropy
exchange between neighboring flow lines, $\partial_A\, (\tilde n\,
s\, U^A) = 0$. Since $U^A\, \partial_A s(\alpha^i ) = 0$, the
perfect fluid is locally adiabatic; instead for an isentropic fluid
we have $\partial_A s = 0$, namely $s = const.$.
\medskip

Since $U^A$ is a unit 4-vector, we have the following expression for
the particle number density and for the unit 4-velocity in terms of
the material current

\bea
  \tilde n(\tau, \vec \sigma) &=& {{|J|}\over {(1 + n)\, \sqrt{\gamma}}}
(\tau ,\vec \sigma ) =  {{\sqrt{\sgn\, {}^4g_{AB}(\tau, \vec
 \sigma)\, J^A(\alpha^i(\tau, \vec \sigma))\, J^B(\alpha^i(\tau, \vec \sigma))}}\over
 {|{}^4g(\tau, \vec \sigma)|}} =\nonumber \\
 &=& {1\over {\sqrt{\gamma(\tau, \vec \sigma)}}}\, \sqrt{(J^{\tau})^2
 - {}^3g_{rs}\, {{J^r + n^r\, J^{\tau}}\over {1 + n}}\,
 {{J^s + n^s\, J^{\tau}}\over {1 + n}}}(\tau, \vec \sigma),\nonumber \\
 &&{}\nonumber \\
 U^A(\tau, \vec \sigma) &=& {{J^A(\alpha^i(\tau, \vec \sigma))}\over
 {\sqrt{\sgn\, {}^4g_{EF}(\tau, \vec \sigma)\, J^E(\alpha^i(\tau,
 \vec \sigma))\, J^F(\alpha^i(\tau, \vec \sigma))}}}.
 \label{3.3}
\eea

\noindent {\it This unit 4-velocity is in general different from the
unit normal $l^A(\tau, \vec \sigma)$ to the 3-space $\Sigma_{\tau}$
and therefore in general it is not surface forming}.
\medskip

The action given in Ref.\cite{9} for an (isentropic if $s = const.$)
perfect fluid with equation of state $\rho = \rho(\tilde n, s)$
\footnote{As shown in Ref.\cite{9}, in the case that the entropy has
the form $s = s(\alpha^i)$ the resulting Euler-Lagrange equations
imply the standard Euler equations obtained from the conservation of
the energy-momentum tensor. See also Ref.\cite{10}.}, can be
reformulated as a parametrized Minkowski theory \cite{10,12}  if
written in the form

\bea
 S_{fluid}&=& \int d\tau d^3\sigma\, L({}^4g_{AB}(\tau ,\vec \sigma
 ),\alpha^i(\tau ,\vec \sigma )) =\nonumber \\
 &=& -\int d\tau d^3\sigma\, ((1 + n)\, \sqrt{\gamma})(\tau ,\vec \sigma )\, \rho
 \Big( {{|J(\alpha^i(\tau ,\vec \sigma ))|} \over  {((1 + n)\,
 \sqrt{\gamma})(\tau, \vec \sigma )}},\,  s({\bf \alpha}^i
 (\tau ,\vec \sigma )) \Big) =\nonumber \\
 &=&- \int d\tau d^3\sigma\,
 ((1 + n)\, \sqrt{\gamma})(\tau ,\vec \sigma ) \nonumber \\
 &&\rho \Big({1\over {\sqrt{\gamma (\tau ,\vec \sigma )}}}\, \sqrt{\Big[
 (J^{\tau})^2 - {}^3g_{uv}\, {{J^{u} + n^{u}\,
 J^{\tau}}\over {1 + n}}\, {{J^{v} + n^{v}\, J^{\tau}}\over {1 + n}}\Big]
 (\tau ,\vec \sigma)},\,  s(\alpha^i(\tau ,\vec \sigma )) \Big).\nonumber \\
 &&{}
 \label{3.4}
  \eea

\noindent Here $S_{fluid}$ is a functional of the fluid and  of the
embedding $z^{\mu}(\tau, \vec \sigma)$. See Ref.\cite{12} for the
general treatment of the parametrized Minkowski theories, for the
gauge equivalence of the descriptions in different admissible 3+1
splittings of Minkowski space-time and for rest-frame instant form
description. Here we only emphasize the properties of the fluid
developed in Ref.\cite{10} in view of the extension to general
relativity.
\medskip

The pressure of the fluid is

\beq
 p(\tilde n, s)\, =\, \tilde n\, {{\partial\, \rho(\tilde n, s)}\over {\partial \tilde
 n}}{|}_s - \rho(\tilde n, s).
 \label{3.5}
 \eeq

\medskip

The canonical momenta of the fluid are

 \bea
 \Pi_i(\tau, \vec \sigma) &=& - {{\delta\, S_{fluid}}\over {\delta\, \partial_{\tau}\,
 \alpha^i(\tau, \vec \sigma)}} =  {{Y^{r}(\tau, \vec \sigma)}\over
  {X(\tau, \vec \sigma)}}\, T_{ri}(\tau, \vec \sigma)\,
  \Big({{\partial\, \rho(\tilde n, s)}\over {\partial\,
  \tilde n}}\Big){|}_{\tilde n =
  {{X}\over {\sqrt{\gamma}}}}(\tau, \vec \sigma),\nonumber \\
  &&{}\nonumber \\
  &&with\,\, the\,\, notations\nonumber \\
  &&{}\nonumber \\
  &&X = |J| = \sqrt{(J^{\tau})^2 - {}^3g_{uv}\, Y^u\, Y^v} = \sqrt{\gamma}\, \tilde n,\qquad
  Y^u = {{J^u + n^u\, J^{\tau}}\over {1 + n}},\nonumber \\
  &&{}\nonumber \\
   {\cal T}_{ri}\, &=&{1\over 2}\, {}^3g_{rs}\,
 \epsilon^{suv}\, \epsilon_{ijk}\, \partial_{u}\, \alpha^j\,
  \partial _{v}\, \alpha^k,\qquad {\cal T}_{ri}\, \partial_n\, \alpha^i =
  {}^3g_{rn}\, J^{\tau}, \nonumber \\
  ({\cal T}^{-1})^{is} &=& {{{}^3g^{sn}\, \partial_n\, \alpha^i}\over
  {J^{\tau}}},\qquad  \partial_n\, \alpha^i\, \Pi_i = J^{\tau}\, {}^3g_{nr}\,
  {{Y^r}\over X}\, \Big({{\partial\, \rho(\tilde n, s)}\over {\partial\,
  \tilde n}}\Big){|}_{\tilde n =
  {{X}\over {\sqrt{\gamma}}}}, \nonumber \\
  &&{}\nonumber \\
  \Rightarrow&& Y^r =  \Pi_i\, ({\cal T}^{-1})^{ir}\,  {{X}\over
  {\Big({{\partial\, \rho(\tilde n, s)}\over {\partial\,
  \tilde n}}\Big){|}_{\tilde n =  {{X}\over {\sqrt{\gamma}}}}  }} =
   {1\over {J^{\tau}}}\, {}^3g^{rs}\, \partial_s\, \alpha^i\, \Pi_i\,
 {{X}\over  {\Big({{\partial\, \rho(\tilde n, s)}\over {\partial\,
  \tilde n}}\Big){|}_{\tilde n =  {{X}\over {\sqrt{\gamma}}}}}}
  = Y^r(X, \Pi_i).\nonumber \\
  &&{}
 \label{3.6}
 \eea

\noindent The matrix ${\cal T}_{ri}$ is equal to $\sum_s\,
{}^3g_{rs}\, (ad\, {\cal J}_{is})$, where $ad\, {\cal J}_{ir} =
J^{\tau}\, {\cal J}^{-1}_{ir}$  is the adjoint matrix of the
Jacobian ${\cal J} = ({\cal J}_{ir} = \partial_{r}\, \alpha^i)$ of
the transformation from the Lagrangian coordinates $\alpha^i(\tau
,\vec \sigma )$ to the Eulerian ones $\vec \sigma$ on
$\Sigma_{\tau}$ (see Ref.\cite{11} and Section III). We use the
notations $J^{\tau} = (det\, {\cal J})$, $\sum_i\, {\cal I}_{ir}\,
{\cal I}^{-1}_{is} = \delta_{rs}$, $\sum_r\, {\cal I}_{ir}\, {\cal
I}^{-1}_{jr} = \delta_{ij}$, $\sum_r\, ({\cal T}^{-1})^{ir}\, {\cal
T}_{rj} = \delta^i_j$, $\sum_i\, {\cal T}_{ri}\, ({\cal
T}^{-1})^{is} = \delta_r^s$.

\medskip

{\it The main problem of this approach is to invert these equations
to get the velocities $\partial_{\tau}\, \alpha^i(\tau, \vec
\sigma)$ in terms of the momenta $\Pi_i(\tau, \vec \sigma)$, i.e. to
get the unit 4-velocity $U^A(\tau, \vec \sigma)$ and then the
energy-momentum tensor and the Hamiltonian in terms of the momenta}.
\medskip

We have to find the solution $X$ of the equation $X^2 + {}^3g_{rs}\,
Y^r(X, \Pi_i)\, Y^{s}(X, \Pi_i) = (J^{\tau})^2$ with $Y^{ r}(X,
\Pi_i)$ given by the last line of Eq.(\ref{3.6}). This equation can
be rewritten in the following form

\bea
 &&X^2\, \Big[(J^{\tau})^2 + A^2\, \Big({{\partial\, \rho(\tilde n, s)}\over {\partial\,
  \tilde n}}\Big)^{-2}{|}_{\tilde n =  {{X}\over {\sqrt{\gamma}}}}\Big] =
  (J^{\tau})^4,\qquad A^2 = \sum_{rs}\, {}^3g^{rs}\, \partial_r\, \alpha^i\, \Pi_i\, \partial_s\,
  \alpha^j\, \Pi_j.\nonumber \\
  &&{}
 \label{3.7}
 \eea

\noindent For every equation of state $\rho = \rho(\tilde n,s)$ the
solution of this equation has the form $X = \tilde X(\Pi_i) =
\sqrt{\gamma}\, \tilde n = F(\sqrt{\gamma}, A^2,
(J^{\tau})^2)[\rho]$, with all the dependence upon the fluid momenta
in the quadratic form $A^2$ and with an extra dependence on
$J^{\tau} = det\, (\partial_r\, \alpha^i)$. See Appendix B for the
known solutions of Eq.(\ref{3.7}).

\medskip

Once $X = \tilde X(\Pi_i)$ is known, we have

\bea
 Y^r &=& Y^r(X, \Pi_i) = {\tilde Y}^r(\Pi_i),\qquad
  J^r = (1 + n)\, {\tilde Y}^r(\Pi_i) - n^r\, J^{\tau},
  \qquad \partial_{\tau}\, \alpha^i = - {{{\tilde J}^r(\Pi_i)}\over
 {J^{\tau}}}\, \partial_r\, \alpha^i,\nonumber \\
 &&{}\nonumber \\
 U^{\tau} &=& {\tilde U}^{\tau}(\Pi_i) = {{ J^{\tau}}\over
 {(1 + n)\, \tilde X(\Pi_i)}},\qquad U^r = {\tilde U}^r(\Pi_i) =
 {{{ \tilde J}^r(\Pi_i)}\over {(1 + n)\, \tilde X(\Pi_i)}}.\nonumber \\
 &&{}
  \label{3.8}
  \eea

\medskip

The energy-momentum tensor of the fluid (see Eq.(4.2) of
Ref.\cite{10}) is

\bea
 T^{AB}(\tau ,\vec \sigma )[\alpha ]&=& - \Big[ {2\over
{\sqrt{g}}}\, {{\delta S}\over {\delta\, {}^4g_{AB}}}\Big] (\tau
,\vec \sigma ) =\nonumber \\
 &=&\Big[  \rho\, {}^4g^{AB} - \tilde n\, {{\partial \rho}\over
 {\partial\, \tilde n}}{|}_s\, ({}^4g^{AB} - {{J^A\, J^B}\over
 {{}^4g_{CD}\, J^C\, J^D}})\Big] (\tau ,\vec \sigma )=
 \nonumber \\
 &=&\Big[ \epsilon\, \rho\, U^A\, U^B - p\, ({}^4g^{AB} -
 \epsilon\, U^A\, U^B)\Big] (\tau ,\vec \sigma ),\nonumber \\
 &&{}
 \label{3.9}
\eea

\medskip

As shown in Eq.(3.11) of Ref.\cite{2} it is convenient to express
its components $T^{\tau\tau}$ and $T^{\tau r}$  in terms of a mass
density ${\cal M}$ and a mass current density ${\cal M}_r$ in the
following way

\bea
 T^{\tau\tau}(\tau, \vec \sigma) &=& \sgn\, (\rho + p)\,
 (U^{\tau})^2 - p\, {}^4g^{\tau\tau}
 = \Big({{{\cal M}}\over {(1 + n)^2\,
 \sqrt{\gamma}}}\Big)(\tau, \vec \sigma),\nonumber \\
 T^{\tau r}(\tau, \vec \sigma) &=& \sgn\, (\rho + p)\, U^{\tau}\,
 U^r - p\, {}^4g^{\tau r} = \Big({{(1 + n)\, {}^3g^{rs}\, {\cal M}_s
 - n^r\, {\cal M}}\over {(1 + n)^2\, \sqrt{\gamma}}}\Big)(\tau, \vec \sigma),
 \label{3.10}
 \eea

\noindent As shown in Ref.\cite{10}, by using Eqs.(\ref{3.5}) and
(\ref{3.6}), the mass density and the mass current density of the
fluid have the expression

\bea
 {\cal M}(\tau, \sigma) &=& \Big(\sqrt{\gamma}\, (1 + n)^2\,
 T^{\tau\tau}\Big)(\tau, \vec \sigma) =\nonumber \\
 &=&  \Big({{{}^3g_{rs}\, Y^r(X, \Pi_i)\, Y^s(X, \Pi_i)}\over
 {X}}\, {{\partial\, \rho(\tilde n, s)}\over {\partial\,
 \tilde n}}{|}_{\tilde n = {X\over {\sqrt{\gamma}}}} +
 \sqrt{\gamma}\, \rho({X\over {\sqrt{\gamma}}},s)\Big)(\tau,
 \vec \sigma) =\nonumber \\
 &=&\Big({{{}^3g^{rs}\, \partial_r\, \alpha^i\, \partial_s\, \alpha^j\,
 \Pi_i\, \Pi_j\,\, X}\over {(J^{\tau})^2\, {{\partial\, \rho(\tilde n, s)}\over
 {\partial\, \tilde n}}{|}_{\tilde n = {X\over {\sqrt{\gamma}}}}}}
 + \sqrt{\gamma}\, \rho({X\over {\sqrt{\gamma}}}, s)\Big)(\tau, \vec \sigma),\nonumber \\
 &&{}\nonumber \\
 {\cal M}_r(\tau, \sigma) &=& \Big(\sqrt{\gamma}\, (1 + n)\,
 {}^3g_{rs}\, \Big[ T^{\tau s} + n^s\, T^{\tau\tau} \Big]\Big)(\tau,
 \vec \sigma) =  \Big(\partial_r\, \alpha^i\, \Pi_i \Big)(\tau,
 \vec \sigma).\nonumber \\
 &&{}
 \label{3.11}
 \eea

\noindent {\it The mass current density has a universal Hamiltonian
expression independent from the equation of state}. Instead one can
get the mass density ${\cal M}(\tau, \vec \sigma)$ explicitly in
terms of the momenta only for those equations of state (like the
dust and the photon gas) allowing to find an explicit solution of
Eq.(\ref{3.7}). Therefore in the rest of the paper we shall consider
only the coupling of the dust to ADM tetrad gravity.
\medskip

Let us remark that Eqs.(\ref{2.8})-(\ref{2.9}) imply that the
energy-momentum tensor (\ref{3.9}) of a perfect fluid looks like the
one of a viscous fluid to the Eulerian observers ($U^A = \Gamma(w)\,
(l^A + m^A)$)

\bea
 T^{AB} &=& \sgn\, \tilde \rho\, l^A\, l^B - \tilde p\, ({}^4g^{AB} -
 \sgn\, l^A\, l^B) + l^A\, q^B + l^B\, q^A + \pi^{AB},
 \label{3.12}
 \eea

\noindent where $\tilde \rho = \Gamma^2(w)\,\rho$, $\tilde p =
\Gamma^2(w)\,p$, $q^A = \sgn\, \Gamma^2(w)\, (\rho + p)\, m^A$ and
$\pi^{AB} = \sgn\, \Gamma^2(w)\, (\rho + p)\, m^A\, m^B +
(\Gamma^{2}(w) - 1)\,p {}^4g^{AB}$. For a viscous fluid $\tilde
\rho$ is the matter energy density, $\tilde p$ the effective
isotropic pressure (equilibrium pressure plus the associated bulk
viscosity), $q^A$ the total energy flux vector and $\pi^{AB}$ the
symmetric trace-free anisotropic stress tensor.

\subsection{The Perfect Fluid Coupled to Gravity and the Bianchi
Identities for the Energy-Momentum Tensor}

When the perfect fluid is coupled to gravity, we have to reinterpret
the metric appearing in the action (\ref{3.14}) as the 4-metric
describing the gravitational field. If we add the ADM action to
Eq.(\ref{3.14}) (see the next Section for the case of dust)), then
we get the Einstein equations ${}^4G^{AB} = {{8\pi\, G}\over
{c^4}}\, T^{AB}$ with the energy-momentum tensor of Eqs.(\ref{3.10})
and (\ref{3.11}). Then the Bianchi identities imply ${}^4\nabla_A\,
T^{AB} \equiv 0$. As shown in Ref.\cite{11}, these identities may be
written in the form

 \bea
 &&U^B\, \Big(\,{}^4\nabla_A\, T^A{}_B\,\Big) = 0,\nonumber \\
  &&{}\nonumber\\
 &&\Big(\,\delta_A{}^B - U_A\, U^B\,\Big)\, {}^4\nabla_C\, T^C{}_B
 = 0.
 \label{3.13}
 \eea

The first line of Eqs.(\ref{3.13}) implies the equation

 \beq
 U^A\, \frac{\partial\rho}{\partial\sigma^A} + (\rho + p)\, {}^4\nabla_A\,
 U^A = 0,
 \label{3.14}
 \eeq

\noindent whose final form

 \beq
  \frac{\partial\rho}{\partial \tilde{n}}\, {}^4\nabla_A(\,
  \tilde{n}\, U^A\,) + \frac{\partial\rho}{\partial
 s}\, U^A\, \frac{\partial s}{\partial\sigma^A} = 0,
 \label{3.15}
 \eeq

\noindent is obtained by using Eq.(\ref{3.5}). This equation is
automatically satisfied when the particles number conservation law
${}^4\nabla_A(\, \tilde{n}\, U^A\,) = 0$ and the entropy
conservation law $U^A\,\frac{\partial s}{\partial\sigma^A} = 0$ are
satisfied.

\medskip

Only three of the four equations in the second line of
Eqs.(\ref{3.13}) are independent: they can be rewritten in the form

\beq
 U^B\, {}^4\nabla_B\, U_A = \frac{1}{\rho + p}\, \Big(\,\delta_A{}^B
 - U_A\, U^B\,\Big)\, \frac{\partial p}{\partial\sigma^B}.
  \label{3.16}
   \eeq

\noindent and turn out to be the {\em relativistic Euler equations}.
They allow to express the {\em acceleration} $a_{(U)A} = U^B\,
{}^4\nabla_B\, U_A$ as a function of  ''internal forces'' depending
on the pressure ( for the dust we will get $a_{(U)A} = 0$ in
Eq.(\ref{4.18}))

 \beq
 a_{(U)A} = \frac{1}{\rho + p}\,\Big(\,\delta_A{}^B - U_A\, U^B\,\Big)\,
 \frac{\partial p}{\partial\sigma^B}.
 \label{3.17}
  \eeq

\bigskip

Finally, by using the notations of Appendix A for tetrad gravity and
Eqs.(3.11) of Ref.\cite{2}, the identities ${}^4\nabla_A\, T^{AB}
\equiv 0$ may be written in  the following form

\bea
 &&\Big[\partial_{\tau} - \sum_{ar}\, {}^3e^r_{(a)}\, n_{(a)}\,
 \partial_r - (1 + n)\, {}^3K\Big]\, (({}^3e)^{-1}\, {\cal M})
 +{{1 + n}\over {{}^3e}}\, \sum_s\, \partial_s\, \Big[\,
 \sum_{ar}\, {}^3e^s_{(a)}\, {}^3e^r_{(a)}\, {\cal M}_r\Big]
 +\nonumber \\
 &&+ 2\, ({}^3e)^{-1}\, \sum_{ars}\, \partial_s n\, {}^3e^s_{(a)}\,
 {}^3e^r_{(a)}\, {\cal M}_r - (1 + n)\, \sum_{abrsuv}\, {}^3K_{rs}\, {}^3e^r_{(a)}\,
 {}^3e^s_{(b)}\, {}^3e^u_{(a)}\, {}^3e^v_{(b)}\,
 T_{uv} \equiv 0,\nonumber \\
 &&\nonumber\\
 &&{}\nonumber \\
 &&\Big[\partial_{\tau} - \sum_{as}\, {}^3e^s_{(a)}\, n_{(a)}\,
 \partial_s\Big]\, (({}^3e)^{-1}\, {\cal M}_r) + \partial_r\, n\, ({}^3e)^{-1}\,
 {\cal M} -\nonumber \\
 &&- \sum_{as}\, \partial_r\, ({}^3e^s_{(a)}\, n_{(a)})\, ({}^3e)^{-1}\,
 {\cal M}_s - (1 + n)\, {}^3K\, ({}^3e)^{-1}\, {\cal M}_r -\nonumber \\
 &&-\sum_{auv}\, \partial_u\, n\, {}^3e^u_{(a)}\,{}^3e^v_{(a)}\, T_{rv} +
  (1 + n)\, \Big[{1\over {{}^3e}}\, \sum_{auv}\, \partial_u\,
 \Big({}^3e\, {}^3e^u_{(a)}\, {}^3e^s_{(a)}\, T_{sr}\Big) -\nonumber \\
 &&- {1\over 2}\, \sum_{abcuvmn}\, \partial_r\, ({}^3e_{(c)u}\, {}^3e_{(c)v})\,
 {}^3e^u_{(a)}\, {}^3e^v_{(b)}\, \, {}^3e^m_{(a)}\, {}^3e^n_{(b)}\,
 T_{mn}\Big] \equiv 0.
 \label{3.18}
\eea

\medskip

Since Eqs.(\ref{2.9}) imply the following form of Eqs.(\ref{3.11})

\bea
 ({}^3e)^{-1}\, {\cal M} &=& \Gamma^2(w)\, (\rho + p) - p,\nonumber \\
 &&{}\nonumber\\
 ({}^3e)^{-1}\, {\cal M}_r &=& - \Gamma^2(w)\, (\rho + p)\,w_r,
 \label{3.19}
 \eea

\noindent the use of Eqs.(\ref{3.9}) allows  to get the following
expression of Eqs.(\ref{3.18}) as equations for $w_r$ and $\rho$

\bea
 &&\left(
 \frac{\partial\, w_r}{\partial\, \tau} - \sum_s\, n^s\, \frac{\partial\, w_r}
 {\partial\, \sigma^s} - \sum_s\, w_s\, \frac{\partial\, n_s}{\partial\, \sigma^r}
 \right) - (1 + n)\, \sum_u\, w^u\, {}^3\nabla_u\, w_r - \frac{\partial\, n}{\partial\,
 \sigma^r} +\nonumber\\
 &&{}\nonumber\\
 &+&w_r\, \Big(\sum_u\, w^u\, \frac{\partial\, n}{\partial\, \sigma^u} + (1 + n)\,
 \sum_{uv}\, w^u\, w^v\, {}^3K_{uv}\,\Big) +\nonumber\\
 &&\nonumber\\
 &+&\frac{1}{\Gamma^2(w)\,(\rho + p)}\,\left[\,-(1 + n)\,
 \frac{\partial\, p}{\partial\, \sigma^r} + w_r\,\left(
 \frac{\partial\, p}{\partial\, \tau} - \sum_u\, n^u\, \frac{\partial\,
  p}{\partial\, \sigma^u} \right)\right] \equiv 0,\nonumber \\
 &&{}\nonumber\\
 &&{}\nonumber\\
 &&\left[
 \frac{\partial}{\partial\, \tau} - \sum_s\, \Big((1 + n)\,w^s + n^s\Big)\,
 \frac{\partial}{\partial\, \sigma^s} \right]\,\rho -\nonumber \\
 &-& (\sum_{uv}\, {}^3g_{uv}\, w^u\, w^v)\,\left[
 \frac{\partial}{\partial\, \tau} - \sum_s\, n^s\,\frac{\partial}{\partial\,
 \sigma^s}\right]\,p + (1 + n)\, \sum_s\, w^s\,\frac{\partial\, p}{\partial\,
 \sigma^s} +\nonumber\\
 &&\nonumber\\
 &-&(\rho + p)\,(1 + n)\,\Big[\,\left(\,{}^3K - \sum_{rs}\,\,w^r\, w^s\,
 {}^3K_{rs}\,\right) +\, \sum_r\, {}^3\nabla_r\, w^r\, \Big] \equiv 0,
 \label{3.20}
  \eea

\noindent where ${}^3\nabla$ is the covariant derivative inside the
3-spaces with 3-metric ${}^3g_{rs}=\sum_a\,e_{(a)r}e_{(a)s}$ and
$n^r = \sum_s\,{}^3g^{rs}\, n_s = \sum_a\, n_{(a)}\, {}^3e^r_{(a)}$
are the shift functions.

\vfill\eject

\section{The Dust in ADM Tetrad Gravity}

In this Section we shall consider the dust coupled to ADM tetrad
gravity \cite{1} in the globally hyperbolic, asymptotically
Minkowskian space-times with boundary conditions killing the
supertranslations and reducing the asymptotic symmetries to the ADM
Poincare' group. The notations for the gravitational field are
defined in Appendix A.

\subsection{The Action of the Dust coupled to ADM Tetrad Gravity}

Let us consider  an isentropic perfect fluid, a dust with $p = 0$,
$s = const.$, and equation of state $\rho = \mu\, \tilde n =
{{\mu}\over {(1 + n)\, \sqrt{\gamma}}}\, \sqrt{\sgn\, {}^4g_{AB}\,
J^A\, J^B}$. In this case the chemical potential $\mu$ is the rest
mass-energy of a fluid particle: $\mu = m\, c$. The full action of
ADM tetrad gravity coupled to a dust (with positive energy) is

\bea
 S &=& S_{ADM} + S_{dust} =\qquad\qquad
 \mu = m\, c, \nonumber \\
 &&{}\nonumber \\
 &=& S_{ADM} -  \int d\tau d^3\sigma\, m\, c\,
 \sqrt{\sgn\, {}^4g_{AB}\, J^A\, J^B}(\tau,
 \vec \sigma),
 \label{4.1}
 \eea

\noindent with $S_{ADM}$ given in Ref.\cite{1}. The gravitational
action depends upon 16 tetradic gravitational configuration
variables, which are defined in Appendix A. At the Hamiltonian level
there are 16 momenta as shown in the canonical basis of
Eq.(\ref{a1}).

\bigskip

The dust momentum conjugate to ${\alpha}^{i}$ is (${\cal T}_{ri}$ is
defined in Eq.(\ref{3.6}) with $\partial\, \rho/\partial\, \tilde n
= \mu$)

\bea
 {\Pi}_i(\tau, \vec \sigma)&=& - {{\delta\, S_{ADM}}\over {\delta\,
 \partial_{\tau}\alpha^i(\tau, \vec \sigma)}} =
   \mu \, {{Y^{r}(\tau, \vec \sigma)}\over
  {X(\tau, \vec \sigma)}}\, {\cal T}_{ri}(\tau, \vec \sigma).\nonumber \\,
 &&{}\nonumber \\
   \label{4.2}
    \eea

\medskip

The  following Poisson brackets are assumed

\beq
 \lbrace \alpha^i(\tau ,\vec \sigma ), \Pi_j(\tau ,{\vec
 \sigma}^{'})\rbrace = - \delta^i_j\, \delta^3(\vec \sigma, {\vec
 \sigma}^{'}).
  \label{4.3}
 \eeq
\bigskip

As shown in Ref. \cite{1} there are the following ten primary first
class constraints involving only the gravitational field.

\bea
 &&\pi_{\varphi_{(a)}}(\tau ,\vec \sigma )\,
\approx\, 0,\qquad \pi_n(\tau ,\vec \sigma )\, \approx\, 0,\qquad
 \pi_{n_{(a)}}(\tau ,\vec \sigma
)\, \approx\, 0,\nonumber \\
 &&{}^3M_{(a)}(\tau ,\vec \sigma )
= \epsilon_{(a)(b)(c)}\, {}^3e_{(b)r} (\tau ,\vec \sigma )\,
{}^3\pi^r_{(c)}(\tau ,\vec \sigma )\, \approx\, 0.
 \label{4.4}
\eea
\medskip

Moreover there are the following four secondary first class
constraints

\bea
 {\cal H}(\tau ,\vec \sigma )&=& {\cal H}_{grav}(\tau ,\vec \sigma )
 +  {\cal M}(\tau ,\vec \sigma )\approx 0,\nonumber \\
 &&{}\nonumber \\
 {\cal H}_{(a)}(\tau ,\vec \sigma )&=& {\cal H}_{grav\, (a)}(\tau ,\vec \sigma )
 + {}^3e^r_{(a)}(\tau, \vec \sigma)\, {\cal M}_r(\tau ,\vec \sigma)
 \approx 0.\nonumber \\
 &&{}
 \label{4.5}
 \eea

\medskip

They are the super-Hamiltonian and super-momentum constraints, which
depend also on the mass density ${\cal M}$ and  on the mass current
density ${\cal M}_r = \partial_r\, \alpha^i\, \Pi_i$ defined in
Eq.(\ref{3.11}) (see Ref.\cite{1} for the gravitational parts ${\cal
H}_{grav}(\tau ,\vec \sigma )$, ${\cal H}_{grav\, (a)}(\tau ,\vec
\sigma )$). Eqs.(\ref{4.5}) show the necessity of having the
Hamiltonian expression of the mass density ${\cal M}$ by solving
Eq.(\ref{3.7}).

\bigskip

For the dust we get the following solution of Eq.(\ref{3.7}) (see
Appendix B)

\bea
 X &=& \sqrt{\gamma}\, \tilde n =
 \sqrt{\gamma}\, {{\rho}\over {\mu}} = {{\mu\, (J^{\tau})^2}\over
 {\sqrt{\mu^2\, (J^{\tau})^2 + A^2}}} =
  \frac{\mu\, det^2\, (\partial_r\, \alpha^i)}
 {\sqrt{\mu^2\, det^2\, (\partial_r\, \alpha^i)
  + \sum_{uvija}\, {}^3e^u_{(a)}\, {}^3e^v_{(a)}\, \partial_u\, \alpha^i\, \partial_v\,
 \alpha^j\, {\Pi}_{i}\, {\Pi}_{j} }},\nonumber \\
  &&{}\nonumber \\
  Y^r &=&  det\, (\partial_r\, \alpha^i)\, {{ \sum_{sia}\, {}^3e^r_{(a)}\,
  {}^3e^s_{(a)}\, \partial_s\, \alpha^i\, \Pi_i}\over
  {\sqrt{\mu^2\, det^2\, (\partial_r\, \alpha^i) + \sum_{uvija}\,
  {}^3e^u_{(a)}\, {}^3e^v_{(a)}\, \partial_u\, \alpha^i\, \partial_v\,
 \alpha^j\, {\Pi}_{i}\, {\Pi}_{j} }}},\nonumber \\
 &&{}\nonumber \\
 J^r &=& (1 + n)\, Y^r - {}^3e^r_{(a)}\, n_{(a)}\, J^{\tau}
 =\nonumber \\
 &=&  det\, (\partial_r\, \alpha^i)\, \sum_a\, {}^3e^r_{(a)}\, \Big[{{(1 + n)\,
  \sum_{si}\, {}^3e^s_{(a)}\, \partial_s\, \alpha^i\, \Pi_i}\over
  {\sqrt{\mu^2\, det^2\, (\partial_r\, \alpha^i) + \sum_{uvija}\,
  {}^3e^u_{(a)}\, {}^3e^v_{(a)}\, \partial_u\, \alpha^i\, \partial_v\,
 \alpha^j\, {\Pi}_{i}\, {\Pi}_{j} }}} -  n_{(a)}\Big],\nonumber \\
 J^{\tau} &=& det\, (\partial_r\, \alpha^i).
 \label{4.6}
  \eea

\noindent Consequently, we can get the expression of the velocities
of the Lagrangian coordinates in terms of the momenta by using
Eq.(\ref{3.1}) (whose validity in general relativity will be shown
in Eq.(\ref{4.14}))

\bea
 \partial_{\tau}\, \alpha^i &=&  - \frac{\sum_r\, J^{r}\, \partial_{r}\, \alpha^i}
 {J^{\tau}} =  \sum_r\, \frac{\Big(J^{\tau}\, \sum_a\, {}^3e^r_{(a)}\, n_{(a)} -
 (1 + n)\, Y^{r}\Big)\, \partial_{r}\, \alpha^i}{J^{\tau}} =\nonumber \\
 &&{}\nonumber \\
 &=& \sum_{ra}\, \partial_{r}\, \alpha^i\,  {}^3e^r_{(a)}\, \Big[n_{(a)} -
 {{(1 + n)\,  \sum_{si}\, {}^3e^s_{(a)}\, \partial_s\, \alpha^i\, \Pi_i}\over
  {\sqrt{\mu^2\, (J^{\tau})^2 + \sum_{uvija}\,
  {}^3e^u_{(a)}\, {}^3e^v_{(a)}\, \partial_u\, \alpha^i\, \partial_v\,
 \alpha^j\, {\Pi}_{i}\, {\Pi}_{j} }  }}\Big],\nonumber \\
 &&{}
 \label{4.7}
  \eea

\noindent so that the unit 4-velocity of the dust becomes
(${}^4g_{AB}\, U^A\, U^B = \sgn$; in the last line the 4-velocity is
decomposed on the contra-variant ortho-normal tetrads $l^A(\tau,
\vec \sigma)$, ${}^4{\buildrel \circ \over {\bar E}}^A_{(a)}(\tau,
\vec \sigma)$ carried by the Eulerian observers as shown in
Eqs.(\ref{2.8}) and (\ref{2.9}))

\bea
 U^{\tau} &=& {{J^{\tau}}\over {(1 + n)\, X}} =
  {1\over {\mu\, (1 + n)\, J^{\tau}}}\,
 \sqrt{\mu^2\, (J^{\tau})^2 + \sum_{uvija}\,
  {}^3e^u_{(a)}\, {}^3e^v_{(a)}\, \partial_u\, \alpha^i\, \partial_v\,
 \alpha^j\, {\Pi}_{i}\, {\Pi}_{j} },\nonumber \\
 &&{} \nonumber \\
 U^r &=& {{J^r}\over {(1 + n)\, X}} =
  {{\sum_a\, {}^3e^r_{(a)}}\over {\mu\, (1 + n)\, J^{\tau}
 }}\, \Big[(1 + n)\, \sum_{si}\, {}^3e^s_{(a)}\, \partial_s\,
 \alpha^i\, \Pi_i -\nonumber \\
 &-& n_{(a)}\, \sqrt{\mu^2\, (J^{\tau})^2 + \sum_{uvija}\,
  {}^3e^u_{(a)}\, {}^3e^v_{(a)}\, \partial_u\, \alpha^i\, \partial_v\,
 \alpha^j\, {\Pi}_{i}\, {\Pi}_{j} }\Big],\nonumber \\
 &&{}\nonumber \\
 &&{}\nonumber \\
 U_{\tau} &=& {}^4g_{\tau A}\, U^A = {{\sgn\, (1 + n)}\over {\mu\, J^{\tau}
 }}\, \Big(\sqrt{\mu^2\, (J^{\tau})^2 + \sum_{uvija}\,
  {}^3e^u_{(a)}\, {}^3e^v_{(a)}\, \partial_u\, \alpha^i\, \partial_v\,
 \alpha^j\, {\Pi}_{i}\, {\Pi}_{j} } -\nonumber \\
 &-& \sum_{ari}\, {{n_{(a)}}\over {1 + n}}\, {}^3e^r_{(a)}\,
 \partial_r\, \alpha^i\, \Pi_i\Big),\nonumber \\
 &&{}\nonumber \\
 U_r &=& {}^4g_{rA}\, U^A = - \sgn\, \Gamma(w)\, w_r = - {1\over {\mu\, J^{\tau}}}\,
 \sum_i\, \partial_r\, \alpha^i\, \Pi_i,\nonumber \\
 &&{}\nonumber \\
 U^A &=& {1\over 2}\, (1 + n)\, U^{\tau}\, \Big[l^A +
 \sum_a\, {{{\bar n}_{(a)} + 2\, \sum_r\, {{U^r}\over {U^{\tau}}}\,
 {}^3{\bar e}_{(a)r}}\over {1 + n}}\, {}^4{\buildrel \circ
 \over {\bar E}}^A_{(a)}\Big].
 \label{4.8}
 \eea

\bigskip

For the dust we have the following Hamiltonian expression of the
energy-momentum tensor (\ref{3.9}), of the energy density $\bar \rho
= \sgn\, \rho = \sgn\, \mu\, \tilde n$ and of the mass density
(\ref{3.11})

\begin{eqnarray*}
 T^{AB} &=&  \sgn\, \mu\, \tilde n\, U^A\, U^B\, {\buildrel {def}\over =}
 \bar \rho\, U^A\, U^B,\nonumber \\
 &&{}\nonumber \\
  {\cal M} &=& \sqrt{\gamma}\, (1 + n)^2\, T^{\tau\tau} =
 (1 + n)^2\, \sqrt{\gamma}\, \bar \rho\, (U^{\tau})^2
 = \sqrt{\mu^2\, \Big(J^{\tau}\Big)^2
  + \sum_{uvija} {}^3e_{(a)}^u\, {}^3e_{(a)}^v\, \partial_u\, \alpha^i\, \partial_v\,
  \alpha^j\, \Pi_i\, \Pi_j },\nonumber \\
  \bar \rho &=& {{T^{\tau\tau}}\over {(U^{\tau})^2}} = {{{\cal
  M}}\over {\sqrt{\gamma}\, (1 + n)^2\, (U^{\tau})^2}},\nonumber \\
 &&{}\nonumber \\
 {\cal M}_r &=&\sqrt{\gamma}\, (1 + n)\, \sum_s\, {}^3g_{rs}\, (T^{\tau s}
 + n^s\, T^{\tau\tau}) =     \partial_r\, \alpha^i\, \Pi_i =
 - \mu\, J^{\tau}\, U_r,
 \end{eqnarray*}

 \bea
 T^{rs} &=&  \bar \rho\, U^r\, U^s  =\nonumber \\
 &=&  {{\sgn}\over {(1 + n)^2\, \sqrt{\gamma}}}\, {{\sum_{ab}\, {}^3e^r_{(a)}\,
 {}^3e^s_{(b)}}\over {\sqrt{\mu^2\, \Big(J^{\tau}\Big)^2
  + \sum_{uvija}\, {}^3e_{(a)}^u\, {}^3e_{(a)}^v\, \partial_u\, \alpha^i\, \partial_v\,
  \alpha^j\, \Pi_i\, \Pi_j }}}\nonumber \\
 &&\Big[(1 + n)\, \sum_{ui}\, {}^3e^u_{(a)}\, \partial_u\, \alpha^i\, \Pi_i
 + n_{(a)}\, \sqrt{\mu^2\, \Big(J^{\tau}\Big)^2
  +  \sum_{uvija}\, {}^3e_{(a)}^u\, {}^3e_{(a)}^v\, \partial_u\, \alpha^i\, \partial_v\,
  \alpha^j\, \Pi_i\, \Pi_j }\Big]\nonumber \\
 &&\Big[(1 + n)\, \sum_{vj}\, {}^3e^v_{(b)}\, \partial_v\, \alpha^j\, \Pi_j
 + n_{(b)}\, \sqrt{\mu^2\, \Big(J^{\tau}\Big)^2
  + \sum_{uvija}\, {}^3e_{(a)}^u\, {}^3e_{(a)}^v\, \partial_u\, \alpha^i\, \partial_v\,
  \alpha^j\, \Pi_i\, \Pi_j }\Big].\nonumber \\
  &&{}
  \label{4.9}
   \eea

Let us remark that for the dust, having $p = 0$, Eqs.(\ref{2.9}) and
(\ref{3.19}) imply ${\cal M} = \tilde \phi\, \Gamma^2(w)\, \bar \rho
= \mu\, J^{\tau}\, \Gamma(w)$ and ${\cal M}_r = - \tilde \phi
\Gamma^2(w)\, \bar \rho\, w_r = - {\cal M}\, w_r$.

\medskip

The Hamilton equations for the dynamical matter, given in the next
Subsection, imply the identities  ${}^4\nabla_A\, T^{AB} \equiv 0$,
consequence of Einstein equations due to the Bianchi identities,
whose expression was given in Eqs.(\ref{3.18}) or (\ref{3.20}). For
the dust  Eqs.(\ref{3.20}) become ($\rho = \sgn\, \bar \rho$; $n^r =
\sum_a\, {\bar n}_{(a)}\, {}^3{\bar e}^r_{(a)}$, ${}^3K_{rs} =
{1\over 3}\, {}^3K\, {}^3g_{rs} + \sum_{ab}\, \sigma_{(a)(b)}\,
{}^3{\bar e}_{(a)r}\, {}^3{\bar e}_{(b)s}$ with the notations of
Appendix A)

 \bea
  &&\left(
 \frac{\partial\, w_r}{\partial\, \tau} - \sum_s\, n^s\,
 \frac{\partial\, w_r}{\partial\, \sigma^s} - \sum_s\, w_s\,
 \frac{\partial\, n_s}{\partial\, \sigma^r}\right) - (1 + n)\,
 \sum_u\, w^u\, {}^3\nabla_u\, w_r - \frac{\partial\, n}{\partial\,
 \sigma^r} +\nonumber\\
 &&{}\nonumber\\
 &&\qquad+ w_r\, \Big(\sum_u\, w^u\, \frac{\partial\, n}{\partial\,
 \sigma^u} + (1 + n) \sum_{uv}\, w^u\, w^v\, {}^3K_{uv}\,\Big) \equiv
 0, \nonumber \\
 &&{}\nonumber\\
 &&{}\nonumber\\
 &&\left[
 \frac{\partial}{\partial\, \tau} - \sum_s\, \Big((1 + n)\, w^s + n^s\Big)\,
 \frac{\partial}{\partial\, \sigma^s} \right]\, \rho -\nonumber \\
 &-&(1 + n)\, \rho\, \Big[\,\left(\,{}^3K - \, \sum_{rs}\, w^r\, w^s\,
 {}^3K_{rs}\,\right) + \sum_r\,{}^3\nabla_r\, w^r\, \Big] \equiv 0.
 \label{eq.rho.dust}
 \eea

At the Hamiltonian level they are a consequence of the Hamilton
equations for the dust, given in Eqs.(\ref{4.13}).

\subsection{The Dust in the York Canonical Basis}

In the York canonical basis (\ref{a2}) we have ($\phi = {\tilde
\phi}^{1/6} = (\sqrt{\gamma})^{1/6}$) \medskip

\begin{eqnarray*}
 \begin{minipage}[t]{4 cm}
\begin{tabular}{|ll|ll|l|l|l|l|} \hline
$\varphi_{(a)}$ & $\alpha_{(a)}$ & $n$ & ${\bar n}_{(a)}$ &
$\theta^r$ & $\tilde \phi$ & $R_{\bar a}$ & $\alpha^i$\\ \hline
$\pi_{\varphi_{(a)}} \approx0$ &
 $\pi^{(\alpha)}_{(a)} \approx 0$ & $\pi_n \approx 0$ & $\pi_{{\bar n}_{(a)}} \approx 0$
& $\pi^{(\theta )}_r$ & $\pi_{\tilde \phi}$ & $\Pi_{\bar a}$ & $\Pi_i$ \\
\hline
\end{tabular}
\end{minipage}\nonumber \\
 &&{}
 \end{eqnarray*}

\begin{eqnarray*}
  X &=& \tilde \phi\, \tilde n =
 \tilde \phi\, {{\rho}\over {\mu}} = \frac{\mu\, (J^{\tau})^2}
 {\sqrt{\mu^2\, (J^{\tau})^2
  + {\tilde \phi}^{-2/3}\, \sum_{auvij}\, Q_a^{-2}\, V_{ua}\, V_{va}\,
  \partial_u\, \alpha^i\, \partial_v\,
 \alpha^j\, {\Pi}_{i}\, {\Pi}_{j} }},\nonumber \\
  &&{}\nonumber \\
  Y^r &=&  J^{\tau}\, {{
 {\tilde \phi}^{-2/3}\, \sum_{bsi}\, Q_b^{-2}\, V_{rb}\, V_{sb}\,
  \partial_s\, \alpha^i\, \Pi_i}\over
  {\sqrt{\mu^2\, (J^{\tau})^2 +
 {\tilde \phi}^{-2/3}\, \sum_{auvij}\, Q_a^{-2}\, V_{ua}\, V_{va}\,
  \partial_u\, \alpha^i\, \partial_v\,
 \alpha^j\, {\Pi}_{i}\, {\Pi}_{j} }}},\nonumber \\
 &&{}\nonumber \\
 J^r &=& (1 + n)\, Y^r - \sum_a\, {}^3e^r_{(a)}\, n_{(a)}\, J^{\tau}
 =  J^{\tau}\, {\tilde \phi}^{-1/3}\,
 \sum_b\, Q_b^{-1}\, V_{rb}\nonumber \\
 && \Big[{{(1 + n)\,
 {\tilde \phi}^{-1/3}\, Q_b^{-1}\, \sum_{si}\, V_{sb}\,
 \partial_s\, \alpha^i\, \Pi_i}\over
  {\sqrt{\mu^2\, (J^{\tau})^2 +
  {\tilde \phi}^{-2/3}\, \sum_{auvij}\, Q_a^{-2}\, V_{ua}\, V_{va}\,
  \partial_u\, \alpha^i\, \partial_v\,
 \alpha^j\, {\Pi}_{i}\, {\Pi}_{j} }}} -
   n_{(b)}\Big],\nonumber \\
 J^{\tau} &=& det\, (\partial_r\, \alpha^i),\nonumber \\
 {\cal T}_{ri} &=& {1\over 2}\, {\tilde \phi}^{2/3}\, \sum_{asuvjk}\,
 Q_a^2\, V_{ra}\, V_{sa}\, \epsilon^{suv}\, \epsilon_{ijk}\,
 \partial_u\, \alpha^j\, \partial_v\, \alpha^k,\nonumber \\
 &&{}\nonumber \\
  \partial_{\tau}\, \alpha^i &=&  - \frac{\sum_r\, J^{r}\, \partial_{r}\, \alpha^i}
 {J^{\tau}} =  \frac{\sum_r\, \Big(\sum_a\,
 {}^3e^r_{(a)}\, n_{(a)}\, J^{\tau} - (1 + n)\, Y^{r}\Big)\,
 \partial_{r}\, \alpha^i}{J^{\tau}} =\nonumber \\
 &&{}\nonumber \\
 &=&   {\tilde \phi}^{-1/3}\, \sum_{rb}\, \partial_{r}\, \alpha^i\,
 Q_b^{-1}\, V_{rb}\, \Big[n_{(b)} -
 {{(1 + n)\, {\tilde \phi}^{-1/3}\, Q_b^{-1}\, \sum_{si}\, V_{sb}\,
 \partial_s\, \alpha^i\, \Pi_i}\over
  {\sqrt{\mu^2\, (J^{\tau})^2 +
 {\tilde \phi}^{-2/3}\, \sum_{auvij}\, Q_a^{-2}\, V_{ua}\, V_{va}\,
 \partial_u\, \alpha^i\, \partial_v\,
 \alpha^j\, {\Pi}_{i}\, {\Pi}_{j} }  }}\Big],
 \end{eqnarray*}

  \bea
 {\cal M} &=& \sqrt{\mu^2\, (J^{\tau})^2 + {\tilde \phi}^{-2/3}\,
 \sum_{arsij}\, Q_a^{-2}\, V_{ra}\, V_{sa}\, \partial_r\, \alpha^i\,
 \partial_s\, \alpha^j\, \Pi_i\, \Pi_j},\nonumber \\
 {\cal M}_r &=&  \sum_i\,  \partial_r\, \alpha^i\, \Pi_i,\nonumber \\
 &&{}\nonumber \\
 T^{rs} &=&  {{\sgn\, {\tilde \phi}^{-5/3}}\over {(1 + n)^2}}\, \sum_{ab}\,
 {{Q_a^{-1}\, Q_b^{-1}\, V_{ra}\, V_{sb}}\over {\sqrt{\mu^2\,
 \Big(J^{\tau}\Big)^2 + {\tilde \phi}^{-2/3}\, \sum_{cuvij}\,
 Q_c^{-2}\, V_{uc}\, V_{vc}\, \partial_u\, \alpha^i\, \partial_v\,
  \alpha^j\, \Pi_i\, \Pi_j } }}\nonumber \\
 &&\Big[(1 + n)\, {\tilde \phi}^{-1/3}\, Q_a^{-1}\, \sum_{mi}\, V_{ma}\,
 \partial_m\, \alpha^i\, \Pi_i
 + {\bar n}_{(a)}\, \sqrt{\mu^2\, \Big(J^{\tau}\Big)^2
  +  {\tilde \phi}^{-2/3}\, \sum_{cuvij}\, Q_c^{-2}\, V_{uc}\, V_{vc}\,
  \partial_u\, \alpha^i\, \partial_v\,
  \alpha^j\, \Pi_i\, \Pi_j }\Big]\nonumber \\
 &&\Big[(1 + n)\, {\tilde \phi}^{-1/3}\, Q_b^{-1}\, \sum_{nj}\,
 V_{nb}\, \partial_n\, \alpha^j\, \Pi_j
 + {\bar n}_{(b)}\, \sqrt{\mu^2\, \Big(J^{\tau}\Big)^2
  +  {\tilde \phi}^{-2/3}\, \sum_{cuvij}\, Q_c^{-2}\, V_{uc}\, V_{vc}\,
  \partial_u\, \alpha^i\, \partial_v\,
  \alpha^j\, \Pi_i\, \Pi_j }\Big].\nonumber \\
  &&{}
 \label{4.10}
 \eea

\bigskip

The unit 4-velocity (\ref{4.6}) and the energy density $\bar \rho$
of Eq.(\ref{4.9}) of the dust are

\begin{eqnarray*}
  U^{\tau} &=& {1\over {\mu\, (1 + n)\, J^{\tau}}}\,
 \sqrt{\mu^2\,  (J^{\tau})^2 + {\tilde \phi}^{-2/3}\,
 \sum_{arsij}\, Q_a^{-2}\, V_{ra}\, V_{sa}\, \partial_r\, \alpha^i\,
 \partial_s\, \alpha^j\, \Pi_i\, \Pi_j},\nonumber \\
 U^r &=&  {{{\tilde \phi}^{-1/3}\, \sum_b\, Q_b^{-1}\, V_{rb}}\over
 {\mu\, (1 + n)\, J^{\tau}}}\, \Big[(1 + n)\,
 {\tilde \phi}^{-1/3}\, Q_b^{-1}\, \sum_{si}\, V_{sb}\, \partial_s\, \alpha^i\,
 \Pi_i\,   \nonumber \\
 &-& {\bar n}_{(b)}\, \sqrt{\mu^2\, (J^{\tau})^2 + {\tilde \phi}^{-2/3}\,
 \sum_{arsij}\, Q_a^{-2}\, V_{ra}\, V_{sa}\, \partial_r\, \alpha^i\,
 \partial_s\, \alpha^j\, \Pi_i\, \Pi_j}\Big],
 \end{eqnarray*}

\bea
 U_{\tau} &=& {}^4g_{\tau\tau}\, U^{\tau} + {}^4g_{\tau s}\, U^s =
 \sgn\, \Big([(1 + n)^2 - \sum_a\, {\bar n}^2_{(a)}]\, U^{\tau}
 - {\tilde \phi}^{1/3}\, \sum_{as}\, Q_a\, {\bar n}_{(a)}\,
 V_{sa}\, U^s\Big) =\nonumber \\
 &=& {{\sgn\, (1 + n)}\over {\mu\, J^{\tau}}}\, \Big[
 \sqrt{\mu^2\, (J^{\tau})^2 + {\tilde \phi}^{-2/3}\,
 \sum_{arsij}\, Q_a^{-2}\, V_{ra}\, V_{sa}\, \partial_r\, \alpha^i\,
 \partial_s\, \alpha^j\, \Pi_i\, \Pi_j} -\nonumber \\
 &-& {\tilde \phi}^{-1/3}\, \sum_a\, {{{\bar n}_{(a)}}\over {1 + n}}\,
 Q_a^{-1}\, \sum_{si}\, V_{sa}\, \partial_s\, \alpha^i\, \Pi_i\Big],
 \nonumber \\
 U_r &=& {}^4g_{r\tau}\, U^{\tau} + {}^4g_{rs}\, U^s = - \sgn\,
 {\tilde \phi}^{1/3}\, \sum_a\, V_{ra}\, Q_a\, \Big({\bar n}_{(a)}\,
 U^{\tau} + {\tilde \phi}^{1/3}\, Q_a\, \sum_s\, V_{sa}\, U^s\Big) =
 \nonumber \\
 &=& - {1\over {\mu\, J^{\tau}}}\, \sum_i\, \partial_r\, \alpha^i\, \Pi_i
 = - {{{\cal M}_r}\over {\mu\, J^{\tau}}},\nonumber \\
 \bar \rho &=& {{\mu^2\, (J^{\tau})^2\, {\tilde \phi}^{-1}}\over
 {\sqrt{\mu^2\, (J^{\tau})^2 + {\tilde \phi}^{-2/3}\,
 \sum_{arsij}\, Q_a^{-2}\, V_{ra}\, V_{sa}\, \partial_r\, \alpha^i\,
 \partial_s\, \alpha^j\, \Pi_i\, \Pi_j}}}.
 \label{4.11}
 \eea

\subsection{The Hamilton Equations in the York Canonical Basis}

The Hamilton equations of the gravitational field in Schwinger time
gauges are generated by the following Dirac Hamiltonian (see
Eq.(3.48) of Ref. \cite{2} for $H_{grav}$), which depends upon
matter only through the mass density and the mass current density

\bea
 H_D&=& {1\over c}\, {\hat E}_{ADM} + \int d^3\sigma\, \Big[ n\, {\cal H}
 - {\bar n}_{(a)}\, {\tilde {\bar {\cal H}}}_{(a)}\Big](\tau ,\vec
\sigma )  +\nonumber \\
 &+&\int d^3\sigma\, \Big[\lambda_n\, {\tilde \pi}^n+\lambda^{\vec
n}_{(a)}\, {\tilde \pi}^{\vec n}_{(a)}\Big](\tau ,\vec \sigma )
=\nonumber \\
 &&{}\nonumber \\
 &=& H_{grav} + \int d^3\sigma\, \Big[(1 + n)\, {\cal M}\Big](\tau ,\vec
 \sigma ) - \int\, d^3\sigma\, \sum_r\, n^r(\tau, \vec \sigma)\,
 {\cal M}_r(\tau, \vec \sigma) =\nonumber \\
 &=& H_{grav} + \int d^3\sigma\, \Big[(1 + n)\, {\cal M}\Big](\tau ,\vec
 \sigma ) -\nonumber \\
 &-& \int\, d^3\sigma\, \sum_a\, \Big({\bar n}_{(a)}\,\,
 {\tilde \phi}^{-1/3}\,  Q_a^{-1}\, \sum_r\, V_{ra}(\theta^i)\, {\cal M}_r
 \Big)(\tau ,\vec \sigma). \nonumber \\
 &&{}
 \label{4.12}
 \eea
\medskip

\noindent and are explicitly given in Section IV of Ref.\cite{2}.
The expressions of the super-Hamiltonian and super-momentum
constraints in the York canonical basis are given in Eqs. (3.45) and
(3.42) of Ref.\cite{2}.
\medskip

The Hamilton equations for the dust are

\begin{eqnarray*}
 \partial_{\tau}\, \alpha^i(\tau, \vec \sigma) &\cir& - \int
 d^3\sigma_1\, [1 + n(\tau, {\vec \sigma}_1)]\, {{\delta
 {\cal M}(\tau, {\vec \sigma}_1)}\over {\delta\, \Pi_i(\tau, \vec \sigma)}}
 +\nonumber \\
 &+& \int d^3\sigma_1\, [{\tilde \phi}^{-1/3}\,\sum_a\, {\bar n}_{(a)}\,
 Q_a^{-1}\, \sum_r\, V_{ra}](\tau, {\vec \sigma}_1)\, {{\delta\, {\cal M}_r(\tau,
 {\vec \sigma}_1)}\over {\delta\, \Pi_i(\tau, \vec \sigma)}}
 =\nonumber \\
 &=&- \Big((1 + n)\, {\tilde \phi}^{-2/3}\, {{\sum_{arsj}\, Q_a^{-2}\, V_{ra}\,
 V_{sa}\, \partial_r\, \alpha^i\, \partial_s\, \alpha^j\, \Pi_j}\over
 {\sqrt{\mu^2\, (J^{\tau})^2 + {\tilde \phi}^{-2/3}\,
 \sum_{buvmn}\, Q_b^{-2}\, V_{ub}\, V_{vb}\, \partial_u\, \alpha^m\,
 \partial_v\, \alpha^n\, \Pi_m\, \Pi_n}}} -\nonumber \\
 &-& {\tilde \phi}^{-1/3}\, \sum_a\, {\bar n}_{(a)}\, Q_a^{-1}\,
 \sum_r\, V_{ra}\, \partial_r\, \alpha^i
 \Big)(\tau, \vec \sigma),\nonumber \\
 &&{}\nonumber \\
 \partial_{\tau}\, \Pi_i(\tau, \vec \sigma) &\cir&
  \int d^3\sigma_1\, [1 + n(\tau, {\vec \sigma}_1)]\, {{\delta
 {\cal M}(\tau, {\vec \sigma}_1)}\over {\delta\, \alpha^i(\tau, \vec \sigma)}}
 -\nonumber \\
 &-& \int d^3\sigma_1\, [{\tilde \phi}^{-1/3}\, \sum_a\, {\bar n}_{(a)}\,
 Q_a^{-1}\, \sum_r\, V_{ra}](\tau, {\vec \sigma}_1)\, {{\delta\, {\cal M}_r(\tau,
 {\vec \sigma}_1)}\over {\delta\, \alpha^i(\tau, \vec \sigma)}}
 =\nonumber \\
 \end{eqnarray*}

 \bea
 &=&- \sum_r\, \partial_r\, \Big({{1 + n}\over
 {\sqrt{\mu^2\, (J^{\tau})^2 + {\tilde \phi}^{-2/3}\,
 \sum_{buvmn}\, Q_b^{-2}\, V_{ub}\, V_{vb}\, \partial_u\, \alpha^m\,
 \partial_v\, \alpha^n\, \Pi_m\, \Pi_n}}}\nonumber \\
 &&\Big[{1\over 2}\, \mu^2\, J^{\tau}\, \sum_{uvjk}\, \epsilon^{ruv}\, \epsilon_{ijk}\,
 \partial_u\, \alpha^j\, \partial_v\, \alpha^k +
  {\tilde \phi}^{-2/3}\, \sum_{asj}\, Q_a^{-2}\, V_{ra}\, V_{sa}\,
 \partial_s\, \alpha^j\, \Pi_j\, \Pi_i\Big] -\nonumber \\
 &-& {\tilde \phi}^{-1/3}\, \sum_a\, {\bar n}_{(a)}\, Q_a^{-1}\,
 \Pi_i \Big)(\tau, \vec \sigma).
 \label{4.13}
 \eea

\medskip

The first half of these equations allows to check the validity at
the Hamiltonian level of the basic defining property of the
Lagrangian (comoving) scalar fields $\alpha^i(\tau, \vec \sigma)$
used as coordinates for the fluid flux lines

\beq
 U^A(\tau, \vec \sigma)\, {}^4\nabla_A\, \alpha^i(\tau, \vec \sigma)
 = U^A(\tau, \vec \sigma)\, \partial_A\, \alpha^i(\tau, \vec \sigma)
 \cir 0.
 \label{4.14}
 \eeq

\medskip

In accord with Eq.(\ref{4.2}) the inversion of the first equation
gives (${\cal I}_{ir} = \partial_r\, \alpha^i$ from Eq.(\ref{3.6}))

\medskip

\bea
 \Pi_i &=& {{\mu\, J^{\tau}\, {\tilde \phi}^{2/3}\,
 \sum_{rscj}\, {\cal I}^{-1}_{ir}\, Q_c^2\, V_{rc}\, V_{sc}\,
 {\cal I}^{-1}_{js}\, [\partial_{\tau}\, \alpha^j -
 {\tilde \phi}^{-1/3}\, \sum_{au}\, {\bar n}_{(a)}\, Q_a^{-1}\,
 V_{ua}\, \partial_u\, \alpha^j]}\over {\sqrt{(1 + n)^2 -
 {\tilde \phi}^{2/3}\, \sum_b\, \Big[Q_b\, \sum_{sk}\, V_{sb}\,
 {\cal I}^{-1}_{ks}\, [\partial_{\tau}\, \alpha^k - {\tilde \phi}^{-1/3}\,
 \sum_{ev}\, {\bar n}_{(e)}\, Q_e^{-1}\, V_{ve}\, \partial_v\,
 \alpha^k]\Big]^2}}},\nonumber \\
 &&{}\nonumber \\
 &&{}\nonumber \\
 &&{\cal I}^{-1}_{ir} = {1\over {2\, J^{\tau}}}\, \epsilon^{ruv}\,
 \epsilon_{ijk}\, \partial_u\, \alpha^j\, \partial_v\,
 \alpha^k,\nonumber \\
 &&{}\nonumber \\
 &&{}\nonumber \\
 &&{1\over {\sqrt{\mu^2\, (J^{\tau})^2 + {\tilde \phi}^{-2/3}\,
 \sum_{buv}\, Q_b^{-2}\, V_{ub}\, V_{vb}\, \partial_u\, \alpha^m\,
 \partial_v\, \alpha^n\, \Pi_m\, \Pi_n}}} = \nonumber \\
 &&{{\sqrt{(1 + n)^2 -
 {\tilde \phi}^{2/3}\, \sum_b\, \Big[Q_b\, \sum_{sk}\, V_{sb}\,
 {\cal I}^{-1}_{ks}\, [\partial_{\tau}\, \alpha^k - {\tilde \phi}^{-1/3}\,
 \sum_{ev}\, {\bar n}_{(e)}\, Q_e^{-1}\, V_{ve}\, \partial_v\,
 \alpha^k]\Big]^2}}\over {\mu\, J^{\tau}\, (1 + n)}}.\nonumber \\
 &&{}
 \label{4.15}
 \eea

\bigskip

By putting this expression for $\Pi_i(\tau, \vec \sigma)$ in the
second set of Hamilton equations, we can get the second order
equations of motion for the dust variables $\alpha^i(\tau, \vec
\sigma)$: they have the form of three conserved currents
$\partial_A\, j_i^A(\tau, \vec \sigma) \cir 0$. These conservation
laws imply that the quantities $\int d^3\sigma\, \Pi_i(\tau, \vec
\sigma)$ are constant of the motion and generate the transformation
$\alpha^i(\tau, \vec \sigma)\, \rightarrow\, \alpha^i(\tau, \vec
\sigma) + const.$ leaving invariant the Lagrangian (\ref{4.1}).
\bigskip

Finally  the Hamilton-Dirac Equation (\ref{4.14}) for the dust are
equivalent to the 4-dimensional, manifestly covariant, equations

 \beq
 U^A\, {}^4\nabla_A\, U_B \cir 0,
 \label{4.16}
 \eeq

\noindent implying that the flux lines are geodesics. For the dust
with null pressure Eq.(\ref{4.16}) is equivalent to eq.(\ref{3.16}).

\subsection{The 3-Orthogonal Schwinger Time Gauges}

The restriction of the Hamilton equations to the family of
(non-harmonic) 3-orthogonal Schwinger time gauges, where
$\theta^i(\tau, \vec \sigma) \approx 0$ so that the 3-metric is
diagonal (${}^3g_{rs} = {\tilde \phi}^{2/3}\, Q_r^2\, \delta_{rs}$
from Eq.(\ref{a3})), is given in Section II of Ref.\cite{3}. The
restriction to these gauges of Eqs.(\ref{4.13}) is

\begin{eqnarray*}
  {\cal M} &=& \sqrt{\mu^2\, (J^{\tau})^2 + {\tilde \phi}^{-2/3}\,
 \sum_a\, Q_a^{-2}\, (\sum_i\, \partial_a\, \alpha^i\, \Pi_i)^2},\nonumber \\
 {\cal M}_r &=&  \sum_i\,  \partial_r\, \alpha^i\, \Pi_i,\nonumber \\
 &&{}\nonumber \\
 T^{rs} &=&  {{\sgn\, {\tilde \phi}^{-5/3}}\over {(1 + n)^2}}\,
 {{Q_r^{-1}\, Q_s^{-1}}\over {\sqrt{\mu^2\,
 \Big(J^{\tau}\Big)^2 + {\tilde \phi}^{-2/3}\, \sum_c\,
 Q_c^{-2}\,  (\sum_i\, \partial_c\, \alpha^i\, \Pi_i)^2 } }}\nonumber \\
 &&\Big[(1 + n)\, {\tilde \phi}^{-1/3}\, Q_a^{-1}\, \sum_i\,
 \partial_a\, \alpha^i\, \Pi_i
 + {\bar n}_{(a)}\, \sqrt{\mu^2\, \Big(J^{\tau}\Big)^2
  +  {\tilde \phi}^{-2/3}\, \sum_c\, Q_c^{-2}\,
  (\sum_i\, \partial_c\, \alpha^i\, \Pi_i)^2 }\Big]\nonumber \\
 &&\Big[(1 + n)\, {\tilde \phi}^{-1/3}\, Q_b^{-1}\, \sum_j\,
  \partial_b\, \alpha^j\, \Pi_j
 + {\bar n}_{(b)}\, \sqrt{\mu^2\, \Big(J^{\tau}\Big)^2
  +  {\tilde \phi}^{-2/3}\, \sum_c\, Q_c^{-2}\,
  (\sum_j\, \partial_c\, \alpha^j\, \Pi_j)^2 }\Big],\nonumber \\
  &&{}\nonumber \\
  U^{\tau} &=& {1\over {\mu\, (1 + n)\, J^{\tau}}}\, \sqrt{\mu^2\,
 \Big(J^{\tau}\Big)^2 + {\tilde \phi}^{-2/3}\, \sum_c\,
 Q_c^{-2}\,  (\sum_i\, \partial_c\, \alpha^i\, \Pi_i)^2 },
  \nonumber \\
  U^r &=& {{{\tilde \phi}^{-1/3}\, Q_r^{-1}}\over {\mu\,
  (1 + n)\, J^{\tau}}}\, \Big[(1 + n)\, {\tilde \phi}^{-1/3}\,
  \sum_i\, Q_r^{-1}\, \partial_r\, \alpha^i\, \Pi_i -\nonumber \\
 &-&{\bar n}_{(r)}\, \sqrt{\mu^2\, (J^{\tau})^2 + {\tilde \phi}^{-2/3}\,
 \sum_a\, Q_a^{-2}\, (\sum_i\, \partial_a\, \alpha^i\, \Pi_i)^2},
 \end{eqnarray*}

\begin{eqnarray*}
 U_{\tau} &=& {{\sgn\, (1 + n)}\over {\mu\, J^{\tau}}}\, \Big[
 \sqrt{\mu^2\, (J^{\tau})^2 + {\tilde \phi}^{-2/3}\,
 \sum_{aij}\, Q_a^{-2}\, \partial_a\, \alpha^i\,
 \partial_a\, \alpha^j\, \Pi_i\, \Pi_j} -\nonumber \\
 &-& {\tilde \phi}^{-1/3}\, \sum_a\, {{{\bar n}_{(a)}}\over {1 + n}}\,
 Q_a^{-1}\, \sum_{i}\,  \partial_a\, \alpha^i\, \Pi_i\Big],
 \nonumber \\
 U_r &=& - \sgn\, \Gamma(w)\, w_r = - {1\over {\mu\, J^{\tau}}}\, \sum_i\, \partial_r\,
 \alpha^i\, \Pi_i,
 \end{eqnarray*}

\bea
  \partial_{\tau}\, \alpha^i(\tau, \vec \sigma) &\cir&-
 \Big((1 + n)\, {\tilde \phi}^{-2/3}\, {{\sum_{aj}\, Q_a^{-2}\,
 \partial_a\, \alpha^i\, \partial_a\, \alpha^j\, \Pi_j}\over
 {\sqrt{\mu^2\, (J^{\tau})^2 + {\tilde \phi}^{-2/3}\,
 \sum_{bmn}\, Q_b^{-2}\,  \partial_b\, \alpha^m\,
 \partial_b\, \alpha^n\, \Pi_m\, \Pi_n}}} -\nonumber \\
 &-& {\tilde \phi}^{-1/3}\, \sum_a\, {\bar n}_{(a)}\, Q_a^{-1}\,
  \partial_a\, \alpha^i
 \Big)(\tau, \vec \sigma),\nonumber \\
 &&{}\nonumber \\
 \Pi_i &=&  {{\mu\, J^{\tau}\, {\tilde \phi}^{2/3}\,
 \sum_{cj}\, {\cal I}^{-1}_{ic}\, Q_c^2\,
 {\cal I}^{-1}_{jc}\, [\partial_{\tau}\, \alpha^j -
 {\tilde \phi}^{-1/3}\, \sum_{a}\, {\bar n}_{(a)}\, Q_a^{-1}\,
 \partial_a\, \alpha^j]}\over {\sqrt{(1 + n)^2 -
 {\tilde \phi}^{2/3}\, \sum_b\, \Big[Q_b\, \sum_{k}\,
 {\cal I}^{-1}_{kb}\, [\partial_{\tau}\, \alpha^k - {\tilde \phi}^{-1/3}\,
 \sum_{e}\, {\bar n}_{(e)}\, Q_e^{-1}\, \partial_e\,
 \alpha^k]\Big]^2}}},\nonumber \\
 &&{}\nonumber \\
 \partial_{\tau}\, \Pi_i(\tau, \vec \sigma) &\cir&-
 \sum_r\, \partial_r\, \Big({{1 + n}\over
 {\sqrt{\mu^2\, (J^{\tau})^2 + {\tilde \phi}^{-2/3}\,
 \sum_{bmn}\, Q_b^{-2}\,  \partial_b\, \alpha^m\,
 \partial_b\, \alpha^n\, \Pi_m\, \Pi_n}}}\nonumber \\
 &&\Big[{1\over 2}\, \mu^2\, J^{\tau}\,
 \sum_{uvjk}\, \epsilon^{ruv}\, \epsilon_{ijk}\,
 \partial_u\, \alpha^j\, \partial_v\, \alpha^k +
  {\tilde \phi}^{-2/3}\, Q_r^{-2}\, \sum_j\,
 \partial_r\, \alpha^j\, \Pi_j\, \Pi_i\Big] -\nonumber \\
 &-& {\tilde \phi}^{-1/3}\, \sum_a\, {\bar n}_{(a)}\, Q_a^{-1}\,
 \Pi_i \Big)(\tau, \vec \sigma).
 \label{4.17}
 \eea

\subsection{The Hamiltonian Post-Minkowskian Linearization}

We now remember the results of Refs.\cite{3,4} on the Hamiltonian
Post-Minkowskian linearization of ADM tetrad gravity with dynamical
matter in the 3-orthogonal Schwinger time gauges by using the dust
as matter. The asymptotic Minkowski 4-metric at spatial infinity is
used as an {\it asymptotic background}. Post-Newtonian expansions
are avoided by introducing a ultraviolet cutoff  $M\, c^2$ for the
energy of  the matter.

\medskip

The basic assumption is that on each instantaneous 3-space
$\Sigma_{\tau}$ we have the following limitation of the
a-dimensional configurational tidal variables $R_{\bar a}$ in the
York canonical basis

\bea
 &&| R_{\bar a}(\tau ,\vec \sigma ) = R_{(1)\bar a}(\tau,
 \vec \sigma) |  = O(\zeta) << 1,\nonumber \\
 &&{}\nonumber \\
 &&|\partial_u\, R_{\bar a}(\tau ,\vec \sigma )| \sim {1\over L}
 O(\zeta),\qquad |\partial_u\, \partial_v\, R_{\bar a}(\tau ,\vec
 \sigma )| \sim {1\over {L^2}} O(\zeta),\nonumber \\
 && |\partial_{\tau}\, R_{\bar a}| = {1\over L}\, O(\zeta),\qquad
 |\partial^2_{\tau}\, R_{\bar a}| = {1\over {L^2}}\, O(\zeta),\qquad
 |\partial_{\tau}\, \partial_u\, R_{\bar a}| = {1\over {L^2}}\, O(\zeta),
 \nonumber \\
 &&{}\nonumber \\
 &&\Rightarrow\,\, Q_a(\tau, \vec \sigma) = e^{\sum_{\bar a}\,
 \gamma_{\bar aa}\, R_{\bar a}(\tau, \vec \sigma)} = 1 +
 \Gamma^{(1)}_a(\tau, \vec \sigma) + O(\zeta^2),\nonumber \\
 &&\qquad \Gamma_a^{(1)} = \sum_{\bar a}\, \gamma_{\bar aa}\,
 R_{\bar a},\qquad \sum_a\, \Gamma^{(1)}_a = 0,\qquad R_{\bar a} =
 \sum_a\, \gamma_{\bar aa}\, \Gamma_a^{(1)},
 \label{4.18}
 \eea

\noindent where $L$ is a {\it big enough characteristic length
interpretable as the reduced wavelength $\lambda / 2\pi$ of the
resulting GW's}. Therefore the tidal variables $R_{\bar a}$ are
slowly varying over the length $L$ and times $L/c$. This also
implies that the Riemann tensor ${}^4R_{ABCD}$, the Ricci tensor
${}^4R_{AB}$ and the scalar 4-curvature ${}^4R$ behave as ${1\over
{L^2}}\, O(\zeta)$. Also the intrinsic 3-curvature scalar of the
instantaneous 3-spaces $\Sigma_{\tau}$, given in Eqs.(\ref{2.5}), is
of order ${1\over {L^2}}\, O(\zeta)$. To simplify the notation we
use $R_{\bar a}$ for $R_{(1)\bar a}$ in the rest of the
paper.\medskip

For the other gravitational variables we make the following
consistent assumptions (the notation $f_{(k)} = O(\zeta^k)$ is used)

\bea
 \tilde \phi &=& 1 + 6\, \phi_{(1)} + O(\zeta^2),\nonumber \\
 N &=& 1 + n = 1 + n_{(1)} + O(\zeta^2),\nonumber \\
 {\bar n}_{(a)} &=& {\bar n}_{(1)(a)} + O(\zeta^2),\nonumber \\
 &&{}\nonumber \\
 &&\Downarrow\nonumber \\
 &&{}\nonumber \\
 {}^4g_{\tau\tau} &=& \sgn\, \Big(1 + 2\, n_{(1)}\Big) + O(\zeta^2),\nonumber \\
 {}^4g_{\tau r} &=& - \sgn\, {\bar n}_{(1)(r)} +
 O(\zeta^2),\nonumber \\
 {}^4g_{rs} &=& - \sgn\, {}^3g_{rs} = - \sgn\,
 \Big(1 + 2\, (\Gamma_r^{(1)} + 2\, \phi_{(1)})\Big)\,
 \delta_{rs} + O(\zeta^2).
 \label{4.19}
 \eea

Moreover we have (see Appendix A for the shear $\sigma_{(a)(b)}$ of
the Eulerian observers)

\bea
  {{8\pi\, G}\over {c^3}}\, \Pi_{\bar a}(\tau, \vec \sigma)\, &=&
  {{8\pi\, G}\over {c^3}}\, \Pi_{(1) \bar a}(\tau, \vec \sigma) =
 {1\over L}\, O(\zeta) \cir
 \Big[\partial_{\tau}\, R_{\bar a} - \sum_a\, \gamma_{\bar aa}\,
 \partial_a\, {\bar n}_{(1)(a)}\Big](\tau, \vec \sigma) + {1\over L}\,
 O(\zeta^2),\nonumber \\
 &&{}\nonumber \\
 &&\sigma_{(a)(a)} = \sigma_{(1)(a)(a)} = - {{8\pi\, G}\over
 {c^3}}\, \sum_{\bar a}\, \gamma_{\bar aa}\, \Pi_{(1) \bar a} +
 {1\over L}\, O(\zeta^2),\nonumber \\
 &&{}\nonumber \\
  &&\sigma_{(a)(b)}{|}_{a\not= b} = \sigma_{(1)(a)(b)}{|}_{a\not= b}
 = {1\over L}\, O(\zeta),\nonumber \\
 &&\Rightarrow\quad {{8\pi\, G}\over {c^3}}\, \pi_i^{(\theta)} =
 {1\over L}\, O(\zeta^2) = \sum_{a\not= b}\, (\Gamma^{(1)}_a -
 \Gamma^{(1)}_b)\, \epsilon_{iab}\, \sigma_{(1)(a)(b)} + {1\over
 L}\, O(\zeta^3),\nonumber \\
 &&{}\nonumber \\
  &&{}^3K = {{12\pi\, G}\over {c^3}}\, \pi_{\tilde \phi} =
 {}^3K_{(1)} = {{12\pi\, G}\over {c^3}}\, \pi_{(1) \tilde \phi} =
 {1\over L}\, O(\zeta),\nonumber \\
 &&{}\nonumber \\
 &&\Downarrow\nonumber \\
 &&{}\nonumber \\
 {}^3K_{rs} &=& {}^3K_{(1)rs} = {1\over L}\, O(\zeta) =\nonumber \\
 &=& (1 - \delta_{rs})\,
 \sigma_{(1)(r)(s)} + \delta_{rs}\, \Big[{1\over 3}\, {}^3K_{(1)} -
 \partial_{\tau}\, \Gamma_r^{(1)} + \sum_a\, (\delta_{ra} - {1\over 3})\,
 \partial_a\, {\bar n}_{(1)(a)}\Big] + {1\over L}\, O(\zeta^2).\nonumber \\
 &&{}
 \label{4.20}
 \eea

\medskip

For the matter we must have

\bea
  {\cal M}(\tau, \vec \sigma) &=& {\cal
 M}_{(1)}(\tau, \vec \sigma) + {\cal R}_{(2)}(\tau, \vec \sigma),\nonumber \\
 &&{}\nonumber \\
 && m_i = M\, O(\zeta),\qquad
 \int d^3\sigma\, {\cal M}_{(1)}(\tau, \vec
 \sigma) = Mc\, O(\zeta),\qquad \int d^3\sigma\, {\cal R}_{(2)}(\tau, \vec \sigma)
 = Mc\, O(\zeta^2),\nonumber \\
 &&{}\nonumber \\
 &&{}\nonumber \\
 {\cal M}_r(\tau, \vec \sigma) &=&
 {\cal M}_{(1)r}(\tau, \vec \sigma),\qquad
 \int d^3\sigma\, {\cal M}_{(1)r}(\tau, \vec \sigma) =
 Mc\, O(\zeta),
 \label{4.21}
 \eea

\noindent where $M$ is the ultraviolet cutoff. We have ${\cal
M}_{(1)}(\tau, \vec \sigma) = {{Mc}\over {L^3}}\, O(\zeta)$, where
the length $L$ is the wavelength of gravitational waves as shown in
Ref.\cite{3}.

\bigskip

For the dust we have $\alpha^i(\tau, \vec \sigma) = O(1)$ and that
the mass scale $m = {{\mu}\over c}$ must satisfy $m = M\, O(\zeta)$.
Therefore the dust momenta are first order quantities $\Pi_i(\tau,
\vec \sigma) = \Pi_{(1)i}(\tau, \vec \sigma)$ such that $\int
d^3\sigma\, {\cal M}_{(1)r}(\tau, \vec \sigma) = \int d^3\sigma\,
\Big(\sum_i\, \partial_r\, \alpha^i\, \Pi_i\Big)(\tau, \vec \sigma)
= Mc\, O(\zeta)$. These conditions also imply $\int d^3\sigma\,
{\cal M}_{(1)}(\tau, \vec \sigma) = Mc\, O(\zeta)$.

\bigskip

As a consequence we have the following results for the
energy-momentum tensor and for the 4-velocity of the dust (see Refs.
\cite{3,4} for the expression of the generators of the asymptotic
ADM Poincare' group)

\bea
   {\cal M} &=& {\cal M}_{(1)} + {{Mc}\over {l^3}}\, O(\zeta^2),\nonumber \\
 &&{\cal M}_{(1)} = \sqrt{\mu^2\, (J^{\tau})^2 +
 \sum_a\, (\sum_i\, \partial_a\, \alpha^i\, \Pi_i)^2},\nonumber \\
 &&{}\nonumber \\
 {\cal M}_r &=&  {\cal M}_{(1)r} = -  \partial_r\, \alpha^i\, \Pi_i
 + {{Mc}\over {L^3}}\, O(\zeta^2),\nonumber \\
 &&{}\nonumber \\
 T_{(1)}^{rs} &=& - {{\sgn\, \sum_{ij}\, \partial_r\, \alpha^i\, \Pi_i\,
 \partial_s\, \alpha^j\, \Pi_j }\over {\sqrt{\mu^2\, \Big(J^{\tau}\Big)^2
 +  \sum_c\,  (\sum_i\, \partial_c\, \alpha^i\, \Pi_i)^2 } }} + \frac{Mc}{L^3}\,O(\zeta^2),\nonumber \\
  &&{}\nonumber \\
 &&\partial_{\tau}\, {\cal M}_{(1)} + \partial_r\, {\cal M}_{(1)r} =
 0 + \frac{Mc}{L^4}\,O(\zeta^2),\nonumber \\
 &&\partial_{\tau}\, {\cal M}_{(1)r} + \partial_s\, T^{rs}_{(1)} = 0
 + \frac{Mc}{L^4}\,O(\zeta^2),\nonumber \\
  &&{}\nonumber \\
  U^{\tau} &=& {1\over {\mu\,  J^{\tau}}}\, \sqrt{\mu^2\,
 \Big(J^{\tau}\Big)^2 +  \sum_c\, (\sum_i\, \partial_c\, \alpha^i\,
 \Pi_i)^2 } + O(\zeta), \nonumber \\
  U^r &=&- {1\over {\mu\, J^{\tau}}}\, \partial_r\, \alpha^i\,
  \Pi_i + O(\zeta).
 \label{4.22}
 \eea

\bigskip

From Refs.\cite{3,4} we have the following solutions of the
linearized equations for the gravitational field

\bea
  \phi_{(1)}(\tau, \vec \sigma) \, &\cir& \Big[- {{2\pi\, G}\over
 {c^3}}\, {1\over {\triangle}}\, {\cal M}_{(1)} +
 {1\over 4}\, \sum_c\, {{\partial_c^2}\over {\triangle}}\,
 \Gamma_c^{(1)}\Big](\tau, \vec \sigma),\nonumber \\
 &&{}\nonumber \\
 n_{(1)}(\tau, \vec \sigma)\, &\cir& \Big[{{4\pi\, G}\over {c^3}}\,
 {1\over {\triangle}}\, \Big({\cal M}_{(1)} +
 \sum_a\, T_{(1)}^{aa}\Big) - {1\over {\triangle}}\, \partial_{\tau}\,
 {}^3K_{(1)} \Big](\tau, \vec \sigma),\nonumber \\
 &&{}\nonumber \\
  {\bar n}_{(1)(a)}(\tau, \vec \sigma) \, &\cir& \Big[{{\partial_a}\over
 {\triangle}}\, {}^3K_{(1)} + {{4\pi\, G}\over {c^3}}\, {1\over
 {\triangle}}\, \Big(4\, {\cal M}_{(1)a} -
 {{\partial_a}\over {\triangle}}\,\, \sum_c\,
 \partial_c\, {\cal M}_{(1)c}\Big) +\nonumber \\
 &+& {1\over 2}\, \partial_{\tau}\, {{\partial_a}\over {\triangle}}\,
 \Big(4\, \Gamma_a^{(1)} -
 \sum_c\,  {{\partial_c^2}\over {\triangle}}\, \Gamma_c^{(1)}\Big)
 \Big](\tau, \vec \sigma),\nonumber \\
 &&{}\nonumber \\
  \sigma_{(1)(a)(b)}{|}_{a \not= b}\, &\cir& {1\over 2}\,
 \Big(\partial_a\, {\bar n}_{(1)(b)} + \partial_b\, {\bar
 n}_{(1)(a)}\Big)(\tau, \vec \sigma).
 \label{4.23}
 \eea

\medskip

The retarded solution for the tidal variables (the gravitational
waves) and for the TT 3-metric are (a multipolar expansion has been
used; $M_{\bar a\bar b} = \delta_{\bar a\bar b} - \sum_a\,
\gamma_{\bar aa}\, {{\partial_a^2}\over {\triangle}}\, \Big(2\,
\gamma_{\bar ba} - {1\over 2}\, \sum_b\, \gamma_{\bar bb}\,
{{\partial_b^2}\over {\triangle}}\Big)$; $q^{uv|\tau\tau}$ is the
quadrupole; see Ref.\cite{3} for the tensors $\Lambda_{rsuv}$ and
${\cal P}_{rsuv}$)

\begin{eqnarray*}
 {}^4h^{TT}_{(1)rs}(\tau, \vec \sigma) &\cir& - \sgn\, {{4\, G}\over
 {c^3}}\,  \int d^3\sigma_1\,  d^3\sigma_2\, \sum_{uv}\,
 d^{TT}_{rsuv}({\vec \sigma}_1 - {\vec \sigma}_2)\,\,
 {{ T_{(1)}^{uv}(\tau - |\vec \sigma - {\vec \sigma}_1|, {\vec \sigma}_2)
 }\over {|\vec \sigma - {\vec \sigma}_1|}} =\nonumber \\
 &&{}\nonumber \\
 &=& - \sgn\, {{2\, G}\over {c^3}}\, \sum_{uv}\, \Lambda_{rsuv}(n)\,
 {{\partial^2_{\tau}\, q^{uv|\tau\tau}((\tau - |\vec \sigma|))}\over
 {|\vec \sigma|}} + (higher\, multipoles) + O(1/r^2),
 \end{eqnarray*}

\begin{eqnarray*}
 R_{\bar a}(\tau, \vec \sigma) &=& \sum_a\, \gamma_{\bar aa}\,
 \Gamma^{(1)}_a(\tau, \vec \sigma) \cir \qquad [\Gamma^{(1)}_a(\tau,
 \vec \sigma) = \sum_{\bar a}\, \gamma_{\bar aa}\, R_{\bar a}(\tau,
 \vec \sigma)]\nonumber \\
 &\cir& - {2\, G\over {c^2}}\, \sum_{ab}\, \gamma_{\bar aa}\, {\tilde
 M}^{-1}_{ab}(\vec \sigma)\,  \int d^3\sigma_1\, \int d^3\sigma_2\,  \sum_{uv}\,
 d^{TT}_{bbuv}({\vec \sigma}_1 - {\vec \sigma}_2)\nonumber\\
 &&{{ T_{(1)}^{uv}(\tau - |\vec \sigma - {\vec \sigma}_1|, {\vec \sigma}_2)
 }\over {|\vec \sigma - {\vec \sigma}_1|}}
 +O(\zeta^2) =\nonumber \\
 &=& - {G\over {c^3}}\, \sum_{ab}\, \gamma_{\bar aa}\, {\tilde
 M}^{-1}_{ab}(\vec \sigma)\, {{\sum_{uv}\, {\cal P}_{bbuv}\,
 \partial^2_{\tau}\, q^{uv | \tau\tau}(\tau - |\vec \sigma|)}\over
 {|\vec \sigma|}} + (higher\, multipoles) +O(1/r^2),
 \end{eqnarray*}

\bea
 q^{uv | \tau\tau}(\tau - |\vec \sigma|) &=& \int d^3\sigma_1\,
 \sigma_1^u\, \sigma_1^v\, {\cal M}_{(1)}(\tau - |\vec \sigma|,
 {\vec \sigma}_1).
 \label{4.24}
 \eea

The tidal momenta can be obtained from Eq.(\ref{4.20})

\bigskip

The linearization of Eqs.(\ref{4.17}) is \medskip

\begin{eqnarray*}
 \partial_{\tau}\, \alpha^i &=& \sum_a\, \Big(1 + n_{(1)} - 4\,
 \phi_{(1)} - 2\, \Gamma_a^{(1)}\Big)\, {{\partial_a\, \alpha^i\,
 \partial_a\, \alpha^j\, \Pi_j}\over {\sqrt{\mu^2\, (J^{\tau})^2 +
 \sum_b\, (\partial_b\, \alpha^k\, \Pi_k)^2}}} + \sum_a\,
 \partial_a\, \alpha^i\, {\bar n}_{(1)(a)},\nonumber \\
 &&{}\nonumber \\
 \partial_{\tau}\, \Pi_i\, &\cir& \sum_r\, \partial_r\, \Big({1\over
 {\sqrt{\mu^2\, (J^{\tau})^2 + \sum_b\, (\partial_b\, \alpha^k\,
 \Pi_k)^2}}}\, \Big[{1\over 2}\, \mu^2\, J^{\tau}\, \epsilon^{ruv}\,
 \epsilon_{ijk}\, \partial_u\, \alpha^j\, \partial_v\, \alpha^k +
 \sum_j\, \partial_r\, \alpha^j\, \Pi_j\, \Pi_i\Big] -\nonumber \\
 &-& \sum_a\, {\bar n}_{(1)(a)}\, \Pi_i \Big),
 \end{eqnarray*}

\bea
  \Pi_i &=& {{\mu\, J^{\tau}}\over {\sqrt{1 - \sum_b\, (\sum_j\,
  {\cal I}^{-1}_{jb}\, \partial_{\tau}\, \alpha^j)^2}}}\,
  \sum_{cj}\, {\cal I}^{-1}_{ic}\, {\cal I}^{-1}_{jc}\,
   \partial_{\tau}\, \alpha^j + O(\zeta^2), \mbox{ since}\, \mu = mc = Mc\,O(\zeta).
 \nonumber \\
 &&{}
 \label{6.14}
 \eea

By putting the solution (\ref{4.24}) for the gravitational waves
into these equations we get an integral differential equation for
the dust.

\bigskip

The HPM linerization of the Bianchi identities (\ref{eq.rho.dust})
is ($w_r = {\cal M}_{(1)r} / {\cal M}_{(1)} = O(1)$)

 \bea
  &&\left( \frac{\partial\,
 w_r}{\partial\, \tau} - \sum_s\, {\bar
 n}_{(1)(s)}\,\frac{\partial\,
 w_r}{\partial\, \sigma^s} - \sum_s\, w_s\, \frac{\partial\,
 {\bar n}_{(1)(s)}}{\partial\, \sigma^r}
 \right) -\nonumber \\
 &&- (1 + n_{(1)})\, \sum_u\, w^u\, {}\partial_u\, w_r
 - 2\, \sum_u\, (w_u)^2\, \partial_r\, \phi_{(1)} - \sum_s\,(w_s)^2\,
 \partial_r\, \Gamma^{(1)}_s +\nonumber\\
 &&{}\nonumber\\
 &&\qquad- \frac{\partial\, n_{(1)}}{\partial\, \sigma^r} + w_r\,
 \Big(\sum_u\, w^u\, \frac{\partial\, n_{(1)}}{\partial\,
 \sigma^u} + (1 + n)\, \sum_{uv}\, w^u\, w^v\, {}^3K_{(1)uv}\,\Big)
 \equiv 0,\nonumber \\
 &&{}\nonumber\\
 &&{}\nonumber\\
 &&\left[
 \frac{\partial}{\partial\, \tau} - \sum_s\,
 {\bar n}_{(1)(s)}\, \frac{\partial}{\partial\,
 \sigma^s}\right]\, \rho \equiv 0.
 \label{eq.rho.dust.HPM}
 \eea

Let us remark  that in the non relativistic limit
$U^A\mapsto(1,U^r_{(nr)})$, $w_r\mapsto-U^r_{(nr)}$ and
$\rho\mapsto\rho_{(nr)}$, where $\rho_{(nr)}$ is the non
relativistic mass density. Therefore, in the non relativistic limit,
the previous equations become the {\em non relativistic Euler's
equations} (with $n_{(1)}$ being the Newtonian gravitational
potential) and the {\em mass conservation equation}

 \bea
 && \frac{\partial\, U^r_{(nr)}}{\partial\, \tau} - \sum_s\, U^s_{(nr)}\,
 \frac{\partial\, U^r_{(nr)}}{\partial\, \sigma^s} = -
 \frac{\partial\, n_{(1)}}{\partial\, \sigma^r},\nonumber \\
 &&{}\nonumber \\
 &&\left[ \frac{\partial}{\partial\, \tau} + \sum_s\, U^s_{(nr)}\,
 \frac{\partial}{\partial\, \sigma^s}\right]\, \rho_{(nr)} = 0.
 \label{aaa}
 \eea

  \vfill\eject

\section{Kinematical and Dynamical Aspects of the Dust}

In this Section we study the congruence of the dust flux lines, its
connection with the skew congruence associated with the 3+1
splitting and the Eulerian point of view for the dust. Then we will
define the acceleration, the expansion, the shear and the vorticity
of the dust flux lines. Finally we will face the problem of how to
select the subset of the irrotational motions of the dust and we
will study the connection between the resulting dust 3-spaces and
the 3-spaces of the Eulerian observers of the 3+1 splitting.

\subsection{The Congruence of Dust Flux Lines}

As said in Section IIC there are two congruences of time-like
observers associated with each 3+1 splitting of space-time, i.e.
with every global non-inertial frame. Their unit 4-velocities are
$l^{\mu}(\tau, \vec \sigma) = \Big(z^{\mu}_A\, l^A\Big)(\tau, \vec
\sigma)$ and $v^{\mu}(\tau, \vec \sigma) = {{z^{\mu}_{\tau}}\over
{\sqrt{\sgn\, {}^4g_{\tau\tau}}}}(\tau, \vec \sigma) =
\Big(z^{\mu}_A\, v^A\Big)(\tau, \vec \sigma)$.\medskip

When the dust (or every type of perfect fluid) is present we also
have the congruence of the time-like flux lines with unit time-like
4-velocity $U^{\mu}(\tau, \vec \sigma) = z^{\mu}_A(\tau, \vec
\sigma)\, U^A(\tau, \vec \sigma)$, which in general is not
surface-forming.\medskip

The flux line of the dust emanating from ${\vec \sigma}_o$ at $\tau
= 0$ are denoted

\beq
 \zeta^{\mu}_{(0, {\vec \sigma}_o)}(\lambda) = z^{\mu}\Big(\tau(\lambda) =
 \zeta^{\tau}_{(0, {\vec \sigma}_o)}(\lambda);\,  \Sigma^r(\tau(\lambda),
 {\vec \sigma}_o) = \zeta^r_{(0, {\vec \sigma}_o)}(\lambda)\Big).
 \label{5.1}
 \eeq

\noindent Here $\lambda$ is an affine parameter, $\zeta^A_{(0, {\vec
\sigma}_o)}(\lambda) = \Big(\tau(\lambda); \vec
\Sigma(\tau(\lambda), {\vec \sigma}_o)\Big)$, $\tau(0) = 0$, $\vec
\Sigma(0, {\vec \sigma}_o) = {\vec \sigma}_o$, $\zeta^A_{(0, {\vec
\sigma}_o)}(0) = (0, {\vec \sigma}_o)$. This equation defines {\it
the 3-coordinates $ \Sigma^r(\tau, {\vec \sigma}_o)$ identifying the
location at time $\tau$ of the flux line emanating from $\sigma^r_o$
at $\tau = 0$}.\medskip

The flux lines are the integral curves of the unit time-like
4-velocity, namely they are the solutions of the equations

\bea
 &&{{d\, \zeta^A_{(0, {\vec \sigma}_o)}(\lambda)}\over {d\, \lambda}} =
 U^A\Big(\tau = \tau(\lambda) = \zeta^{\tau}_{(0, {\vec
 \sigma}_o)}(\lambda);\, \sigma^r = \Sigma^r(\tau(\lambda),
 {\vec \sigma}_o) = \zeta^r_{(0, {\vec \sigma}_o)}(\lambda)\Big).
 \nonumber \\
 &&{}
 \label{5.2}
 \eea

\bigskip

If we would impose the three gauge fixings on the gravitational
field

\bea
 U^r(\tau, \vec \sigma) &\approx& 0,\nonumber \\
 &&{}\nonumber \\
 \Downarrow && (\ref{4.11}),\nonumber \\
 &&{}\nonumber \\
 {\bar n}_{(a)}(\tau, \vec \sigma) &\approx& - {{(1 + n)\, {\tilde \phi}^{-1/3}\, Q_a^{-1}\,
 \sum_{si}\, V_{sa}\, \partial_s\, \alpha^i\, \Pi_i}\over { \sqrt{\mu^2\,
 (J^{\tau})^2 + {\tilde \phi}^{-2/3}\, \sum_{arsij}\, Q_a^{-2}\,
 V_{ra}\, V_{sa}\, \partial_r\, \alpha^i\, \partial_s\, \alpha^j\,
 \Pi_i\, \Pi_j} }}(\tau, \vec \sigma)),\nonumber \\
 &&{}\nonumber \\
 \alpha^i(\tau, \vec \sigma) &\approx& \alpha^i(\vec \sigma),\quad
 from\quad Eq.(\ref{4.14}),
 \label{5.3}
 \eea

\noindent  we would get $U^{\mu}(\tau, \vec \sigma) = v^{\mu}(\tau,
\vec \sigma)$, namely the fluid 4-velocity would coincide with the
4-velocity of the observers of the skew congruence of the 3+1
splitting, whose world-line passing through $\sigma^r_o$ is now also
a flux line with {\it constant} 3-coordinate $\sigma_o^r$

\bea
 x^{\mu}_{{\vec \sigma}_o}(\tau(\lambda)) &=& \zeta^{\mu}_{(0, {\vec
 \sigma}_o)}(\lambda) = z^{\mu}(\tau(\lambda), {\vec \sigma}_o),
 \nonumber \\
 &&{}\nonumber \\
 \Rightarrow&& \zeta^r_{(0, {\vec \sigma}_o)}(\lambda) = \Sigma^r(\tau(\lambda),
 {\vec \sigma}_o) = \sigma_o^r.
 \label{5.4}
 \eea

\noindent Since the conditions (\ref{5.3}) determine the inertial
shift functions, it means that there is a choice of 3-coordinates
$\theta^i(\tau, \vec \sigma)$ on the 3-spaces $\Sigma_{\tau}$
implying this coincidence \footnote{As said in Appendix A, the
$\tau$-preservation of the three gauge fixing for the angles
$\theta^i$ implies three equations for the determination of the
shift functions.}.\medskip

For $\tau = \lambda$ Eqs. (\ref{5.3}) are consistent with the first
half of the Hamilton equations (\ref{4.13}). In this case we have
$\Sigma^r(\tau, {\vec \sigma}_o) = \sigma_o^r$ and $\alpha^i(\tau,
\vec \sigma) = f^i(\vec \sigma)$.
\bigskip

The comoving coordinates of the gauges (\ref{5.3}) are usually used
in {\it cosmology} when the matter is dust. See for instance
Ref.\cite{17} for the dust and Refs. \cite{7,18} for more general
fluids.

\subsection{Other Aspects of the Dust Flux Lines}

As shown in Ref.\cite{11}, instead of the (comoving) Lagrangian
coordinates $\alpha^i(\tau, \vec \sigma)$ describing the fluid in
the 3-spaces $\Sigma_{\tau}$, we can use the set of  Eulerian
coordinates $\Sigma^r(\tau, {\vec \sigma}_o)$ appearing in the
definition (\ref{5.1}) of the fluid flux lines.\medskip

If we identify the affine parameter $\lambda$ with the time $\tau$
of the observer origin of the radar 4-coordinates, i.e. if we put
$\lambda = \tau$, the flux lines are described by the radar
4-coordinates $\zeta^A_{(0, {\vec \sigma}_o)}(\tau) = \Big(\tau;\,
\vec \Sigma(\tau, {\vec \sigma}_o)\Big)$.\medskip

Since Eq.(\ref{4.14}), i.e. $U^A(\tau, \vec \sigma)\, \partial_A\,
\alpha^i(\tau, \vec \sigma) = 0$, imply that the scalar fields
$\alpha^i(\tau, \vec \sigma)$ can be interpreted as {\it labels}
assigned to the flux lines (they are constant along the flux lines),
we can write

\beq
 \alpha^i\Big(\tau, \vec \Sigma(\tau, {\vec \sigma}_o)\Big) =
 \alpha^i(0, {\vec \sigma}_o)\, {\buildrel {def}\over =}\,
 \alpha^i_o({\vec \sigma}_o).
 \label{5.5}
 \eeq

 \noindent The quantities $\alpha^i_o({\vec \sigma}_o)$ are Cauchy
data on the Cauchy surface $\Sigma_{\tau = 0}$ at $\tau = 0$, whose
3-coordinates are denoted $\sigma^r_o = \Sigma^r(0, {\vec
\sigma}_o)$. If we invert the relation $\sigma^r = \Sigma^r(\tau,
{\vec \sigma}_o)$ to get $\sigma^r_o = g^r_{\vec \Sigma}(\tau,
\sigma)$, we have

\beq
 \alpha^i(\tau, \vec \sigma) = \alpha^i(0,\, {\vec g}_{\vec \Sigma}(\tau,
 \vec \sigma)) = \alpha^i_o({\vec g}_{\vec \Sigma}(\tau, \vec \sigma)).
 \label{5.6}
 \eeq

Since the conserved particle number is ${\cal N} =
\int_{V_{(\alpha)}(\tau)}\, d^3\sigma\, J^{\tau}(\alpha^i(\tau, \vec
\sigma))$, see after Eq.(\ref{3.2}), on $\Sigma_{\tau = 0}$ the
initial number density is

\beq
 {\tilde n}_o({\vec \sigma}_o) = J^{\tau}(\alpha^i(0, \vec \sigma)) =
 det\, \Big({{\partial\, \alpha^i_o({\vec \sigma}_o)}\over {\partial\,
 \sigma_o^r}}\Big).
 \label{5.7}
 \eeq

\noindent If the fluid has compact support we can take
$\alpha^i_o({\vec \sigma}_o) = \sigma_o^{r = i}$, so that ${\tilde
n}_o({\vec \sigma}_o) = 1$.

\subsection{The Eulerian Canonical Coordinates}

The flux lines of Eq.(\ref{5.1}), $\zeta^A_{(0, {\vec
\sigma}_o)}(\tau) = \Big(\tau,\, \vec \Sigma(\tau, {\vec
\sigma}_o)\Big)$ with $\lambda = \tau$, are identified by the
3-coordinates $\sigma^r = \Sigma^r(\tau, {\vec \sigma}_o)$, where
$\sigma_o^r$ are the 3-coordinates on the Cauchy surface
$\Sigma_{\tau = 0}$.\medskip

As shown in Ref.\cite{11}, there is a point canonical transformation
allowing to pass from the (comoving) Lagrangian canonical
coordinates $\alpha^i(\tau, \vec \sigma)$, $\Pi_i(\tau, \vec
\sigma)$, to a set of Eulerian canonical coordinates $\Sigma^r(\tau,
{\vec \sigma}_o)$, $K^r(\tau, {\vec \sigma}_o)$. Therefore the new
fluid coordinates coincide with the 3-coordinates $\sigma_o^r$ (and
not with $\alpha^i_o({\vec \sigma}_o)$) on the Cauchy surface
$\Sigma_{\tau = 0}$ and for $\tau > 0$ they label the flux lines
with $\Sigma^r(\tau, {\vec \sigma}_o)$ instead that with
$\alpha^i(\tau, \vec \sigma) = \alpha^i_o({\vec g}_{\vec
\Sigma}(\tau, {\vec \sigma}_o))$. The initial number density
${\tilde n}_o({\vec \sigma}_o)$ is assumed to be a known function
given as part of the Cauchy data. Also $\alpha^i_o({\vec \sigma}_o)$
is known from the Cauchy data.
\bigskip

The relation between the old coordinates and the new ones is
obtained by rewriting the relation $\alpha^i(\tau, \vec \sigma) =
\alpha^i_o({\vec g}_{\vec \Sigma}(\tau, {\vec \sigma}_o))$ in the
following form

\beq
 \alpha^i(\tau, \vec{\sigma}) = \int
 d^3\sigma_o\, \det\left(\frac{\partial
 \Sigma^u}{\partial\sigma_o^v}\right)(\tau, {\vec \sigma}_o)\,
 \delta^3(\sigma^r - \Sigma^r(\tau, \vec{\sigma_o}))\,
 \alpha^i_o(\vec{\sigma_o}).
 \label{5.8}
 \eeq

\medskip

By using $\alpha^i(\tau, \vec \Sigma(\tau, {\vec \sigma}_o)) =
\alpha^i_o({\vec \sigma}_o)$, it can be shown that the inverse of
Eq.(\ref{5.8}) is

\beq
 \Sigma^r(\tau, \vec{\sigma}_o) = \int
 d^3\sigma\, \det\left(\frac{\partial\, \alpha^i}{\partial\, \sigma^u}\right)
 (\tau, \vec \sigma)\, \delta^3(\alpha^i_o(\vec{\sigma}_o) -
 \alpha^i(\tau,\vec{\sigma}))\, \sigma^r.
 \label{5.9}
 \eeq

\medskip

The generating functional and the new momenta of the point canonical
transformation are

\bea
  \Phi\Big[\Sigma^r, \Pi_i\Big] &=& \int
 d^3\sigma\, \Pi_i(\tau, \vec{\sigma})\, \int d^3\sigma_o\,
 \det\left(\frac{\partial \Sigma^u}{\partial\sigma^v_o}\right)(\tau,
 {\vec \sigma}_o)\, \delta^3(\sigma^r - \Sigma^r(\tau, \vec{\sigma_o}))\,
 \alpha^i_o(\vec{\sigma_o}),\nonumber \\
 &&{}\nonumber \\
 &&\Downarrow\nonumber \\
 &&{}\nonumber \\
  K_r(\tau, \vec{\sigma}_o)&=&\frac{\delta\Phi}{\delta\Sigma^r(\tau, \vec{\sigma}_o)}
 =\nonumber \\
 &=& - \det\left(\frac{\partial \Sigma^u}{\partial\sigma^v_o}\right)(\tau, {\vec \sigma}_o)\,
 \frac{\partial\sigma_o^u}{\partial\Sigma^r}(\tau, {\vec \sigma}_o)\,
 \frac{\partial\alpha_o^i(\vec{\sigma}_o)}{\partial\sigma_o^u}\,
  \int d^3\sigma\, \Pi_i(\tau, \vec{\sigma})\, \delta^3(\sigma^s -
  \Sigma^s(\tau,\vec{\sigma}_o)).\nonumber \\
  &&{}
 \label{5.10}
 \eea

 \medskip

Instead for the inverse canonical transformation we have the
following generating functional

\bea
  \Phi^{\,\prime}\Big[\alpha^i, K_r\Big] &=& \int
 d^3\sigma_o\, K_r(\tau, \vec{\sigma})\, \int
 d^3\sigma\, \det\left(\frac{\partial \alpha^i}{\partial\sigma^u}\right)
 (\tau, \vec \sigma)\, \delta^3(\alpha^i_o(\vec{\sigma}_o) -
 \alpha^i(\tau, \vec{\sigma}))\, \sigma^r,\nonumber \\
 &&{}\nonumber \\
 &&\Downarrow\nonumber \\
 &&{}\nonumber \\
 \Pi_i(\tau,\vec{\sigma})&=&\frac{\delta\Phi^{\,\prime}}{\delta\alpha^i(\tau,
 \vec{\sigma})}=\nonumber \\
  &=& - \det\left(\frac{\partial \alpha}{\partial\sigma}\right)(\tau, \vec \sigma)\,
 \frac{\partial\sigma^r}{\partial\alpha^i}(\tau, \vec \sigma)\, \int
 d^3\sigma_o\, K_r(\tau, \vec{\sigma}_o)\,
 \delta^3(\alpha_o^i(\sigma_o) - \alpha(\tau,
 \vec{\sigma})).\nonumber \\
 &&{}
 \label{5.11}
 \eea

\noindent It can be checked that these expressions for the momenta
are one the inverse of the other.

\bigskip

By using the notations introduced after Eq.(\ref{3.6}), i.e. ${\cal
I}_{ir}(\tau, \vec \sigma) = \partial_r\, \alpha^i(\tau, \vec
\sigma)$, we get $\sum_r {\cal I}_{ir}(\tau, \vec{\Sigma}(\tau,
\vec{\sigma}_o)) \, \frac{\partial\Sigma^{r}(\tau, \vec{\sigma}_o)}
{\partial\sigma^{s}_o} = \frac{\partial\alpha_o^i(\vec{\sigma}_o)}
{\partial\sigma^{s}_o}$ with $\alpha^i_o({\vec \sigma}_o)$ defined
in Eq.(\ref{5.5}). Since we have $0 = {{d\, \alpha^i_o({\vec
\sigma}_o)}\over {d\tau}} = \frac{d\alpha^i(\tau, \vec{\Sigma}(\tau,
\vec{\sigma}_o))}{d\tau} = \frac{\partial\, \alpha^i(\tau,
\vec{\Sigma}(\tau, \vec{\sigma}_o))}{\partial\, \tau} + \sum_r\,
{\cal I}_{ir}(\tau, \vec{\Sigma}(\tau, \vec{\sigma}_o))\,
\frac{\partial\, \Sigma^{r}(\tau, \vec{\sigma}_o)}{\partial\,
\tau}$, we get from Eq.(\ref{3.2}) ($\partial_{\tau}\, \alpha^i = -
\sum_r\, {{J^r}\over {J^{\tau}}}\, {\cal I}_{ir}$) the following
results

\bea
  \frac{\partial\, \Sigma^{r}(\tau, \vec{\sigma}_o)} {\partial\, \tau}
 &=& - \sum_i\, {\cal
 I}_{ir}^{-1}(\tau,\vec{\Sigma}(\tau,\vec{\sigma}_o))\,
 \frac{\partial\, \alpha^i(\tau, \vec{\Sigma}(\tau,
 \vec{\sigma}_o))}{\partial\, \tau} = \frac{J^{r}(\tau,
 \vec{\Sigma}(\tau, \vec{\sigma}_o))} {J^\tau(\tau,
 \vec{\Sigma}(\tau, \vec{\sigma}_o))},\nonumber \\
 &&{}\nonumber \\
  J^\tau(\tau, \vec{\Sigma}(\tau, \vec{\sigma}_o)) &=& - \det\,
 (\partial_r\, \alpha^i) (\tau, \vec{\Sigma}(\tau, \vec{\sigma}_o)) = -
 {\det}^{-1}\left(\frac{\partial\, \Sigma^u}{\partial\, \sigma_o^v}\right)(\tau,
 {\vec \sigma}_o)\, \det\left(\frac{\partial\,
 \alpha^i_o(\vec{\sigma}_o)}{\partial\, \sigma_o^s}\right)
 =\nonumber\\
  &=& {\tilde n}_o(\vec{\sigma}_o)\, {\det}^{-1}
 \left(\frac{\partial\, \Sigma^u}{\partial\, \sigma_o^v}\right)(\tau, {\vec
 \sigma}_o),\quad \Rightarrow\quad J^{\tau}(\tau, \vec \sigma) =
  \Big(\left.{\tilde n}(\vec{\sigma}_o)\,{\det}^{-1}\left(\frac{\partial\,
 \Sigma^r}{\partial\, \sigma_o^s}\right)\Big)
 \right|_{\vec{\sigma}_o = \vec{g}_\Sigma(\tau, \vec{\sigma})},
 \nonumber \\
 &&{}\nonumber \\
  J^{r}(\tau, \vec{\Sigma}(\tau, \vec{\sigma}_o)) &=&
 {\tilde n}_o(\vec{\sigma}_o)\, {\det}^{-1}
 \left(\frac{\partial\, \Sigma^u}{\partial\, \sigma_o^v}\right)(\tau, {\vec
 \sigma}_o) \, \frac{\partial\, \Sigma^{r}(\tau, {\vec
 \sigma}_o)}{\partial\, \tau}.\nonumber \\
 &&{}
 \label{5.12}
 \eea

\bigskip

As shown in Ref.\cite{11}, these results imply the following form of
the mass density of the dust

\bea
 {\cal M}(\tau, \vec \sigma) &=&
  \Big[\left.{\det}^{-1}\left(\frac{\partial\,
 \Sigma^r(\tau, {\vec \sigma}_o)}{\partial\, \sigma_o^s}\right)\,
 \sqrt{\mu^2\,{\tilde n}_o^2(\vec{\sigma}_o)+
 {}^3g^{rs}(\tau, \vec{\Sigma}(\tau, \vec{\sigma}_o))\,
 K_r(\tau, \vec{\sigma}_o)\,K_s(\tau, \vec{\sigma}_o)}
 \Big]\right|_{\vec{\sigma}_o = \vec{g}_\Sigma(\tau, \vec{\sigma})}
 =\nonumber \\
 &=&\int d^3\sigma_o\, \delta^3(\sigma^r - \Sigma^r(\tau,
 \vec{\sigma}_o))\, \sqrt{\mu^2\, {\tilde n}_o^2(\vec{\sigma}_o)
 + {}^3g^{rs}(\tau, \vec{\Sigma}(\tau, \vec{\sigma}_o))\,
 K_r(\tau, \vec{\sigma}_o)\, K_s(\tau, \vec{\sigma}_o)}.\nonumber \\
 &&{}
 \label{5.13}
 \eea

\medskip

 Instead for the mass current density we get the following result
 valid for any kind of perfect fluid (for the dust we have ${\cal M}_r =
 - \mu\, J^{\tau}\, U_r$)

\bea
 {\cal M}_r(\tau, \vec \sigma) &=& \sum_i\, \partial_r\, \alpha^i(\tau, \vec{\sigma})\,
 \Pi_i(\tau, \vec{\sigma}) = - \Big[\left.{\det}^{-1}\left(
 \frac{\partial\, \Sigma^u(\tau, {\vec \sigma}_o}{\partial\, \sigma_o^v)}\right)\,
 K_r(\tau, \vec{\sigma}_o)\Big]\right|_{\vec{\sigma}_o = \vec{g}_\Sigma(\tau,
 \vec{\sigma})},\nonumber \\
 &&{}\nonumber \\
 \Rightarrow&& U_r(\tau, \vec \sigma) = {{K_r(\tau, {\vec g}_{\vec \Sigma}(\tau,
 \vec \sigma)}\over {\mu\, {\tilde n}_o({\vec g}_{\vec \Sigma}(\tau, \vec \sigma))}}.
 \label{5.14}
 \eea

\subsection{Acceleration, Expansion, Shear and Vorticity of the
Dust}

The covariant derivative of the dust 4-velocity $U^A(\tau, \vec
\sigma)$, whose Hamiltonian expression is given in Eqs.(\ref{4.11}),
allows to find the associated acceleration $a^A_{(U)}$, expansion
$\theta_{(U)}$, shear $\sigma_{(U)AB}$ and vorticity
$\omega_{(U)AB}$ of the dust. The analogous quantities for the
Eulerian observers, associated with the 3+1 splitting and having the
unit normal $l^A(\tau, \vec \sigma)$ to the 3-spaces
$\Sigma_{\tau}$, are given in Eqs.(\ref{a5})-(\ref{a10}) of Appendix
A.\medskip

For the dust we get

\bea
 {}^4\nabla_A\, \sgn\, U_B &=& \sgn\, U_A\, a_{(U)B} +
 \sigma_{(U)AB} + {1\over 3}\, \theta_{(U)}\, {}^3h_{(U)AB} -
 \omega_{(U)AB},\qquad {}^3h_{(U)AB} = {}^4g_{AB} - \sgn\, U_A\, U_B,
 \nonumber \\
 &&{}\nonumber \\
 a^A_{(U)} &=& U^B\, {}^4\nabla_B\, U^A \cir 0,\qquad  a_{(U)A} =
 {}^4g_{AB}\, a^A_{(U)},\qquad a^A_{(U)}\, U_A = 0,\nonumber \\
 &&{}\nonumber \\
 \theta_{(U)} &=& {}^4\nabla_A\, U^A,\nonumber \\
 &&{}\nonumber \\
 \sigma_{(U)AB} &=& \sigma_{(U)BA} = - {{\sgn}\over 2}\, (a_{(U)A}\,
 U_B + a_{(U)B}\, U_A) + {{\sgn}\over 2}\, ({}^4\nabla_A\, U_B + {}^4\nabla_B\, U_A)
 - {1\over 3}\, \theta_{(U)}\, {}^3h_{(U)AB} \cir,\nonumber \\
  &&\cir {{\sgn}\over 2}\, ({}^4\nabla_A\, U_B + {}^4\nabla_B\, U_A)
 - {1\over 3}\, \theta_{(U)}\, {}^3h_{(U)AB},
  \qquad\qquad \sigma_{(U)AB}\, U^B = 0,\nonumber \\
 &&{}\nonumber \\
 \omega_{(U)AB} &=& - \omega_{(U)BA}  = \eta_{ABCD}\, \omega^C_{(U)}\,
 U^D =\nonumber \\
 &=& - {{\sgn}\over 2}\, (a_{(U)A}\,
 U_B - a_{(U)B}\, U_A) - {{\sgn}\over 2}\, ({}^4\nabla_A\, U_B -
 {}^4\nabla_B\, U_A) \cir - {{\sgn}\over 2}\, ({}^4\nabla_A\, U_B -
 {}^4\nabla_B\, U_A),\nonumber \\
 &&\quad\omega^A_{(U)} = {1\over 2}\, \eta^{ABCD}\, \omega_{(U)BC}\, U_D,
 \qquad \omega_{(U)AB}\, U^B = 0,\qquad \omega^A_{(U)}\, U_A = 0.\nonumber \\
 &&{}
 \label{5.15}
 \eea

\bigskip

We have explicitly shown the consequences of the fact that the
equations of motion (\ref{4.16}) imply the vanishing of the
acceleration $a^A_{(U)}(\tau, \vec \sigma) \cir 0$.
\bigskip

In many contexts, especially in cosmology, one considers
irrotational dust, having $\omega_{(U)AB}(\tau, \vec \sigma) = 0$.
In this case the unit time-like 4-velocity $U^A(\tau, \vec \sigma)$
is surface forming, namely there is a 3+1 splitting of space-time
whose 3-spaces $\Sigma_{(U)\tau}$ are orthogonal to $U^A(\tau, \vec
\sigma)$. In these cases there should exist gauges of canonical
gravity in which the congruence of Eulerian observers coincides with
the dust congruence of observers.
\bigskip

However our dynamical dust has in general non-vanishing vorticity.
If we try to impose the condition $\omega_{(U)AB}(\tau, \vec \sigma)
= 0$ by asking to have $U_A(\tau, \vec \sigma) \approx l_A(\tau,
\vec \sigma) = \sgn\, \Big(1 + n(\tau, \vec \sigma)\Big)\, (1; 0)$,
we get the conditions $U_r(\tau, \vec \sigma) \approx 0$. But from
Eqs.(\ref{4.12}) this implies $\sum_i\, \partial_r\, \alpha^i(\tau,
\vec \sigma)\, \Pi_i(\tau, \vec \sigma) \approx 0$, i.e.
$\Pi_i(\tau, \vec \sigma) \approx 0$ since $J^{\tau}(\tau, \vec
\sigma) \not= 0$. Then the second of Eqs.(\ref{4.13}) gives
$\sum_r\, \partial_r\, \Big((1 + n)\, \epsilon^{ruv}\,
\epsilon_{ijk}\, \partial_u\, \alpha^j\, \partial_v\,
\alpha^k\Big)(\tau, \vec \sigma) \approx 0$, i.e. three conditions
on the lapse function implying $n = n(\tau)$.

\bigskip

To understand better this restriction, in the next Subsection we
will explore how to implement the condition $\omega_{(U)AB}(\tau,
\vec \sigma) = 0$ as a restriction on the space of solution of the
dust motion. Then, after this identification, we will explore which
gauge-fixing on canonical gravity will imply $U^A \approx l^A$ for
the irrotational motions of the dust.

\subsection{The Subset of Irrotational Dust Motions}
.

From Eqs.(\ref{5.15}) we get that the conditions
$\omega_{(U)AB}(\tau, \vec \sigma) \approx 0$  are implied by
$\omega_{(U)rs}(\tau, \vec \sigma) \approx 0$, because we have
$\omega_{(U)\tau r}(\tau, \vec \sigma) = \sum_s
\Big(\omega_{(U)rs}\, {{U^s}\over {U^{\tau}}}\Big)(\tau, \vec
\sigma) \approx 0$.\medskip

Therefore we must study $\omega_{(U)rs} \cir - {{\sgn}\over 2}\,
({}^4\nabla_r\, U_s - {}^4\nabla_s\, U_r) = - {{\sgn}\over 2}\,
(\partial_r\, U_s - \partial_s\, U_r) {\buildrel {def}\over =}\, -
{{\sgn}\over 2}\, \Omega_{(U)rs} \approx 0$ as a restriction on the
solutions for the dust motion.\medskip

Let us remark that this restriction is preserved in time by the
Hamilton equations, because Eqs.(\ref{4.12}), (\ref{4.13}) and
(\ref{4.11}) imply $\frac{\partial\, \Omega_{(U)rs}(\tau,
\vec{\sigma})}{\partial\, \tau} \cir \{ \Omega_{(U)rs}(\tau,
\vec{\sigma}),H_D \}=\left[ -\frac{\partial}{\partial\,
\sigma^r}\left(\Omega_{(U)su}\, \frac{U^u}{U^\tau}\right) +
\frac{\partial}{\partial\, \sigma^s}\left(\Omega_{(U)ru}\,
\frac{U^u}{U^\tau}\right) \right](\tau, \vec{\sigma}) \approx 0$.
Therefore it is enough to impose the condition $\Omega_{(U)rs}(0,
\vec \sigma) \approx 0$ on the Cauchy surface $\Sigma_{\tau = 0}$.
This also shows that these conditions can be interpreted as
first-class constraints to be added by hand to select the family of
irrotational motions of the fluid.
\bigskip

Let us remark that $\Omega_{(U)rs} \approx 0$ are only two
independent conditions, because we have the identity $\partial_r\,
\Omega_{(U)uv} + \partial_u\, \Omega_{(U)vr} + \partial_v\,
\Omega_{(U)ru} = 0$.\bigskip

If we make the following non-covariant decomposition of $U_r(\tau,
\vec \sigma)$ \footnote{Compare with the analogous decomposition of
the electro-magnetic vector potential in Refs. \cite{10}, \cite{3},
where it is shown that one must firstly fix the 3-coordinates on the
3-space $\Sigma_{\tau}$ and then make the non-covariant
decomposition.} ($\triangle = \sum_r\, \partial_r^2$; the function
$c(\tau)$ is required by the boundary conditions at spatial infinity
and will make the gauge fixing later introduced in Eq.(\ref{5.25})
well defined)

\bea
 U_r(\tau, \vec \sigma) &=& \partial_r\, S(\tau, \vec \sigma) +
 U_{\perp\, r}(\tau, \vec \sigma),\nonumber \\
 &&{}\nonumber \\
 &&S(\tau, \vec \sigma) = c(\tau) + \sum_r\, {{\partial_r}\over {\triangle}}\,
 U_r(\tau, \vec \sigma) = c(\tau) - {1\over {\mu}}\, \sum_{ri}\,
 {{\partial_r}\over {\triangle}}\, {{\partial_r\, \alpha^i\,
 \Pi_i}\over {J^{\tau}}}(\tau, \vec \sigma), \nonumber \\
 &&U_{\perp\, r}(\tau, \vec \sigma) = - \sum_v\, {{\partial_v}\over {\triangle}}\,
 \Omega_{(U)vr}(\tau, \vec \sigma),\qquad \sum_r\,
 \partial_r\, U_{\perp\, r}(\tau, \vec \sigma) = 0,
 \label{5.16}
 \eea

\noindent the two conditions of vanishing vorticity are

\beq
 U_{\perp\, r}(\tau, \vec \sigma) \approx 0,\qquad \Rightarrow\quad
 U_r(\tau, \vec \sigma) \approx \partial_r\, S(\tau, \vec \sigma).
 \label{5.17}
 \eeq

\medskip

These are the two first-class constraints to be added by hand to
eliminate the states of motion of the dust with non-zero vorticity.
They show that two of the three pairs of canonical variables
$\alpha^i(\tau, \vec \sigma)$, $\Pi_i(\tau, \vec \sigma)$,
describing the dust are associated with vorticity. The addition of
two suitable gauge fixing constraints (whose form is not known), so
to get two pairs of second class constraints, would allow to go to
Dirac brackets and to describe the irrotational dust with only one
pair of canonical variables.\medskip

Since $S(\tau, \sigma)$ and $U_{\perp\, r}(\tau, \vec \sigma)$ have
non-trivial Poisson brackets (i.e. $\{S(\tau, \vec{\sigma}), S(\tau,
\vec{\sigma}^{\,\prime})\} \neq 0$, $\{\Omega_{(U)rs}(\tau,
\vec{\sigma}), \Omega_{(U)uv}(\tau, \vec{\sigma}^{\,\prime})\} \neq
0$, $\{S(\tau, \vec{\sigma}), \Omega_{(U)uv}(\tau,
\vec{\sigma}^{\,\prime})\} \neq 0$), it is not easy to identify the
two gauge fixings to be added. This is not easy also in the Eulerian
point of view, where Eqs.(\ref{5.14}) and (\ref{5.12}) imply
$U_r(\tau, \vec \sigma) = {{K_r(\tau, {\vec g}_{\vec \Sigma}(\tau,
\vec \sigma))}\over {\mu\, {\tilde n}_o({\vec g}_{\vec \Sigma}(\tau,
\vec \sigma))}}$. The presence of the initial number density and of
the function ${\vec g}_{\vec \Sigma}(\tau, \vec \sigma)$ imply a
complicated expression for $U_{\perp\, r}(\tau, \vec \sigma)$, so
that the two gauge fixings will be complicated functions of
$\Sigma^r(\tau, {\vec \sigma}_o)$.
\bigskip

However it is possible to identify the reduced phase space of the
irrotational dust also without knowing the gauge fixings explicitly.
Eqs.(\ref{4.11}) imply

\bea
 &&\{U_r(\tau, \vec{\sigma}), U_s(\tau, \vec{\sigma}^{\,\prime})\} =
 -\left(\frac{\Omega_{(U)rs}}{\mu\, \det(\partial_u\,
 \alpha^j)}\right)(\tau, \vec{\sigma})\, \delta^3(\sigma^v -
 \sigma^{\,\prime\, v}),\nonumber \\
 &&{}\nonumber \\
 &&\{\det(\partial_u\, \alpha^i(\tau, \vec{\sigma})), S(\tau,
 \vec{\sigma}^{\,\prime})\}=
 - \frac{1}{\mu}\, \delta^3(\sigma^v - \sigma^{\,\prime\, v}),\nonumber \\
 &&\nonumber \\
 &&\{\det(\partial_r\, \alpha^i(\tau, \vec{\sigma})),
 \Omega_{(U)uv}(\tau, \vec{\sigma}^{\,\prime})\}= 0.
 \label{5.18}
 \eea

As a consequence we get

\bea
 &&\left.\{U_r(\tau, \vec{\sigma}), U_s(\tau, \vec{\sigma}^{\,\prime})\}
 \right|_{\Omega_{(U)rs}=0} = 0,\nonumber \\
  &&\nonumber \\
 &&\left.\{S(\tau, \vec{\sigma}), S(\tau, \vec{\sigma}^{\,\prime})\}
 \right|_{\Omega_{(U)rs}=0} = 0,\nonumber \\
 &&\nonumber \\
 &&\left.\{S(\tau, \vec{\sigma}), \Omega_{(U)uv}(\tau, \vec{\sigma}^{\,\prime})\}
 \right|_{\Omega_{(U)rs}=0} = 0.
 \label{5.19}
 \eea
\medskip

Even if we do not know the expression of the two gauge fixings, it
turns out that the reduced phase space of the irrotational dust is
spanned by the two conjugate canonical variables

\beq
 \nu(\tau, \vec{\sigma}) = \mu\, J^{\tau}(\tau, \vec \sigma) = \mu\,
 \det(\partial_r\, \alpha^i(\tau, \vec{\sigma})),\qquad
 S(\tau,\vec{\sigma})=\sum_r\,\partial_r\, U_r (\tau,
 \vec{\sigma}),
 \label{5.20}
 \eeq

\noindent because, due to Eqs.(\ref{5.18}) and (\ref{5.19}), their
Dirac brackets are

\bea
 &&\{\nu(\tau, \vec{\sigma}), \nu(\tau, \vec{\sigma}^{\,\prime})\}^* =
 \{\nu(\tau, \vec{\sigma}), \nu(\tau, \vec{\sigma}^{\,\prime})\}
 {}_{\Omega_{(U)rs}=0} = 0,\nonumber \\
 &&\{S(\tau, \vec{\sigma}), S(\tau, \vec{\sigma}^{\,\prime})\}^* =
 \{S(\tau, \vec{\sigma}), S(\tau, \vec{\sigma}^{\,\prime})\}
  {}_{\Omega_{(U)rs}=0} = 0,\nonumber \\
 &&\{\nu(\tau, \vec{\sigma}), S(\tau, \vec{\sigma}^{\,\prime})\}^* =
 \{\nu(\tau, \vec{\sigma}), S(\tau, \vec{\sigma}^{\,\prime})\}
 {}_{\Omega_{(U)rs}=0} = - \delta^3(\sigma^r - \sigma^{\,\prime\,
 r}).
 \label{5.21}
 \eea

\medskip

While $\nu(\tau, \vec \sigma)$ describes the numerical density of
the dust (${\cal N} = \frac{1}{\mu}\int d^3\sigma\,\nu(\tau,
\vec{\sigma})$), the function $S(\tau, \vec \sigma)$ allows to put
the dust unit 4-velocity in the form (${}^3g^{rs} = {\tilde
\phi}^{-2/3}\, \sum_a\, Q_a^{-2}\, V_{ra}\, V_{sa}$, $n^r = \sum_a\,
{\bar n}_{(a)}\, Q_a^{-1}\, V_{ra}$)

\bea
  U_\tau(\tau, \vec{\sigma}) &=&
 (1 + n(\tau, \vec{\sigma}))\, \sqrt{1 + {}^3g^{rs}(\tau, \vec{\sigma})\,
 \partial_r\, S(\tau, \vec{\sigma})\, \partial_s\, S(\tau, \vec{\sigma})}
 + n^r(\tau, \vec{\sigma})\, \partial_r\, S(\tau,
 \vec{\sigma}),\nonumber \\
 &&{}\nonumber \\
  U_r(\tau, \vec{\sigma}) &=& \partial_r\, S(\tau,
 \vec{\sigma}),\qquad U_A\, {}^4g^{AB}\, U_B = 1.
 \label{5.22}
 \eea

\medskip

With the substitutions $J^{\tau} = \det(\partial_r\, \alpha^i)
\mapsto \frac{1}{\mu}\,\nu$, $\partial_r\alpha^i\,\Pi_i \mapsto -
\nu\, \partial_r\, S$, the part $\int d^3\sigma\, \Big((1 + n)\,
{\cal M} + n^r\, {\cal M}_r\Big)(\tau, \vec \sigma)$ of the Dirac
Hamiltonian (\ref{4.12}) takes the form

\bea
  H_{dust}^{(R)} &=& \int d^3\sigma\, \nu(\tau, \vec{\sigma})\, \left[
 (1 + n(\tau, \vec{\sigma}))\, \sqrt{1 + {}^3g^{rs}(\tau, \vec{\sigma})\,
 \partial_r\, S(\tau, \vec{\sigma})\,\partial_s\, S(\tau, \vec{\sigma})} +
 n^r(\tau, \vec{\sigma})\, \partial_r\, S(\tau, \vec{\sigma})
 \right].\nonumber \\
 &&{}
 \label{5.23}
 \eea

\medskip

This reduced Hamiltonian gives the following Hamilton equations in
the reduced phase space of the irrotational dust

\bea
  &&\partial_\tau\, S(\tau, \vec{\sigma}) \cir \{S(\tau, \vec{\sigma}), H^{(R)}_{dust}\}^* =
 \Big((1 + n)\, \sqrt{1 + {}^3g^{rs}\, \partial_r\, S\, \partial_s\, S} + n^r\,
 \partial_r\,S\Big)(\tau, \vec{\sigma})
 = U_{\tau}(\tau, \vec{\sigma}),\nonumber \\
 &&\nonumber \\
 &&\partial_\tau\, \nu(\tau, \vec{\sigma}) \cir \{\nu, H^{(R)}_{dust}\}^* = - \Big(\partial_r\, \left[
 \nu\, \left( (1 + n)\, \frac{{}^3g^{rs}\, \partial_s\, S}
 {\sqrt{1 + {}^3g^{rs}\, \partial_r\,
 S\, \partial_s\, S}} + n^r \right)
 \right]\Big)(\tau, \vec{\sigma}),\nonumber \\
 &&{}
 \label{5.24}
 \eea

\noindent which allow to write $U_A(\tau, \vec{\sigma}) \cir
\partial_A\, S(\tau, \vec{\sigma})$ consistently with
Eqs.(\ref{5.22}). The second half of Eqs.(\ref{5.24}) allow to check
the constancy of the particle number: $\partial_\tau\, {\cal N} =
\frac{1}{\mu}\, \int d^3\sigma\, \partial_\tau\, \nu(\tau,
\vec{\sigma}) \cir 0$.
\bigskip

The result $U_A(\tau, \vec{\sigma}) \cir \partial_A\, S(\tau,
\vec{\sigma})$ shows that the unit time-like 4-velocity of the
irrotational dust is orthogonal to the 3-spaces $S(\tau, \vec
\sigma) = const.$ Moreover one can check the validity of
Eq.(\ref{4.16}) in the reduced phase space, namely that the flux
lines are geodesics.\medskip

If we add the constraint

\beq
 \chi(\tau, \vec \sigma) = S(\tau, \vec \sigma) - \tilde S(\tau)
 \approx 0,\quad i.e. \quad c(\tau) \approx \tilde S(\tau),
 \label{5.25}
 \eeq

\noindent implying $U_r(\tau, \vec \sigma) \approx 0$, $U_A(\tau,
\vec \sigma) \approx l_A(\tau, \vec \sigma)$ and $\Pi_i(\tau, \vec
\sigma) \approx  0$ for the irrotational dust in the original phase
space, its preservation in time produces the constraint

\bea
 \partial_{\tau}\, \chi(\tau, \vec
 \sigma) &\cir& \{ \chi(\tau, \vec \sigma), H^{(R)}_{dust} \}^* -
 {{d\,\tilde S(\tau)}\over {d\tau}} = U_{\tau}(\tau, \vec \sigma) -
 {{d\, S(\tau)}\over {d\tau}} \approx\nonumber \\
 &\approx& 1 + n(\tau, \vec \sigma) -
 {{d\, \tilde S(\tau)}\over {d\tau}} \approx 0,\nonumber \\
 &&{}\nonumber \\
 &&\Downarrow\qquad (\ref{5.24})\nonumber \\
 &&{}\nonumber \\
 &&\partial_{\tau}\, \nu(\tau, \vec \sigma) \approx - \sum_r\,
 \partial_r\, \Big(\nu\, n^r\Big)(\tau, \vec \sigma),\qquad n =
 n(\tau).
 \label{5.26}
 \eea

\noindent But this means that there is a gauge fixing for the
inertial gauge variable ${}^3K(\tau, \vec \sigma)$ of canonical
gravity whose $\tau$-preservation implies that the lapse is only a
function of time. As a consequence  there is a 3+1 splitting of
space-time whose 3-spaces coincide with the ones of the irrotational
dust.\medskip

If before the restriction to irrotational dust we had chosen the
3-coordinates $\theta^i(\tau, \vec \sigma)$ such that
Eqs.(\ref{5.3}) hold, i.e. $U^r(\tau, \vec \sigma) \approx 0$, then
Eqs.(\ref{5.26}) imply ${\bar n}_{(a)}(\tau, \vec \sigma) \approx 0$
and $\partial_{\tau}\, \nu(\tau, \vec \sigma) \approx 0$.\medskip

This, with the extra condition $n(\tau) = 0$, is often the starting
point in cosmology with irrotational dust and comoving 3-coordinates
\cite{7,17}. In this formulation the irrotational dust is not
described by canonical coordinates like $\alpha^i(\tau, \vec
\sigma)$, $\Pi_i(\tau, \vec \sigma)$, but only by a function
$\rho^{'}(\tau, \vec \sigma)$ such that the energy-momentum tensor
is $T^{\mu\nu} = \rho^{'}\, U^{\mu}\, U^{\nu}$. This function
satisfies the Bianchi identities $T^{\mu\nu}{}_{;\nu} = 0$, which
imply \footnote{This has the consequence that the mass inside an
averaging volume is constant $M_{\cal V}(\tau) = \int_{\cal V}\,
d^3\sigma [\rho\, \tilde \phi](\tau, \sigma) = M_{\cal V}(\tau_o)$.}
 $\partial_{\tau}\, [\rho^{'}\, \tilde \phi] = 0$ ($\tilde \phi =
\sqrt{det\, {}^3g}$) and the result $\partial_{\tau}\, \rho^{'} =
{}^3K\, \rho^{'}$.\medskip

In our approach we have the function $\bar \rho$ of
Eqs.(\ref{4.11}), whose Hamilton equations are induced by
Eqs.(\ref{4.13}) and by the Hamilton equations for the gravitational
field \cite{2}. By using the Hamiltonian Bianchi identities of
Ref.\cite{2} and by restricting to irrotational dust satisfying both
$U_r \approx 0$ and $U^r \approx 0$ (i.e. with $n \approx {\bar
n}_{(a)} \approx 0$) we recover $\partial_{\tau}\, \bar \rho =
{}^3K\, \bar \rho$.

\bigskip

Finally let us remark that by using  Eqs.(\ref{5.15}) we  could also
study: i)  dust without expansion: it requires the study of the
equation ${}^4\nabla\, U_A \approx 0$; ii) shear-free dust: one
should study the equations ${{\sgn}\over 2}\, ({}^4\nabla_A\, U_B +
{}^4\nabla_B\, U_A) \approx {1\over 3}\, \theta_{(U)}\,
{}^3h_{(U)AB}$.

\subsection{About Irrotational Motions in Arbitrary Perfect Fluids}

Let us make some remarks on the existence of irrotational motions
for arbitrary perfect fluids like the one discussed in Appendix C.
\medskip

In Ref. \cite{19} it is shown that Eqs.(\ref{5.15}) together with
the definition $\Big({}^4\nabla_A\, {}^4\nabla_B\, - {}^4\nabla_B\,
{}^4\nabla_A\Big)\, U_C = - {}^4R_{CDAB}\, U^D$ of the 4-Riemann
tensor imply the following equations for acceleration, expansion,
shear and vorticity ($[AB]$ means anti-symmetrization)

\bea
 &&{}^3h_{(U)A}{}^C\, {}^3h_{(U)B}{}^D\, \Big(U^E\, \nabla_E\,
 \omega_{(U)CD} - \,{}^4\nabla_{\big[\,C}\, a_{(U)D\,\big]}\Big) +\nonumber\\
 &&\qquad + 2\, \sigma_{(U)E\,\big[\,A}\, \omega_{(U)}^E{}_{B\,\big]}
 +\, \frac{2}{3}\, \theta_{(U)}\, \omega_{(U)AB} = 0,\nonumber \\
 &&{}\nonumber \\
 &&{}^4\nabla_{\big[\,C}\, \omega_{(U)AB\,\big]} + {}^4\nabla_{\big[\,C}\,
 a_{(U)A}\, U_{B\,\big]} + a_{(U)\big[\,A}\, \omega_{(U)BC\,\big]} =
 0.\nonumber \\
 &&{}
  \label{5.27}
  \eea

\bigskip

Let us look whether a perfect fluid with arbitrary equation of state
may admit  irrotational motions, namely motions with null vorticity

\beq
 \omega_{(U)AB} \equiv 0.
 \label{5.28}
 \eeq

\medskip

While the second of Eqs.(\ref{5.27}) is identically satisfied, the
first of Eqs.(\ref{5.27}) implies that {\it the neccessary condition
for the existence of irrotational motions} is

\beq
 {}^3h_{(U)A}{}^C\, {}^3h_{(U)B}{}^C\, {}^4\nabla_{\big[\,C}\,
 a_{(U)D\,\big]} \equiv 0.
  \label{5.29}
   \eeq

\noindent Therefore Eq.(\ref{5.29}) is a restriction on the equation
of state of the fluid, giving the pressure as a function of the
energy density. Moreover Eq.(\ref{5.29}) must be compatible with the
relativistic Euler equations (see Eqs. (3.16) and (3.17)).

\medskip

This problem was considered in Ref.\cite{18}, where it was shown
that if for a family of motions the 4-velocity of the fluid admits
the following parametrization in terms of a scalar function $S(\tau,
\vec \sigma)$

\bea
 U_A(\tau, \vec \sigma) &=& \frac{1}{h(\tau, \vec \sigma)}\,
 \nabla_A\, S(\tau, \vec \sigma),\nonumber \\
 &&{}\nonumber \\
 && h(\tau, \vec \sigma) = \sqrt{g^{AB}(\tau, \vec \sigma)\, \nabla_A\,
 S(\tau, \vec \sigma)\, \nabla_B\, S(\tau, \vec \sigma)},
 \label{5.30}
 \eea

\noindent then the acceleration turns out to depend only on the
normalization function $h(\tau, \vec \sigma)$

\bea
 a_{(U)B}(\tau, \vec \sigma) &=& U^A(\tau, \vec \sigma)\, \nabla_A\,
 U_B(\tau, \vec \sigma) =\nonumber \\
 &=& \Big(\delta_B{}^A -\, U_B(\tau, \vec \sigma)\, U^A(\tau, \vec \sigma)\Big)\,
 \nabla_A\, \ln\,h(\tau, \vec \sigma),
 \label{5.31}
 \eea

\noindent and the condition (\ref{5.29}) is satisfied. For the dust
Eqs.(\ref{5.22}) and (\ref{5.24}) show that  Eq.(\ref{5.30}) is
satisfied with $h(\tau, \vec \sigma) = 1$.

\medskip

As shown in Ref.\cite{18}  barotropic fluids, with equation of state
$\rho = \rho (p)$, admit a family of irrotational motions for which
Eqs.(\ref{5.30}) and (\ref{5.31}), and then (\ref{5.29}), are valid
(the function $h$ being proportional to $\Pi(p) = \int {{dp}\over {p
+ \rho(p)}}$).

\vfill\eject

\section{Conclusion}

Brown's formulation \cite{9} of perfect fluids allows to describe
them only in terms of three Lagrangian (comoving) coordinates in
Minkowski space-time by means of an action principle whose
Lagrangian is determined by the equation of state of the fluid. This
action was reformulated as a parametrized Minkowski theory in
Ref.\cite{10} and then studied in the rest-frame instant form of
dynamics. The main drawback of this approach is that we can get an
explicit closed form of the Hamiltonian quantities only for few
physically relevant equations of state, including the dust and the
photon gas, due to the necessity of solving a trascendental equation
(see Appendix B).\medskip

As a consequence, in this paper we studied the coupling of the dust
to ADM tetrad gravity in globally hyperbolic, asymptotically
Minkowskian space-times. We found the Hamiltonian formulation of the
gravitational field with dynamical (not test) dust as matter and we
gave the Hamilton equations of dust in the York canonical basis of
Refs.\cite{1,2}. In this way we can disentangle the inertial gauge
effects of the gravitational field from the tidal ones (the
gravitational waves in the linearized theory). Also the Hamiltonian
Post-Minkowskian linearization \cite{3,4}, avoiding the
Post-Newtonian expansion, is given in the 3-orthogonal Schwinger
time gauges.\medskip

By using radar coordinates adapted to a time-like observer we define
the Hamiltonian theory in global non-inertial frames, centered on
the observer, with well-defined instantaneous 3-spaces, dynamically
determined by Einstein's equations. In the York canonical basis the
basic inertial canonical gauge variables are: i) three angles
describing the freedom in the choice of the 3-coordinates inside the
3-spaces (their gauge fixing determines the shift functions); ii)
the York time (the trace of the extrinsic curvature of the 3-spaces)
describing the general relativistic remnant of the freedom in clock
synchronization (its gauge fixing determines the lapse
function).\medskip

In this framework we have studied the problem of selecting the
subset of the irrotational motions of the dust. For this subset
there are special 3-spaces determined by the dust 4-velocity and we
have studied the problem of which gauge fixing is needed for having
these dust 3-spaces coinciding with the 3-spaces of the global
non-inertial frame. Also the Eulerian point of view \cite{11} was
discussed in the case of dust.\medskip

Since dust is the type of matter used in cosmological models
(usually irrotational dust in comoving coordinates) and since there
are indications that at least part of {\it dark matter} can be
explained as a relativistic inertial effect induced by the York time
\cite{4,20}, the material of this paper is preparatory for the ADM
formulation of cosmology. In particular we want to use this
Hamiltonian description of dust in the framework of {\it
back-reaction} \cite{7,21}, in which an averaging procedure inside
the 3-spaces (leading to a breaking of homogeneity and isotropy)
opens the possibility to describe {\it dark energy}, and all the
associated effects as the accelerated expansion of the universe, as
an effect induced by the non-linearities of Einstein's equations.
Since the spatial average is well defined only for 3-scalar
functions and since most of the quantities describing the
gravitational field and the dust are 3-scalars in the York canonical
basis, we now have a framework for studying the spatial average of
most of the Hamilton equations. Moreover we can study which
cosmological notions have an inertial origin being functionals of
the inertial gauge variable York time. These problems will be faced
in a future paper.\medskip

Moreover we have to study the photon gas in the framework of this
paper. Also we can look for Hamiltonian expressions of the fluid
mass density, whose Lagrangian expression gives approximations to
the equations of state usually used for fluids (an inverse procedure
to the one used in Appendix B). This would open the possibility of
studying {\it compact fluid bodies} without symmetries (relevant for
the description of star) and their multipoles \cite{22,23} at the
Hamiltonian level and to see whether our approach can be useful in
formulating a well-posed Cauchy problem with a free boundary for the
body, a still unsolved problem at the mathematical level \cite{24}.

\vfill\eject

\appendix

\section{Canonical ADM Tetrad Gravity and of the York Canonical
Basis}

In this Appendix we review the formulation of canonical ADM tetrad
gravity developed In Refs.\cite{1} starting from the ADM action
considered as a functional of the cotetrads defined in
Eqs.(\ref{2.1}). Then we introduce the York canonical basis and the
expansion and the shear of the congruence of the Eulerian observers.

\subsection{The Original Canonical Variables of ADM Tetrad Gravity}

As said in ref.\cite{1,2,3} and with the notations of Eqs.
(\ref{2.1})-(\ref{2.5}), in ADM canonical tetrad gravity the 16
configuration variables are: the 3 boost variables
$\varphi_{(a)}(\tau, \vec \sigma)$; the lapse and shift functions
$n(\tau, \vec \sigma)$ and $n_{(a)}(\tau, \vec \sigma)$; the
cotriads ${}^3e_{(a)r}(\tau, \vec \sigma)$. Their conjugate momenta
are $\pi_{\varphi_{(a)}}(\tau, \vec \sigma)$, $\pi_n(\tau, \vec
\sigma)$, $\pi_{n_{(a)}}(\tau, \vec \sigma)$, ${}^3\pi^r_{(a)}(\tau,
\vec \sigma)$. There are 14 first-class constraints: A) the 10
primary constraints of Eqs.(\ref{4.4}): $\pi_{\varphi_{(a)}}(\tau,
\vec \sigma) \approx 0$, $\pi_n(\tau, \vec \sigma) \approx 0$,
$\pi_{n_{(a)}}(\tau, \vec \sigma) \approx 0$ and the 3 rotation
constraints $M_{(a)}(\tau, \vec \sigma) \approx 0$ implying the
gauge nature of the 3 Euler angles $\alpha_{(a)}(\tau, \vec
\sigma)$; B) the 4 secondary super-Hamiltonian and super-momentum
constraints ${\cal H}(\tau, \vec \sigma) \approx 0$, ${\cal
H}_{(a)}(\tau, \vec \sigma) \approx 0$ of Eqs.(\ref{4.5}). As a
consequence there are 14 gauge variables (the {\it inertial
effects}) and two pairs of canonically conjugate physical degrees of
freedom (the {\it tidal effects}, which become the gravitational
waves in the linearized theory).

\bigskip

The basis of canonical variables for this formulation of tetrad
gravity, naturally adapted to 7 of the 14 first-class constraints,
is (see Refs, \cite{1,25} for the direction-independent boundary
conditions at spatial infinity)

\bea
 &&\begin{minipage}[t]{3cm}
 \begin{tabular}{|l|l|l|l|} \hline
 $\varphi_{(a)}$ & $n$ & $n_{(a)}$ & ${}^3e_{(a)r}$ \\ \hline $
 \pi_{\varphi_{(a)}}\, \approx 0$ & $\pi_n\, \approx 0$ &
 $\pi_{n_{(a)}}\, \approx 0 $ & ${}^3{ \pi}^r_{(a)}$
 \\ \hline
 \end{tabular}
 \end{minipage}\nonumber \\
 &&{}\nonumber \\
   &&\lbrace n(\tau ,\vec \sigma ),
 \pi_n(\tau ,{\vec \sigma}^{'} ) \rbrace = \delta^3(\vec \sigma
 ,{\vec \sigma}^{'}),\nonumber \\
 &&\lbrace n_{(a)}(\tau ,\vec
 \sigma ),\pi_{n_{(b)}}(\tau ,{\vec \sigma}^{'} )\rbrace
 =\delta_{(a)(b)} \delta^3(\vec \sigma ,{\vec
 \sigma}^{'}), \nonumber \\
 &&\lbrace \varphi_{(a)}(\tau ,\vec
 \sigma ),\pi_{\varphi_{(b)}} (\tau ,{\vec \sigma}^{'} )\rbrace =
 \delta_{(a)(b)} \delta^3(\vec \sigma ,
 {\vec \sigma}^{'}),\nonumber \\
 &&\lbrace {}^3e_{(a)r}(\tau ,\vec
 \sigma ),{}^3\pi^s_{(b)}(\tau , {\vec \sigma}^{'} )\rbrace
 =\delta_{(a)(b)} \delta^s_r \delta^3(\vec \sigma , {\vec
 \sigma}^{'}).
 \label{a1}
 \eea

\subsection{The York Canonical Basis}

In Ref.\cite{1} we studied a point canonical transformation on the
canonical variables (\ref{a1}), implementing the York map and
adapted to  10 primary first-class constraints (it is a
Shanmugadhasan canonical transformation adapted also to the
constraints $M_{(a)}(\tau, \vec \sigma) \approx 0$). It is based on
the fact that the 3-metric $ {}^3g_{rs}$ is a real symmetric $3
\times 3$ matrix, which may be diagonalized with an {\it orthogonal}
matrix $V(\theta^r)$, $V^{-1} = V^T$ ($\sum_u\, V_{ua}\, V_{ub} =
\delta_{ab}$, $\sum_a\, V_{ua}\, V_{va} = \delta_{uv}$, $\sum_{uv}\,
\epsilon_{wuv}\, V_{ua}\, V_{vb} = \sum_c\, \epsilon_{abc}\,
V_{cw}$), $det\, V = 1$, depending on 3 Euler angles $\theta^r$
\footnote{Due to the positive signature of the 3-metric, we define
the matrix $V$ with the following indices: $V_{ru}$. Since the
choice of Shanmugadhasan canonical bases breaks manifest covariance,
we will use the notation $V_{ua} = \sum_v\, V_{uv}\, \delta_{v(a)}$
instead of $V_{u(a)}$. We use the following types of indices: $a =
1,2,3$ and $\bar a = 1,2$.}. The gauge Euler angles $\theta^r$ give
a description of the 3-coordinate systems on $\Sigma_{\tau}$ from a
local point of view, because they give the orientation of the
tangents to the 3 coordinate lines through each point (their
conjugate momenta are determined by the super-momentum constraints).
We only consider 3-metrics with 3 distinct positive eigenvalues. In
this way we get the following York canonical basis

\bea
 &&\begin{minipage}[t]{3cm}
\begin{tabular}{|l|l|l|l|} \hline
$\varphi_{(a)}$ & $n$ & $n_{(a)}$ & ${}^3e_{(a)r}$ \\ \hline
$\pi_{\varphi_{(a)}} \approx 0$ & $\pi_n \approx 0$ & $
\pi_{n_{(a)}} \approx 0 $ & ${}^3{ \pi}^r_{(a)}$
\\ \hline
\end{tabular}
\end{minipage} \hspace{1cm}\nonumber \\
 &&{}\nonumber \\
 &&\hspace{1cm} {\longrightarrow \hspace{.2cm}} \
\begin{minipage}[t]{4 cm}
\begin{tabular}{|ll|ll|l|l|l|} \hline
$\varphi_{(a)}$ & $\alpha_{(a)}$ & $n$ & ${\bar n}_{(a)}$ &
$\theta^r$ & $\tilde \phi$ & $R_{\bar a}$\\ \hline
$\pi_{\varphi_{(a)}} \approx0$ &
 $\pi^{(\alpha)}_{(a)} \approx 0$ & $\pi_n \approx 0$ & $\pi_{{\bar n}_{(a)}} \approx 0$
& $\pi^{(\theta )}_r$ & $\pi_{\tilde \phi}$ & $\Pi_{\bar a}$ \\
\hline
\end{tabular}
\end{minipage}\nonumber \\
 &&{}
 \label{a2}
 \eea

\bigskip

In the York canonical basis we have (from now on we will use
$V_{ra}$ for $V_{ra}(\theta^n)$ to simplify the notation; we use the
following definitions: $n_{(a)}\, {\buildrel {def}\over =}\,
\sum_b\, R_{(a)(b)}(\alpha_{(c)})\, {\bar n}_{(b)}$, ${}^3e_{(a)r}\,
{\buildrel {def}\over =}\, \sum_b\, R_{(a)(b)}(\alpha_{(c)})\,
{}^3{\bar e}_{(b)r}$, ${}^3e^r_{(a)}\, {\buildrel {def}\over =}\,
\sum_b\, R_{(a)(b)}(\alpha_{(c)})\, {}^3{\bar e}^r_{(b)}$, where
$R_{(a)(b)}(\alpha_{(c)})$ are rotation matrices, $R^T = R^{-1}$)

\begin{eqnarray*}
 {}^4g_{\tau\tau} &=& \sgn\, \Big[(1 + n)^2 - \sum_a\,
 {\bar n}_{(a)}^2\Big],\nonumber \\
 {}^4g_{\tau r} &=& - \sgn\, \sum_a\, {\bar n}_{(a)}\, {}^3{\bar e}_{(a)r} =
 - \sgn\, {\tilde \phi}^{1/3}\, \sum_a\, Q_a\,
 V_{ra}\, {\bar n}_{(a)},\nonumber \\
 {}^4g_{rs} &=& - \sgn\, {}^3g_{rs}
 = - \sgn\, {\tilde \phi}^{2/3}\, \sum_a\, Q^2_a\,
 V_{ra}\, V_{sa},\qquad
 Q_a\, =\, e^{\sum_{\bar a}^{1,2}\, \gamma_{\bar aa}\, R_{\bar a}},\nonumber \\
 &&{}\nonumber \\
 \tilde \phi &=& \phi^6 = \sqrt{\gamma} =
 \sqrt{det\, {}^3g} = {}^3\bar e,\qquad
 {}^3{\bar e}_{(a)r} = {\tilde \phi}^{1/3}\, Q_a\,
 V_{ra},\qquad {}^3{\bar e}^r_{(a)} = {\tilde \phi}^{- 1/3}\, Q^{-1}_a\,
 V_{ra},
 \end{eqnarray*}

 \begin{eqnarray*}
 {}^3\pi^r_{(a)} &=& \sum_b\,
 R_{(a)(b)}(\alpha_{(e)})\, {\bar \pi}^r_{(b)},\nonumber \\
 &&{}\nonumber \\
 {}^3{\bar \pi}^r_{(a)} &\approx& {\tilde \phi}^{-1/3}\, \Big[
 V_{ra}\, Q^{-1}_a\, (\tilde \phi\, \pi_{\tilde \phi} +  \sum_{\bar b}\,
 \gamma_{\bar ba}\, \Pi_{\bar b}) +\nonumber \\
 &+& \sum_{l}^{l \not= a}\, \sum_{twi}\, Q^{-1}_l\, {{V_{rl}\,
 \epsilon_{alt}\, V_{wt}}\over {Q_l\, Q^{-1}_a - Q_a\, Q^{-1}_l
}}\, B_{iw}\, \pi^{(\theta )}_i \Big],
 \end{eqnarray*}

\bea
 \pi^{(\theta)}_i &=& - \sum_{lmra}\, A_{ml}(\theta^n)\,
 \epsilon_{mir}\, {}^3e_{(a)l}\, {}^3{\bar \pi}^r_{(a)},\nonumber \\
  \pi_{\tilde \phi} &=&   {{c^3}\over {12\pi\, G}}\, {}^3K
  \approx {1\over {3\,\, {}^3e}}\, \sum_{ra}\, {}^3{\bar
  \pi}^r_{(a)}\,  {}^3{\bar e}_{(a)r},
  \nonumber \\
  \Pi_{\bar a} &=& \sum_{ra}\, \gamma_{\bar aa}\, {}^3{\bar
  \pi}^r_{(a)}\,  {}^3{\bar e}_{(a)r}.
 \label{a3}
 \eea

\noindent The set of numerical parameters $\gamma_{\bar aa}$
satisfies \cite{1,13} $\sum_u\, \gamma_{\bar au} = 0$, $\sum_u\,
\gamma_{\bar a u}\, \gamma_{\bar b u} = \delta_{\bar a\bar b}$,
$\sum_{\bar a}\, \gamma_{\bar au}\, \gamma_{\bar av} = \delta_{uv} -
{1\over 3}$. Each solution of these equations defines a different
York canonical basis. See Ref.\cite{1} for the SO(3) Cartan matrices
$A_{ij}(\theta^n)$ and $B = A^{-1}$.

 \bigskip

In Eq.(\ref{a3}) the quantity $ {}^3K(\tau, \vec \sigma)$ is the
trace of the extrinsic curvature ${}^3K_{rs}(\tau, \vec \sigma)$ of
the instantaneous 3-spaces $\Sigma_{\tau}$, whose expression in the
York canonical basis is

\bea
  {}^3K_{rs} &\approx&
  - {{4\pi\, G}\over {c^3}}\, {\tilde \phi}^{-1/3}\,
 \Big(\sum_a\, Q^2_a\, V_{ra}\, V_{sa}\, [2\, \sum_{\bar b}\, \gamma_{\bar ba}\,
 \Pi_{\bar b} -  \tilde \phi\, \pi_{\tilde \phi}] +\nonumber \\
 &+& \sum_{ab}\, Q_a\, Q_b\, (V_{ra}\, V_{sb} +
 V_{rb}\, V_{sa})\, \sum_{twi}\, {{\epsilon_{abt}\,
 V_{wt}\, B_{iw}\, \pi_i^{(\theta )}}\over {
 Q_b\, Q^{-1}_a  - Q_a\, Q^{-1}_b}} \Big).
 \label{a4}
 \eea

\bigskip

The {\it York internal extrinsic time} ${}^3K(\tau, \vec \sigma)$ is
the only gauge variable among the momenta: this is a reflex of the
Lorentz signature of space-time, because $\pi_{\tilde \phi}$ and
$\theta^n$ can be used as a set of 4-coordinates \cite{1}. Its
conjugate variable, to be determined by the super-hamiltonian
constraint, is $\tilde \phi = \phi^6 = {}^3\bar e$, which is
proportional to {\it Misner's internal intrinsic time}; moreover
$\tilde \phi$ is the {\it volume density} on $\Sigma_{\tau}$: $V_R =
\int_R d^3\sigma\, \phi^6$, $R \subset \Sigma_{\tau}$. Since we have
${}^3g_{rs} = {\tilde \phi}^{2/3}\, {}^3{\hat g}_{rs}$ with $det\,
{}^3{\hat g}_{rs} = 1$, $\tilde \phi$ is also called the conformal
factor of the 3-metric. The two pairs of 3-scalar canonical
variables $R_{\bar a}$, $\Pi_{\bar a}$, $\bar a = 1,2$, describe the
generalized {\it tidal effects}, namely the independent degrees of
freedom of the gravitational field. In particular the configuration
tidal variables $R_{\bar a}$ depend {\it only on the eigenvalues of
the 3-metric}. They are Dirac observables {\it only} with respect to
the Hamiltonian gauge transformations generated by 10 of the 14
first class constraints.

\medskip

Since the variables $\tilde \phi$  and $\pi_i^{(\theta )}$ are
determined by the super-Hamiltonian and super-momentum constraints,
the {\it arbitrary gauge variables} are $\alpha_{(a)}$,
$\varphi_{(a)}$, $\theta^i$, $\pi_{\tilde \phi}$, $n$ and ${\bar
n}_{(a)}$. As shown in Refs.\cite{1}, they describe the following
generalized {\it inertial effects}:

a) $\alpha_{(a)}(\tau ,\vec \sigma )$ and $\varphi_{(a)}(\tau ,\vec
\sigma )$ are the 6 configuration variables parametrizing the O(3,1)
gauge freedom in the choice of the tetrads in the tangent plane to
each point of $\Sigma_{\tau}$ and describe the arbitrariness in the
choice of a tetrad to be associated to a time-like observer, whose
world-line goes through the point $(\tau ,\vec \sigma )$. They fix
{\it the unit 4-velocity of the observer and the conventions for the
orientation of gyroscopes and their transport along the world-line
of the observer}.

b) $\theta^i(\tau ,\vec \sigma )$  describe the arbitrariness in the
choice of the 3-coordinates in the instantaneous 3-spaces
$\Sigma_{\tau}$ of the chosen non-inertial frame  centered on an
arbitrary time-like observer. Their choice will induce a pattern of
{\it relativistic inertial forces} for the gravitational field ,
whose potentials are the functions $V_{ra}(\theta^i)$ present in the
weak ADM energy $E_{ADM}$ \cite{1,2}.

c) ${\bar n}_{(a)}(\tau ,\vec \sigma )$, the shift functions
appearing in the Dirac Hamiltonian, describe which points on
different instantaneous 3-spaces have the same numerical value of
the 3-coordinates. They are the inertial potentials describing the
effects of the non-vanishing off-diagonal components ${}^4g_{\tau
r}(\tau ,\vec \sigma )$ of the 4-metric, namely they are the {\it
gravito-magnetic potentials} responsible of effects like the
dragging of inertial frames (Lens-Thirring effect) in the
post-Newtonian approximation. The shift functions are determined by
the $\tau$-preservation of the gauge fixings determining the gauge
variables $\theta^i(\tau, \vec \sigma)$.

d) $\pi_{\tilde \phi}(\tau ,\vec \sigma )$, i.e. the York time
${}^3K(\tau ,\vec \sigma )$, describes the non-dynamical
arbitrariness in the choice of the convention for the
synchronization of distant clocks which remains in the transition
from special to general relativity. The York time (being a momentum)
gives rise to a {\it semi-definite negative kinetic term} in the
Dirac Hamiltonian \cite{1,2,3}. This {\it unusual  inertial effect}
is connected to the problem of the relativistic non-dynamical
freedom in the choice of the {\it instantaneous 3-space}, which has
no non-relativistic analogue (in Galilei space-time time is absolute
and there is an absolute notion of Euclidean 3-space). Its effects
are completely unexplored.

e) $n(\tau ,\vec \sigma )$, the lapse function appearing in the
Dirac Hamiltonian, describes the arbitrariness in the choice of the
unit of proper time in each point of the simultaneity surfaces
$\Sigma_{\tau}$, namely how these surfaces are packed in the 3+1
splitting. The lapse function is determined by the
$\tau$-preservation of the gauge fixing for the gauge variable
${}^3K(\tau, \vec \sigma)$.

\subsection{The Expansion and the Shear of the Eulerian Observers.}

Let us now consider the geometrical interpretation of the extrinsic
curvature ${}^3K_{rs}$ of the instantaneous 3-spaces $\Sigma_{\tau}$
in terms of the properties of the surface-forming (i.e.
irrotational) congruence of Eulerian (non geodesic) time-like
observers, whose world-lines have the tangent unit 4-velocity equal
to the unit normal orthogonal to the instantaneous 3-spaces
$\Sigma_{\tau}$. If we use radar 4-coordinates, the covariant unit
normal $\sgn\, l_A = (1 + n)\, (1; 0)$  of Eqs.(\ref{2.4}) has the
following covariant derivative\medskip

\bea
 {}^4\nabla_A\,\, \sgn\, l_B &=& \sgn\, l_A\, {}^3a_B + \sigma_{AB} +
 {1\over 3}\, \theta\, h_{AB} - \omega_{AB} = \sgn\, l_A\, {}^3a_B
 + {}^3K_{AB},\nonumber \\
 &&{}\nonumber \\
 &&{}\nonumber \\
 {}^3K_{AB} &=& {}^3K_{rs}\, {\hat b}^r_A\, {\hat b}^s_B,\qquad
 {\hat b}^r_A = \delta^r_A + {}^3{\bar e}^r_{(a)}\, {\bar n}_{(a)}\,
 \delta^{\tau}_A,\qquad h_{AB} = {}^4g_{AB} - \sgn\, l_A\, l_B.
 \nonumber \\
 &&{}
 \label{a5}
 \eea
\medskip

The quantities appearing in Eqs.(\ref{a5}) are:\medskip

a) the {\it acceleration} of the Eulerian observers

\bea
 {}^3a^A &=& l^B\, {}^4\nabla_B\, l^A =
 {}^4g^{AB}\, {}^3a_B,\qquad {}^3a_A = {}^3a_r\,
  {\hat b}^r_A,\nonumber \\
 &&{}\nonumber \\
 &&{}^3a_r = \partial_r\, ln\, (1 + n) = {}^3a_r\,
 {}^3{\bar e}^r_{(a)}\, {}^3{\bar e}^u_{(a)}\, n_u
 = {}^3a_r\, {}^3{\bar e}^r_{(a)}\, {\bar n}_{(a)},\nonumber \\
 &&{}^3a_r = - \sgn\, {}^3{\bar e}^r_{(a)}\, {}^3{\bar e}^s_{(a)}\,
 {}^3a_s,\qquad {}^3a^{\tau} = 0;
 \label{a6}
 \eea

\medskip

b) the {\it vorticity} or {\it twist} (a measure of the rotation of
the nearby world-lines infinitesimally surrounding the given one),
which is vanishing because the congruence is surface-forming

\bea
 \omega_{AB} &=& - \omega_{BA} =  {{\sgn}\over 2}\,
 (l_A\, {}^3a_B -  l_B\, {}^3a_A) -
 {{\sgn}\over 2}\, ({}^4\nabla_A\,\, l_B -
 {}^4\nabla_B\,\, l_A) = 0,\nonumber \\
 &&{}\nonumber \\
 &&{}\nonumber \\
 &&\omega_{AB}\, l^B = 0,\qquad \omega^A = {1{\sgn}\over 2}\,
{}^4\eta^{ABCD}\, \omega_{BC}\, l_D = 0;
 \label{a7}
 \eea

 \medskip

c) the {\it expansion} \footnote{It measures the average expansion
of the infinitesimally nearby world-lines surrounding a given
world-line in the congruence.}, which coincides with the {\it York
external time}, is proportional in cosmology to the {\it Hubble
parameter} $H$ \footnote{$l$ is a representative length along the
integral curves of ${}^4{\buildrel \circ \over {\bar E}}^A_{(o)}$,
describing the volume expansion (contraction) behavior of the
congruence.} and determines the dimensionless (cosmological) {\it
deceleration parameter} $q = 3\, l^A\, {}^4\nabla_A\, {1\over
{\theta}} - 1 = - 3\, \theta^{-2}\, l^A\, \partial_A\, \theta - 1$,

\bea
 \theta &=& {}^4\nabla_A\,\, l^A
 = - \sgn\, {}^3K = - {{4\pi\, G}\over {c^3}}\,
{{{}^3{\bar e}_{(a)r}\, {}^3{\bar \pi}^r_{(a)}}\over {{}^3\bar e}}
=  - \sgn\, {{12\pi\, G}\over {c^3}}\, \pi_{\tilde \phi},\nonumber \\
 &&{}\nonumber \\
 H &=& {1\over 3}\, \theta = {1\over {l}}\, l^A\,
{}^4\nabla_A\, l =  - \sgn\, {{4\pi\, G}\over {c^3}}\, \pi_{\tilde
\phi},\qquad q = 3\, l^A\, {}^4\nabla_A\, {1\over {\theta}} - 1;
 \label{a8}
 \eea

\medskip

d) the {\it shear} \footnote{It measures how an initial sphere in
the tangent space to the given world-line, which is Lie-transported
along the world-line tangent $l^{\mu}$ (i.e. it has zero Lie
derivative with respect to $l^{\mu}\, \partial_{\mu}$), is distorted
towards an ellipsoid with principal axes given by the eigenvectors
of $\sigma^{\mu}{}_{\nu}$, with rate given by the eigenvalues of
$\sigma^{\mu}{}_{\nu}$.}

\bea
 \sigma_{AB} &=& \sigma_{BA} = - {{\sgn}\over 2}\, ({}^3a_A\,
 l_B + {}^3a_B\, l_A) + {{\sgn}\over
 2}\, ({}^4\nabla_A\, l_B + {}^4\nabla_B\, l_A) -
 {1\over 3}\, \theta\, {}^3h_{AB}
 =\nonumber \\
 &&{}\nonumber \\
 &=& ({}^3K_{rs} - {1\over 3}\, {}^3g_{rs}\, {}^3K)\,
{\hat {\bar b}}^r_A\, {\hat {\bar b}}^s_B,\qquad
 {}^4g^{AB}\, \sigma_{AB} = 0,\qquad \sigma_{AB}\,
 l^B = 0.
 \label{a9}
 \eea

\bigskip

By explicit calculation we get the following components of the shear
along the tetrads (\ref{2.2})
\medskip

\bea
  \sigma_{AB} &=& \sigma_{(\alpha )(\beta )}\, {}^4{\buildrel \circ \over
 {\bar E}}^{(\alpha )}_A\, {}^4{\buildrel \circ \over {\bar E}}^{(\beta )}_B
 = {}^4g_{AC}\, {}^4g_{BD}\, \sigma^{CD}, \qquad
  \sigma_{(\alpha )(\beta )} = \sigma_{AB}\,
 {}^4{\buildrel \circ \over {\bar E}}^A_{(\alpha
 )}\, {}^4{\buildrel \circ \over {\bar E}}^B_{(\beta )},\nonumber \\
 &&{}\nonumber \\
 \sigma_{(o)(o)} &=& 0,\qquad \sigma_{(o)(a)} = 0,\nonumber \\
 \sigma_{(a)(b)} &=& \sigma_{(b)(a)} =
 ({}^3K_{rs} - {1\over 3}\, {}^3g_{rs}\, {}^3K)\,
 {}^3{\bar e}^r_{(a)}\, {}^3{\bar e}^s_{(b)},\qquad \sum_a\,
 \sigma_{(a)(a)} = 0.
 \label{a10}
 \eea

\noindent $\sigma_{(a)(b)}$ depends upon $\theta^r$, $\tilde \phi$,
$R_{\bar a}$, $\pi^{(\theta )}_r$ and $\Pi_{\bar a}$.

\bigskip

As a consequence, by using Eqs.(\ref{a3}) and (\ref{a4}) we have

\medskip

\bea
 \tilde \phi\, \sigma_{(a)(a)} &=& - {{8\pi\, G}\over {c^3}}\,
 \sum_{\bar a}\, \gamma_{\bar aa}\, \Pi_{\bar a},\,
 \rightarrow\, \Pi_{\bar a} = - {{c^3}\over {8\pi\, G}}\, \tilde \phi\,
 \sum_a\, \gamma_{\bar aa}\, \sigma_{(a)(a)},\nonumber \\
 &&{}\nonumber \\
 \tilde \phi\, \sigma_{(a)(b)}{|}_{a \not= b} &=& - {{8\pi\, G}\over
 {c^3}}\, \sum_{tw}\, {{\epsilon_{abt}\, V_{wt}}\over
 {Q_b\, Q_a^{-1} - Q_a\, Q_b^{-1}}}\, \sum_i\, B_{iw}\,
 \pi_i^{(\theta )},\nonumber \\
 &&{}\nonumber \\
 \Rightarrow&& \pi_i^{(\theta )} =  {{c^3}\over {8\pi\, G}}\, \tilde
 \phi\, \sum_{wtab}\, A_{wi}\, V_{wt}\, Q_a\, Q_b^{-1}\, \epsilon_{tab}\,
 \sigma_{(a)(b)}{|}_{a\not= b},\nonumber \\
 &&{}\nonumber \\
 {}^3K_{rs} &=& - {{\sgn}\over 3}\, {}^3g_{rs}\, \theta +
 \sigma_{(a)(b)}\, {}^3{\bar e}_{(a)r}\, {}^3{\bar e}_{(b)s}.
 \label{a11}
 \eea

\bigskip

Therefore the diagonal elements of the shear of the Eulerian
observers describe the tidal momenta $\Pi_{\bar a}$, while the
non-diagonal elements determine the variables $\pi_i^{(\theta )}$,
determined by the super-momentum constraints. Moreover their
expansion $\theta$  is the inertial gauge variable  determining the
non-dynamical part (general relativistic gauge freedom in clock
synchronization) of the shape of the instantaneous 3-spaces
$\Sigma_{\tau}$.

\vfill\eject

\section{The Fluid Velocity in terms of the Fluid Momentum}

Let us restrict the action (\ref{3.4}) to isentropic, $s=const.$,
perfect fluids. As shown in Ref.\cite{10} the Lagrangian and the
fluid momenta have the form \footnote{In general $\mu$ is not the
chemical potential but only a parameter. The quantities $X$, $Y^r$
and ${\cal T}_{ti}$ are defined in Eq.(\ref{3.6}). We have
${{\partial X}\over {\partial N}}{|}_{N = 1 + n} = {{\sum_{rs}\,
{}^3g_{rs}\, Y^{r}\, Y^{s}}\over {(1 + n)\, X}} = {{(J^{\tau})^2 -
X^2}\over {(1 + n)\, X}}$, ${{\partial X}\over {\partial n^{u}}} = -
J^{\tau} {{\sum_s\, {}^3g_{us}\, Y^{s}}\over {(1 + n)\, X}}$.}

\bea
 \tilde n&=&{{|J|}\over {(1 + n)\, \sqrt{\gamma}}} {\buildrel {def}\over =} {X\over {\sqrt{\gamma}}},
 \qquad \rho = \rho (\tilde n) = \rho ({{|J|}\over {(1 + n)\, \sqrt{\gamma}}})
 {\buildrel {def}\over =} \mu\, f({X\over {\sqrt{\gamma}}}),\nonumber \\
 \Rightarrow&& L = - \mu \, (1 + n)\, \sqrt{\gamma} f({X\over
 {\sqrt{\gamma}}}),\nonumber \\
 &&{}\nonumber \\
 \Rightarrow&& \Pi_i(\tau ,\vec \sigma )= {{\partial\, L}\over {\partial\,
 \partial_{\tau}\, \alpha^i}}(\tau, \vec \sigma) =
 - \mu \Big[ {{\partial f(x)}\over
 {\partial x}}{|}_{x={X\over {\sqrt{\gamma}}}}
 {{Y^{r}\, {\cal T}_{ri}}\over X} \Big] (\tau,\vec \sigma ).\nonumber \\
 &&{}
 \label{b1}
  \eea

\medskip

Eq.(\ref{3.7}) becomes

\bea
 &&X^2 \Big[ \mu^2\, (J^{\tau})^2 + A^2 ({{\partial f(x)}\over {\partial
 x}}{|}_{x= {X\over {\sqrt{\gamma}}}})^{-2}\Big]
 = \mu^2 (J^{\tau})^4,\qquad
 A^2 = \sum_{rsij}\, {}^3g^{rs}\, \partial_r\, \alpha^i\, \Pi_i\, \partial_s\,
 \alpha^j\, \Pi_j,\nonumber \\
 &&{}\nonumber \\
 \Rightarrow && X = \sqrt{\gamma}\, \tilde n =
 F(\sqrt{\gamma}, A^2, (J^{\tau})^2)[\rho],
 \nonumber \\
 &&{}\nonumber \\
 {\buildrel {(\ref{3.11})}\over {\Rightarrow}} && {\cal M} =
 {{A^2\, X}\over {\mu\, (J^{\tau})^2\, {{\partial\, f(x)}\over {\partial\,
 x}}{|}_{x = {X\over {\sqrt{\gamma}}}}}} + \mu\, \sqrt{\gamma}\, f({X\over {\sqrt{\gamma}}}).
 \label{b2}
  \eea

\bigskip

Some possible equations of state for barotropic fluids, i.e. with $p
= p(\rho)$, are (in the isentropic case one gets $\rho =\rho (\tilde
n)$ by solving $p(\rho (\tilde n)) = \tilde n {{\partial \rho
(\tilde n)}\over {\partial \tilde n}} - \rho (\tilde n)$; the
definition of the sound velocity is $v_s^2 = c^2\, {{\partial\,
p(\rho)}\over {\partial\, \rho}}$):\hfill\break \hfill\break

1) $p=0$, dust: this implies

\bea
 \rho (\tilde n)&=&\mu\, \tilde n =\mu {X\over {\sqrt{\gamma}}},\quad\quad i.e.
 \qquad f({X\over {\sqrt{\gamma}}}) = {X\over {\sqrt{\gamma}}}, \qquad
{{\partial f({X\over {\sqrt{\gamma}}})}\over {\partial X}}={1\over
{\sqrt{\gamma}}}.\nonumber \\
&&{}
 \label{b3}
  \eea
\medskip

The equation for $X$ and its solution are

\bea
 X^2 && [\mu^2\, (J^{\tau})^2 + A^2] = B^2,\nonumber \\
 &&{}\nonumber \\
 X&=& { {\mu  (J^{\tau})^2}\over {\sqrt{\mu^2\, (J^{\tau})^2 +
 \sum_{rsij}\, {}^3g^{rs}\, \partial_r\, \alpha^i\, \partial_s\, \alpha^j\,
 \Pi_i\, \Pi_j}}},\nonumber \\
 &&{}\nonumber \\
 Y^r&=& -{{|J^{\tau}|\, \sum_{si}\, {}^3g^{rs}\, \partial_s\, \alpha^i\, \Pi_i}\over {
 \sqrt{\mu^2\, (J^{\tau})^2 + \sum_{uvij}\,{}^3g^{uv}\, \partial_u\,
 \alpha^i\, \partial_v\, \alpha^j\, \Pi_i\, \Pi_j   } }},\nonumber \\
 \Rightarrow&& {\cal M} = \sqrt{\mu^2\, (J^{\tau})^2 +
 \sum_{rsij}\, {}^3g^{rs}\, \partial_r\, \alpha^i\, \partial_s\, \alpha^j\,
 \Pi_i\, \Pi_j}.
 \label{b4}
  \eea
\bigskip

2) $p=k \rho (\tilde n) = \tilde n {{\partial \rho (\tilde n)}\over
{\partial \tilde n}}-\rho (\tilde n)$ ($k\not= -1$ because otherwise
$\rho = const.$, $\mu =0$). The previous differential equation for
$\rho (\tilde n)$ implies

\bea
 \rho (\tilde n) &=& \mu {\tilde n}^{k+1}=\mu ({X\over
 {\sqrt{\gamma}}})^{k+1},\quad\quad  i.e.\nonumber \\
 &&{}\nonumber \\
 f({X\over {\sqrt{\gamma}}})&=&({X\over
 {\sqrt{\gamma}}})^{k+1},\quad\quad {{\partial f({X\over
 {\sqrt{\gamma}}})}\over {\partial X}}= {{k+1}\over
 {\sqrt{\gamma}}}({X\over {\sqrt{\gamma}}})^k,
 \label{b5}
  \eea

\noindent [for $k\rightarrow 0$ we recover case 1)]. More in general
one can have $k=k(s)$: this is a non-isentropic perfect fluid with
$\rho =\rho (\tilde n,s)$.\medskip

The equation for $X$ is

\beq
 X^2\, [\mu^2\, (J^{\tau})^2 + {{A^2}\over {(k+1)^2({X\over
 {\sqrt{\gamma}}})^{2k}}}] =B^2.
 \label{b6}
  \eeq
\medskip

\noindent In general this equation cannot be solved explicitly, but
is soluble for $k = {1\over 3}, {1\over 2}, 1, 2, - {1\over 3}, -
{1\over 2}, -2$: the equation for $X$ is linear for $k = 1$,
quadratic for $k = 2, {1\over 2}$, cubic for $k = {1\over 3}, -
{1\over 2}, - 2$ and biquadratic for $k = - {1\over 3}$.
\medskip

By using Section V of Ref.\cite{10} we get the following results for
the cases $k = 1, 2, {1\over 2}, {1\over 3}$ ($A^2$ is defined in
Eq.(\ref{a2})):\bigskip

2a) $k=1$, $p = \rho$, $\rho = \mu\, {\tilde n}^2$

\bea
 X &=& {1\over {2\, \mu\, J^{\tau}}}\, \sqrt{{{4\, \mu^2}\over {\gamma}}\,
 (J^{\tau})^2 + A^2},\nonumber \\
 &&{}\nonumber \\
 {\cal M} &=& {{2\, \gamma^{-1}\, \mu^2\, (J^{\tau})^2 - (\gamma +
 {1\over 2})\, A^2}\over {2\, \mu\, \sqrt{\gamma}\, (J^{\tau})^2}}.
 \label{b7}
 \eea

\medskip

2b) $k = 2$, $p = 2\, \rho$, $\rho = \mu\, {\tilde n}^3$

\bea
 X &=& {{J^{\tau}}\over {\sqrt{2}}}\, \sqrt{1 + \sqrt{1 - {{4\,
 \gamma^2\, A^2}\over {9\, \mu^2\, (J^{\tau})^6}}}},\nonumber \\
 &&{}\nonumber \\
 {\cal M} &=&{{\mu}\over {\gamma}}\, \Big(X^2 + {{\gamma^2\, A^2}\over
 {3\, \mu^2\, (J^{\tau})^2\,\, X}}\Big).
 \label{b8}
 \eea

 \medskip

2c) $k = {1\over 2}$, $p = {1\over 2}\, \rho$, $\rho = \mu\, {\tilde
n}^{3/2}$:

\bea
 X &=&{1\over {\mu}}\, \Big(\sqrt{\mu^2\, (J^{\tau})^2 + {{\gamma\, A^4}\over
 {9\, \mu^2\, (J^{\tau})^4}}} - {{\gamma^{1/2}\, A^2}\over {3\, \mu\,
 (J^{\tau})^2}}\Big),\nonumber \\
 &&{}\nonumber \\
 {\cal M} &=&\mu\, \gamma^{-1/4}\, \sqrt{X}\, \Big(X + {{2\, \gamma^{1/2}\, A^2}\over
 {3\, \mu^2\, (J^{\tau})^2}}\Big).
 \label{b9}
 \eea

 \medskip

2d) photon gas, $k = {1\over 3}$, $p = {1\over 3}\, \rho$, $\rho =
\mu\, {\tilde n}^{4/3}$:

\bea
 X&=&|J^{\tau}| \Big[ - {{3\, A^2\, \gamma^{1/3}}\over {16\, \mu\,
 (J^{\tau})^{8/3}}} + {1\over {2^{1/3}}}\, \Big( 1 - {{27\, A^6\, \gamma}\over
 {2^{11}\, \mu^6\, (J^{\tau})^8}} + \sqrt{1 - {{27\, A^6\, \gamma}\over
 {2^{10}\, \mu^6\, (J^{\tau})^8}} }\,\, \Big)^{1/3} -\nonumber \\
 &-& {1\over {2^{1/3}}}\,
 \Big( -1 + {{27\, A^6\, \gamma}\over {2^{11}\, \mu^6\, (J^{\tau})^8}} + \sqrt{1 -
 {{27\, A^6\, \gamma}\over {2^{10}\, \mu^6\, (J^{\tau})^8}} }\,\, \Big)^{1/3}
 \Big]^{3/2},\nonumber \\
 &&{}\nonumber \\
 {\cal M} &=& \mu\, \gamma^{-1/6}\, X^{2/3}\, \Big(X^{2/3} + {{3\,
 \gamma^{1/3}\, a^2}\over {4\, \mu^2\, (J^{\tau})^2}}\Big).
 \label{b10}
  \eea

\bigskip

3) $p=k \rho^{\gamma}(\tilde n) = \tilde n {{\partial \rho (\tilde
n)}\over {\partial \tilde n}}-\rho (\tilde n)$ ($\gamma \not= 1$)
\cite{26}. It is an isentropic polytropic perfect fluid ($\gamma
=1+{1\over n}$). The differential equation for $\rho (\tilde n)$
implies [$a$ is an integration constant; the chemical potential is
$\mu ={{\partial \rho}\over {\partial \tilde n}}{|}_s$]

\bea
 \rho (\tilde n)&=&{{a \tilde n}\over {[1-k(a \tilde n)^{\gamma -1}]^{ {1\over
{\gamma -1}} } }}= {{a n}\over {[1-k(a n)^{{1\over n}}]^n
}},\quad\quad i.e.
\nonumber \\
 &&{}\nonumber \\
  f({X\over {\sqrt{\gamma}}})&=&{{X\over
{\sqrt{\gamma}}}\over {[1-k(a {X\over {\sqrt{\gamma}}})^{\gamma
-1}]^{ {1\over {\gamma -1}} } }}= {{X\over {\sqrt{\gamma}}}\over
{[1-k(a {X\over {\sqrt{\gamma}}})
^{{1\over n}}]^n }},\nonumber \\
 &&{}\nonumber \\
{{\partial f({X\over {\sqrt{\gamma}}})}\over {\partial X}}&=&{1\over
{\sqrt{\gamma}}}[1-k(a {X\over {\sqrt{\gamma}}})^{\gamma
-1}]^{-{{\gamma}\over {\gamma -1}} }={1\over {\sqrt{\gamma}}} [1-k(a
{X\over {\sqrt{\gamma}}})^{{1\over n}}]^{-(n+1)}.
 \label{b11}
  \eea

\medskip

Instead in Ref.\cite{27,28} a polytropic perfect fluid is defined by
the equation of state

\beq
 \rho (\tilde n, \tilde n\, s) = \mu\, \tilde n + {{k(s)}\over {\gamma
 - 1}}\, (\mu\, \tilde n)^{\gamma},
 \label{b9}
  \eeq

\noindent and has pressure $p=k(s) (\mu\, \tilde n)^{\gamma}=(\gamma
-1)(\rho - \mu\, \tilde n)$. The (in general non explicitly soluble)
equation for $X$ is

\beq
  X^2\, \Big[ \mu^2\, (J^{\tau})^2 + A^2 \Big( 1-k(\mu {X\over
 {\sqrt{\gamma}}})^{{1\over n}}
 \Big)^{2(n+1)}\Big] =B^2.
 \label{b13}
  \eeq

\bigskip

In Ref. \cite{10} there is also a discussion of the relativistic
ideal non-isentropic (Boltzmann) gas \cite{29} ($p= \tilde n k_B T$,
$\rho =m c^2 \tilde n \Gamma (\beta )-p$, $\mu ={{\rho +p}\over
{\tilde n}}=mc^2 \Gamma (\beta )$).

\vfill\eject


\begin{thebibliography}{}

\bibitem{1}D.Alba and L.Lusanna, {\it The York Map as a Shanmugadhasan
Canonical Transformationn in Tetrad Gravity and the Role of
Non-Inertial Frames in the Geometrical View of the Gravitational
Field}, Gen.Rel.Grav. {\bf 39}, 2149 (2007) (arXiv gr-qc/0604086,
v2).

\bibitem{2}D.Alba and L.Lusanna, {\it The Einstein-Maxwell-Particle System
in the York Canonical Basis of ADM Tetrad Gravity: I) The Equations
of Motion in Arbitrary Schwinger Time Gauges.} (arXiv 0907.4087)

\bibitem{3}D.Alba and L.Lusanna, {\it The Einstein-Maxwell-Particle System
in the York Canonical Basis of ADM Tetrad Gravity: II) The Weak
Field Approximation in the 3-Orthogonal Gauges and Hamiltonian
Post-Minkowskian Gravity: the N-Body Problem and Gravitational Waves
with Asymptotic Background.} (arXiv 1003.5143).

\bibitem{4}D.Alba and L.Lusanna, {\it III) The Post-Minkowskian N-Body Problem, its
Post-Newtonian Limit in Non-Harmonic 3-Orthogonal Gauges and Dark
Matter as an Inertial Effect. } (arXiv 1009.1794).


\bibitem{5}J.Isenberg and J.E.Marsden, {\it The York Map is a Canonical
Transformation}, J.Geom.Phys. {\bf 1}, 85 (1984).


\bibitem{6} L.Lusanna and
M.Pauri, {\it General Covariance and the Objectivity of Space-Time
Point-Events}, talk at the Oxford Conference on Spacetime Theory
(2004) (arXiv gr-qc/0503069); {\it Explaining Leibniz equivalence as
difference of non-inertial Appearances: Dis-solution of the Hole
Argument and physical individuation of point-events}, History and
Philosophy of Modern Physics {\bf 37}, 692 (2006) (arXiv
gr-qc/0604087); {\it The Physical Role of Gravitational and Gauge
Degrees of Freedom in General Relativity. I: Dynamical
Synchronization and Generalized Inertial Effects; II: Dirac versus
Bergmann Observables and the Objectivity of Space-Time},
Gen.Rel.Grav. {\bf 38}, 187 and 229 (2006) (arXiv gr-qc/0403081 and
0407007); {\it Dynamical Emergence of Instantaneous 3-Spaces in a
Class of Models of General Relativity}, to appear in the book {\it
Relativity and the Dimensionality of the World}, ed. A. van der
Merwe (Springer Series Fundamental Theories of Physics) (arXiv
gr-qc/0611045).

\bibitem{7}T.Buchert, {\it Dark Energy from Structure: a Status Report},
Gen.Rel.Grav. {\bf 40}, 467 (2008) (arXiv 0707.2153).

\bibitem{8}J.Larena, {\it Spatially Averaged Cosmology in an
Arbitrary Coordinate System}, Phys.Rev. {\bf D79}, 084006 (2009)
(arXiv 0902.3159).


\bibitem{9}J.D.Brown, {\it Action Functionals for Relativistic Perfect
Fluids}, Class.Quantum Grav. {\bf 10}, 1579 (1993).


\bibitem{10}L. Lusanna and D. Nowak-Szczepaniak, {\it  The Rest-Frame Instant Form
 of Relativistic Perfect Fluids with Equation of State $\rho = \rho (\eta, s)$
 and of Nondissipative Elastic Materials.}, Int. J. Mod. Phys. {\bf A15}, 4943
 (2000).

\bibitem{11} D.Alba and L.Lusanna, {\it Generalized Eulerian Coordinates for
Relativistic Fluids: Hamiltonian Rest-Frame Instant Form, Relative
Variables, Rotational Kinematics}  Int.J.Mod.Phys. {\bf A19}, 3025
(2004) (arXiv hep-th/020903).

\bibitem{12}D.Alba and L.Lusanna,  {\it Charged Particles
and the Electro-Magnetic Field in Non-Inertial Frames: I. Admissible
3+1 Splittings of Minkowski Spacetime and the Non-Inertial Rest
Frames},  Int.J.Geom.Methods in Physics {\bf 7}, 33 (2010) (arXiv
0908.0213) and {\it II. Applications: Rotating Frames, Sagnac
Effect, Faraday Rotation, Wrap-up Effect }, Int.J.Geom.Methods in
Physics, {\bf 7}, 185 (2010) (arXiv 0908.0215).

\bibitem{13}L.Lusanna, {\it The Rest-Frame Instant Form of Metric Gravity},
Gen.Rel.Grav. {\bf 33}, 1579 (2001)(arXiv gr-qc/0101048).



\bibitem{14}I.A.Brown, J.Behrend and K.A.Malik, {\it Gauges and
Cosmological Backreaction}, JCAP {\bf 0911}, 027 (2009) (arXiv
0903.3264).\hfill\break
 O.Umeh, J.Larena and C.Clarkson, {\it The Hubble Rate in Averaged
 Cosmology} (arXiv 1011.3959).


\bibitem{15}G.F.R.Ellis and H. van Elst, {\it Cosmological Models},
Cargese Lectures 1998, NATO Adv.Stud.Inst.Ser.C.Math.Phys.Sci. {\bf
541}, 1 (1999) (arXiv gr-qc/9812046).\hfill\break
 C.G.Tsagas, A.Challinor and R.Maartens, {\it Relativistic Cosmology
 and Large-Scale Structure}, Phys.Rep. {\bf 465}, 61 (2008) (arXiv 0705.4397).


\bibitem{16}R.Maartens, {\it Is the Universe Homogeneous?}
(arXiv 1104.1300).\hfill\break
 C.Clarkson and R.Maartens, {\it Inhomogeneity and the Foundations
 of Concordance Cosmology}, Class.Quantum Grav. {\bf 27}, 124008
 (arXiv 1005.2165).





\bibitem{17}T.Buchert, {\it On Average Properties of Inhomogeneous
Fluids in General Relativity: I. Dust Cosmologies}, Gen.Rel.Grav.
{\bf 32}, 105 (2000) (arXiv gr-qc/9906015).





\bibitem{18}T.Buchert, {\it On Average Properties of Inhomogeneous
Fluids in General Relativity: II. Perfect Fluid Cosmologies},
Gen.Rel.Grav. {\bf 32}, 105 (2000) (arXiv gr-qc/9906015).


\bibitem{19}J. Pleba\'nski and A. Krasi\'nski, {\it An Introduction to
General Relativity and Cosmology}, (Cambridge University Press,
Cambridge (UK), 2006), Cap. 15, sect.15.3.



\bibitem{20}L.Lusanna, {\it Dark Matter as a Relativistic Inertial
Effect in Einstein Canonical Gravity?}, talk at {\it Group 28: The
XXVIII International Colloquium on Group-Theoretical Methods in
Physics}, Northumbria University, Newcastle, 26-30 July 2010(arXiv
1011.4908).


\bibitem{21}R.J.van den Hoogen, {\it Averaging Spacetime: Where do we go from here ?},
(arXiv 1003.4020).

\bibitem{22}D.Alba, L.Lusanna and M.Pauri, {\it Multipolar Expansions for Closed
and Open Systems of Relativistic Particles}, J.Math.Phys. {\bf 46},
062505 (2005) (arXiv hep-th/0402181).\hfill\break
 D.Alba, L.Lusanna and M.Pauri, {\it New Directions in
Non-Relativistic and Relativistic Rotational and Multipole
Kinematics for N-Body and Continuous Systems} (2005), in {\it Atomic
and Molecular Clusters: New Research}, ed.Y.L.Ping (Nova Science,
New York, 2006) (arXiv hep-th/0505005).


\bibitem{23}W.G.Dixon, {\it Extended Objects
in General Relativity: their Description and Motion}, in {\it
Isolated Gravitating Systems in General Relativity}, Proc.Int.School
of Phys. Enrico Fermi LXVII, ed. J.Ehlers (North-Holland, Amsterdam,
1979), p. 156.\hfill\break
 W.G.Dixon, {\it Description of Extended Bodies by
Multipole Moments in Special Relativity}, J.Math.Phys. {\bf 8}, 1591
(1967).\hfill\break
 W.G.Dixon, {\it Mathisson's New Mechanics: its Aims and
Realisation}, Acta Physica Polonica B Proc.Suppl. {\bf 1}, 27
(2008).\hfill\break
 J.Ehlers and E.Rundolph, {\it Dynamics of Extended Bodies in
 General Relativity: Center-of-Mass Description and Quasi-Rigidity},
 Gen.Rel.Grav. {\bf 8}, 197 (1977).\hfill\break
 W.Beiglboeck, {\it The Center of Mass in Einstein's Theory of
 Gravitation}, Commun.Math.Phys. {\bf 5}, 106 (1967).\hfill\break
 R.Schattner, {\it The Center of Mass in General Relativity},
 Gen.Rel.Grav. {\bf 10}, 377 (1978); {\it The Uniqueness of the
 Center of Mass in General Relativity}, Gen.Rel.Grav. {\bf 10}, 395
 (1979).\hfill\break
 J.Ehlers and R.Geroch, {\it Equation of Motion of Small Bodies in
 Relativity}, Ann.Phys. {\bf 309}, 232 (2004).\hfill\break
 J.Steinhoff and D.Puetzfeld, {\it Multipolar Equations of Motion
 for Extended Test Bodies in General Relativity} (arXiv 0909.3756).




\bibitem{24}S.Dain and G.Nagy, {\it Initial Data for Fluid Bodies
in General Relativity}, Phys.Rev. {\bf D65}, 084029 (2002) (arXiv
gr-qc/0201091).






\bibitem{25}L.Lusanna and S.Russo, {\it A New Parametrization for Tetrad Gravity},
Gen.Rel.Grav. {\bf 34}, 189 (2002)(arXiv gr-qc/0102074).






\bibitem{26}M.P.Ryan jr and L.C.Shepley, {\it Homogeneous
Relativistic Cosmologies} (Princeton Univ.Press, Princeton, 1975).

\bibitem{27}L.Blanchet, T.Damour and G.Schaefer, {\it Postnewtonian
Hydrodynamics and Postnewtonian Gravitational Wave Generation for
Numerical Relativity}, Mon.Not.R.Astr.Soc. {\bf 242}, 289 (1990).

\bibitem{28}A.M.Anile, {\it Relativistic Fluids and Magnetofluids}
(Cambridge Univ.Press, Cambridge, 1989).

\bibitem{29}W.Israel, {\it Covariant Fluid Mechanics and
Thermodynamics: An Introduction}, in {\it Relativistic Fluid
Dynamics}, eds. A.Anile and Y.Choquet-Bruhat, Lecture Notes in Math.
n. 1385 (Springer, Berlin, 1989).











\end{thebibliography}
\end{document}